\documentclass[fleqn,usenatbib]{mnras}

\usepackage{newtxtext,newtxmath}

\usepackage[T1]{fontenc}

\usepackage[british]{babel}

\usepackage[inline]{enumitem}

\DeclareRobustCommand{\VAN}[3]{#2}
\let\VANthebibliography\thebibliography
\def\thebibliography{\DeclareRobustCommand{\VAN}[3]{##3}\VANthebibliography}

\usepackage{graphicx}	
\usepackage{amsmath}	
\usepackage{mathtools}
\usepackage[nameinlink]{cleveref}

\usepackage{caption}
\usepackage{subcaption}
\usepackage{bm}

\usepackage{adjustbox}
\usepackage{framed}
\usepackage{tcolorbox}

\usepackage{pifont}
\newcommand{\cmark}{\ding{51}}%
\newcommand{\xmark}{\ding{55}}%

\newcommand{\sTheta}{\bm{\upTheta}} 

\graphicspath{{Figures/}}

\title[Bayesian analysis of BFCC data - II]{A Bayesian approach to modelling spectrometer data chromaticity corrected using beam factors - II. Model priors and posterior odds}

\author[Sims et al.]{Peter H. Sims,$^{1,2,3}$\thanks{E-mail: psims3@asu.edu}
Judd D. Bowman,$^{1}$
Steven G. Murray,$^{1,4}$
John P. Barrett,$^{5}$
Rigel C. Cappallo,$^{5}$ \and
Colin J. Lonsdale,$^{5}$
Nivedita Mahesh,$^{6}$
Raul A. Monsalve,$^{1,7,8}$
Alan E. E. Rogers,$^{5}$
Titu Samson,$^{1}$ \and
and Akshatha K. Vydula$^{1}$
\\
$^{1}$School of Earth and Space Exploration, Arizona State University, Tempe, AZ 85287, USA\\
$^{2}$Astrophysics Group, Cavendish Laboratory, J. J. Thomson Avenue, Cambridge CB3 0HE, UK\\
$^{3}$Kavli Institute for Cosmology, Madingley Road, Cambridge CB3 0HA, UK\\
$^{4}$Scuola Normale Superiore (SNS), Piazza dei Cavalieri 7, I-56125 Pisa, PI, Italy\\
$^{5}$MIT Haystack Observatory, Westford, MA 01886-1299, USA \\
$^{6}$Cahill Center for Astronomy and Astrophysics, California Institute of Technology, Pasadena CA 91125, USA\\
$^{7}$Space Sciences Laboratory, University of California Berkeley,
Berkeley, CA 94720, USA \\
$^{8}$Facultad de Ingeniería, Universidad Católica de la Santísima Concepción, Alonso de Ribera 2850, Concepción, Chile
}

\date{Accepted XXX. Received YYY; in original form ZZZ}

\pubyear{2021}

\begin{document}
\label{firstpage}
\pagerange{\pageref{firstpage}--\pageref{lastpage}}
\maketitle

\begin{abstract}
The reliable detection of the global 21-cm signal, a key tracer of Cosmic Dawn and the Epoch of Reionization, requires meticulous data modelling and robust statistical frameworks for model validation and comparison. In Paper I of this series, we presented the Beam-Factor-based Chromaticity Correction (BFCC) model for spectrometer data processed using BFCC to suppress instrumentally induced spectral structure. We demonstrated that the BFCC model, with complexity calibrated by Bayes factor-based model comparison (BFBMC), enables unbiased recovery of a 21-cm signal consistent with the one reported by EDGES from simulated data. Here, we extend the evaluation of the BFCC model to lower amplitude 21-cm signal scenarios where deriving reliable conclusions about a model's capacity to recover unbiased 21-cm signal estimates using BFBMC is more challenging. Using realistic simulations of chromaticity-corrected EDGES-low spectrometer data, we evaluate three signal amplitude regimes -- null, moderate, and high. We then conduct a Bayesian comparison between the BFCC model and three alternative models previously applied to 21-cm signal estimation from EDGES data. To mitigate biases introduced by systematics in the 21-cm signal model fit, we incorporate the Bayesian Null-Test-Evidence-Ratio (BaNTER) validation framework and implement a Bayesian inference workflow based on posterior odds of the validated models. We demonstrate that, unlike BFBMC alone, this approach consistently recovers 21-cm signal estimates that align with the true signal across all amplitude regimes, advancing robust global 21-cm signal detection methodologies.
\end{abstract}



\section{Introduction}
\label{Sec:Intro}

Global 21-cm experiments operating in the frequency range $10 \lesssim \nu \lesssim 230~\mathrm{MHz}$, corresponding to a redshift range\footnote{The 21-cm hyperfine line of neutral hydrogen has a rest-frame frequency of $\nu_{21} \approx 1420.4~\mathrm{MHz}$. Due to the expansion of the Universe, the wavelength of radiation is stretched, which reduces its frequency and establishes a one-to-one mapping between the observation frequency, $\nu_\mathrm{obs}$, and the redshift, $z$, at which the 21-cm line is emitted: $\nu_\mathrm{obs} = \nu_{21}/(1+z)$.} $150 \lesssim z \lesssim 5$, aim to provide conclusive and unbiased measurements of the sky-averaged redshifted 21-cm hyperfine line radiation emitted by neutral hydrogen in the high-redshift Universe. Observations of the 21-cm signal during the Universe's Dark Ages have the potential to provide precision cosmological constraints (e.g. \citealt{2023NatAs...7.1025M,2024MNRAS.527.1461M, 2024MNRAS.529..519G, 2025arXiv250102538N}), to probe directly the initial stages of structure formation, and to characterise the properties of the first stars, proto-galaxies, and accreting black holes during Cosmic Dawn (CD) and the Epoch of Reionisation (e.g. \citealt{2024MNRAS.527..813B, 2024MNRAS.531.1113P, 2024arXiv241108134C, 2025arXiv250218098G}). However, to achieve this, the cosmological 21-cm signal must be extracted from spectrometer data containing astrophysical foreground emission, dominated by synchrotron radiation, that, depending on the frequency range and field observed, is 3--6 orders of magnitude brighter.

The foreground emission is intrinsically spectrally smooth and thus separable from the more spectrally structured global 21-cm signal. However, even low-level spectral structure in the instrumental transfer function mixes the foreground emission into the narrower spectral scales relevant for 21-cm signal detection complicating this separation in the measured data. For such structure to be treated as negligible requires that the instrument transfer function is smooth at the approximately\footnote{The sky-averaged brightness temperature observed by the EDGES~2 instrument, when the Galaxy is low in the beam, is approximately $5000~\mathrm{K}$ at $50~\mathrm{MHz}$ \citep[e.g.][]{2018Natur.555...67B}. For a noise level of $\sim 20~\mathrm{mK}$ -- consistent with that reported in \citet{2018Natur.555...67B} -- this corresponds to a foreground-to-noise ratio of $2.5 \times 10^{5}$.}  $10^{5}:1$ foreground-to-noise ratio in the data, which is challenging to achieve in practice. Consequently, accurately accounting for spectral structure induced by instrumental chromaticity on scales relevant for 21-cm signal detection, remains one of the most significant challenges to achieving robust signal estimates. A primary contributor to this effect is the frequency-dependent weighting of the sky by the instrument beam. The Experiment to Detect the Global Epoch of Reionisation Signature (EDGES; \citealt{2018Natur.555...67B}, hereafter B18) has pioneered a beam-factor-based Chromaticity Correction (BFCC) approach to mitigate this effect by dividing the calibrated spectrometer data by an estimated sky-weighted beam response.

If BFCC perfectly removed instrumental chromaticity, unbiased recovery of the global 21-cm signal would be possible by directly fitting an intrinsic astrophysical model to the corrected data. However, in Paper I of this series (\citealt{2023MNRAS.521.3273S}), it was demonstrated that while BFCC reduces the impact of instrumental chromaticity under realistic assumptions about the spectral structure of foregrounds, the correction is only partial, leaving residual spectral systematics that must be carefully modelled to avoid biases in signal estimation. To account for residual instrumental structure in BFCC data, we derived a flexible closed-form model for beam-factor chromaticity-corrected spectrometer data (hereafter the BFCC model) and demonstrated how to optimise the model's complexity for a given dataset using Bayes factor-based model comparison (BFBMC). Using realistic simulations of time-averaged EDGES data, we showed that when embedding a simulated 21-cm signal consistent with the deep ($A=500^{+500}_{-200}~\mathrm{mK}$) best fitting absorption trough reported in B18, fitting the data with an intrinsic sky model -- one that describes the spectrum in the absence of chromatic effects (hereafter, the Intrinsic model) -- yields biased estimates of the underlying 21-cm signal. In contrast, the BFCC model enables unbiased recovery of the simulated global 21-cm signal.

While the results of Paper I are encouraging, the 21-cm signal reported in B18 is deeper than expected at redshift 17 under standard cosmological assumptions (e.g. B18; \citealt{2018Natur.555...71B}). Additionally, reanalyses of the data in B18, using alternate data models, have since been carried out that suggest either a lower amplitude 21-cm signal than reported in B18 or the absence of a detectable 21-cm signal altogether (e.g., \citealt{2018Natur.564E..32H, 2019ApJ...874..153B, 2019ApJ...880...26S, 2020MNRAS.492...22S, 2021MNRAS.502.4405B,2024arXiv241108134C}). Accounting for this possibility, in this work, we extend our evaluation of the BFCC model to data sets with lower amplitude 21-cm signal and additional comparison models for which it is more challenging to draw reliable conclusions about the model's ability to recover unbiased estimates of the 21-cm signal using BFBMC alone. Specifically, we consider two additional scenarios for the amplitude of the 21-cm signal in the data:
\begin{enumerate}
    \item a null signal, which we use to validate the models within the Bayesian Null-Test-Evidence-Ratio (BaNTER) validation framework (\citealt{2025arXiv250214029S}; hereafter, S25),
    \item a signal with a lower amplitude compared to that reported in B18, consistent with expectations for the 21-cm absorption trough associated with CD at redshift 17 under the standard cosmological assumption that the background brightness temperature at CD is dominated by the Cosmic Microwave Background (CMB) and the minimum hydrogen gas temperature is determined by adiabatic cooling after decoupling from the radiation temperature.
\end{enumerate}
Furthermore, we analyse the data using two additional comparison models for the non-21-cm component of the publicly available EDGES data. This nuisance component may, itself, be composed of multiple subcomponents describing, for example, astrophysical foregrounds, ionospheric effects and instrumental systematics. However, for brevity, here we refer to the non-21-cm component as the `foreground' model with the understanding that they are intended to describe the full range of non-21-cm structure in the data. These models are the linearised physical model (hereafter, 'LinPhys' model) and the more general polynomial foreground model (hereafter, 'MultLin' model) used in the analysis carried out in B18 (see \Cref{Sec:DataModels} for details).

When analysing simulated data, preferred models can be directly evaluated by comparing the input and inferred signal of interest. However, when analysing instrumental data where the detailed structure of the signal is a priori unknown, robust inference requires accounting for both the uncertainty in parameter estimates within models and the uncertainty in the model itself. Ignoring model uncertainty can lead to underestimated uncertainties in inferences and biased conclusions due to improperly weighted model-averaged parameter posteriors. Bayesian model comparison offers a unified and statistically consistent framework to address both sources of uncertainty (e.g., \citealt{1935PCPS...31..203J, 1939thpr.book.....J} and \citealt{KandR}; hereafter, KR95).

In the context of 21-cm cosmology, Bayes-factor-based model comparison distinguishes between models that offer a compact explanation of the data (i.e., a good fit using relatively few effective degrees of freedom) and those that do not. However, it does not distinguish between:
\begin{enumerate}
    \item models in which the 21-cm signal component and the nuisance component (e.g., foregrounds, ionosphere, and instrument systematics) are each accurate, enabling unbiased recovery of the 21-cm signal, and
    \item models in which the nuisance component is inaccurate, but its deficiencies are absorbed by the 21-cm signal model, still yielding a high Bayesian evidence fit -- albeit with a biased signal estimate.
\end{enumerate}

Here, we refer to the 21-cm signal component as the \textit{model of interest}, and to the combination of all other components (e.g., astrophysical foregrounds, ionospheric effects, and residual instrumental systematics) as the \textit{nuisance model}. While the nuisance model may itself have multiple subcomponents, we are concerned only with the possibility that inaccuracies in the nuisance model could correlate with the 21-cm signal model in a way that biases its inference.

Scenario (ii), above, can occur when a composite model has a nuisance model that fails to fully describe the nuisance signal component of the data, but is paired with a 21-cm model capable of fitting the sum of the true 21-cm signal \textit{and} the residual systematics resulting from inaccuracies in the nuisance model. In this case, the composite model can provide a good fit to the data, but the 21-cm signal estimates it yields will be biased by the systematic residuals.

This situation introduces a form of \textit{model-level degeneracy} (hereafter, \textit{model degeneracy}), analogous to parameter degeneracy in conventional parameter estimation, but arising at the level of composite models. In such cases, different combinations of component models may provide similarly good fits to the data in aggregate, even though only some combinations yield accurate and unbiased recovery of the 21-cm signal.

To account for this phenomenon, two categories of composite model comparison were defined in S25. Let the full set of models under consideration be denoted $\bm{\mathcal{M}} = \{ \bm{\mathcal{M}}_\mathrm{inac}, \bm{\mathcal{M}}_\mathrm{ac} \}$, where $\bm{\mathcal{M}}_\mathrm{inac}$ contains models that do not provide accurate or predictive fits to the data, and $\bm{\mathcal{M}}_\mathrm{ac}$ contains models that do. The key distinction between the two categories lies in the nature of the accurate models in $\bm{\mathcal{M}}_\mathrm{ac}$. In \textit{category I} model comparison, all accurate models are only capable of fitting the data when their component sub-models are also accurate - no component can compensate for the inaccuracies of another. In this case, BFBMC is sufficient to distinguish between models that recover the true component signals and those that do not. Conversely, in \textit{category II} model comparison, $\bm{\mathcal{M}}_\mathrm{ac}$ includes models that achieve high evidence fits by absorbing the inaccuracies of one component into another (e.g. a 21-cm signal component fitting residual foreground structure). In such cases, BFBMC alone leads to biased model-averaged inferences, and model validation becomes essential.

To address this issue, S25 introduced the BaNTER validation framework, which uses a Bayesian null test to derive model priors through comparison of a single component model and a composite model for single-component validation data set. This test is designed to identify composite models that, while fitting the data well overall, yield biased signal inferences, enabling these poorly performing models to be downweighted or excluded a priori. Combining the BaNTER validation framework with Bayesian model comparison for observational data allows one to derive the model-validated posterior odds. This approach enables selection for models that are accurate and predictive of the data in aggregate and, crucially for unbiased 21-cm signal recovery, are composed of accurate and predictive component models.

In this work, we apply the BaNTER validation framework to derive model priors and use the model-validated posterior odds to compare the BFCC model with the Intrinsic, LinPhys, and MultLin models. We analyse realistic simulations of BFCC EDGES-low spectrometer data across null, moderate, and high amplitude 21-cm signals and demonstrate that this approach:
\begin{enumerate}
    \item favours the BFCC model over the alternatives, and
    \item in contrast to using BFBMC-alone, reliably recovers 21-cm signal estimates consistent with the true signal in the data across all signal-amplitude regimes.
\end{enumerate}

The remainder of the paper is organised as follows. In \Cref{Sec:BayesianInferenceAndModelValidtionFramework}, we describe the Bayesian inference and BaNTER validation frameworks used to analyse the data and compare models. \Cref{Sec:RealisitcBFCCSimulations} summarises the simulations of BFCC EDGES-low data developed in Paper I and updates to the 21-cm signals included in the simulations in this work. In \Cref{Sec:DataModels}, we describe the models we fit to the data. \Cref{Sec:Results} presents the analysis results of the simulated data sets and compares Bayesian model selection based on BFBMC-alone and the BaNTER-validated posterior-odds as methods for identifying preferred models for recovery of unbiased estimates of the 21-cm signal. In \Cref{Sec:Discussion}, we discuss the performance of the BFCC model relative to the Intrinsic, LinPhys, and MultLin models under different signal amplitude assumptions and emphasise the importance of model validation for deriving reliable inferences from the comparison of composite models for global 21-cm signal data. Finally, in \Cref{Sec:Conclusions}, we summarise our findings and present our conclusions.

\section{Bayesian inference and model validation framework}
\label{Sec:BayesianInferenceAndModelValidtionFramework}

\subsection{Bayesian inference}
\label{Sec:BayesianInference}

\subsubsection{Bayes' theorem}
\label{Sec:BayesTheorem}

Bayesian inference provides a consistent approach to estimate a set of parameters, $\sTheta$, from a model, $\bm{M}$, given a set of data, $\bm{D}$. Using Bayes' theorem we can write the posterior probability density of the parameters of the model as:
\begin{equation}
\label{Eq:BayesEqn}
\mathcal{P}(\sTheta\vert\bm{D},\bm{M}) = \dfrac{\mathcal{P}(\bm{D}\vert\sTheta,\bm{M})\ \mathcal{P}(\sTheta\vert \bm{M})}{\mathcal{P}(\bm{D}\vert \bm{M})} \ .
\end{equation}
Here, $\mathcal{P}(\bm{D}\vert\sTheta,\bm{M}) \equiv \mathcal{L}(\sTheta)$ is the likelihood of the data, $\mathcal{P}(\sTheta\vert \bm{M}) \equiv \pi(\sTheta)$ is the prior probability density of the parameters and $\mathcal{P}(\bm{D}\vert \bm{M}) \equiv \mathcal{Z} = \int\mathcal{L}(\sTheta)\pi(\sTheta)\mathrm{d}^{n}\sTheta$ is the Bayesian evidence, where $n$ is the dimensionality of the parameter space.

\subsubsection{Bayesian model comparison}
\label{Sec:BayesianModelComparison}

Comparison of competing models in the light of observed data is a fundamental scientific goal. When one has a set of models for the data, $\bm{\mathcal{M}} = \{\bm{M}_{1}, \bm{M}_{2}, \cdots, \bm{M}_{N}\}$, preferred models can be determined from their marginal probabilities. Bayes' theorem for the marginal probability of a model gives:
\begin{equation}
\label{Eq:PrMiGivenD}
\mathcal{P}(\bm{M}_{i} \vert \bm{D},\bm{\mathcal{M}}) =  \frac{\mathrm{\mathcal{P}}(\bm{D} \vert \bm{M}_{i},\bm{\mathcal{M}})\mathcal{P}(\bm{M}_{i} \vert \bm{\mathcal{M}})}{\mathrm{\mathcal{P}}(\bm{D}\vert \bm{\mathcal{M}})} \ .
\end{equation}
Here, $\mathcal{P}(\bm{D} \vert \bm{\mathcal{M}}) = \sum_{k=1}^{N} \mathcal{P}(\bm{D}\vert \bm{M}_{k},\bm{\mathcal{M}}) \mathcal{P}(\bm{M}_{k} \vert \bm{\mathcal{M}})$ is the marginal probability of the data over the models and their parameters, $\mathcal{P}(\bm{D} \vert \bm{M}_{i},\bm{\mathcal{M}})$ is the Bayesian evidence of $\bm{M}_{i}$ and $\mathcal{P}(\bm{M}_{i} \vert \bm{\mathcal{M}})$ is the probability of $\bm{M}_{i}$ prior to analysing the data. For brevity, we leave the conditioning of the probability densities on $\bm{\mathcal{M}}$ implicit going forward.

Bayesian methodology addresses model comparison between two possible models, $\bm{M}_{i}$ and $\bm{M}_{j}$,  for a data set, $\bm{D}$, via consideration of $\mathcal{R}_{ij}$, the posterior odds in favour of $\bm{M}_{i}$ over $\bm{M}_{j}$. Using \Cref{Eq:PrMiGivenD} we can write this as:
\begin{align}
\label{Eq:RgivenData}
\mathcal{R}_{ij} &= \dfrac{\mathcal{P}(\bm{M}_{i}\vert\bm{D})}{\mathcal{P}(\bm{M}_{j}\vert\bm{D})} = \dfrac{\mathcal{P}(\bm{D}\vert \bm{M}_{i})\mathcal{P}(\bm{M}_{i})}{\mathcal{P}(\bm{D}\vert \bm{M}_{j})\mathcal{P}(\bm{M}_{j})} \ .
\end{align}
Here, $\mathcal{P}(\bm{M}_{i}\vert\bm{D})$ is the posterior probability of model $\bm{M}_{i}$, $\mathcal{P}(\bm{D}\vert \bm{M}_{i})/\mathcal{P}(\bm{D}\vert \bm{M}_{j}) \equiv \mathcal{B}_{ij}$ is the Bayes factor between the models, $\mathcal{P}(\bm{D}\vert \bm{M}_{i}) \equiv \mathcal{Z}_{i}$ and $\mathcal{P}(\bm{D}\vert \bm{M}_{j}) \equiv \mathcal{Z}_{j}$ are the Bayesian evidences of models $\bm{M}_{i}$ and $\bm{M}_{j}$, respectively, and $\mathcal{P}(\bm{M}_{i})/\mathcal{P}(\bm{M}_{j})$ is the ratio of the prior probabilities of the two models before any conclusions have been drawn from the data.

As the model-prior-weighted average of the likelihood over the parameters priors, the marginal probability of a model is larger if the model is probable a priori and more of its parameter space is likely given the data; it is smaller for a model that is improbable a priori or if large areas of its parameter space have low likelihood values, even if the likelihood function is very highly peaked. It thus represents an updating of one's prior credence in the model, given the data, and automatically incorporates an `Occam penalty' against a more complex theory with a broad parameter space. As such, in absence of an a priori reason to prefer it over a simpler alternative, it will be favoured only if it is significantly better at explaining the data.

In this work, we adopt the mapping of qualitative terms describing the relative preference for one model over another defined in S25\footnote{The mapping of qualitative terms defined in S25 generalises to $\mathcal{R}_{ij}$ the mapping for $\mathcal{B}_{ij}$ introduced in KR95. When model priors are uninformative, it reduces to the model-odds thresholds established in KR95.}. Specifically, we describe posterior odds in the range 1--3 ($0 \le \ln({\mathcal{R}_{ij}}) < 1$) as a weak preference for $\bm{M}_{i}$ over $\bm{M}_{j}$, posterior odds in the range 3--20 ($1 \le \ln({\mathcal{R}_{ij}}) < 3$) as a moderate preference, posterior odds in the range 20--150 ($3 \le \ln({\mathcal{R}_{ij}}) < 5$) as a strong preference, and posterior odds of greater than 150 ($5 \le \ln({\mathcal{R}_{ij}})$) as a decisive preference. When we have no information that lends additional credibility to one model over the other in advance of analysing the data, we set the prior odds to unity; therefore, the posterior odds are equal to the Bayes factor between the models ($\mathcal{R}_{ij}=\mathcal{B}_{ij}$). In this case, for the purpose of defining our qualitative descriptions, the Bayes factor between models takes the place of the posterior odds.

\subsection{BaNTER validation}
\label{Sec:BaNTERValidation}

When the set of competing models under consideration includes at least one model capable of providing accurate and predictive fits to the data using biased component fits, the Bayes factor between such models and those in the subset of interest -- those containing models with accurate and predictive subcomponent models -- is insufficient to distinguish them. Comparison of models for global 21-cm data can fall under this scenario (S25). In such a \textit{category II} model comparison, BFBMC enables one to separate models that are predictive of the data in aggregate from those that are not, but informative prior odds on the models are necessary to separate predictive composite models that also have accurate and predictive component fits to the data from those that do not.

In this work, we use the BaNTER validation framework for composite models introduced in S25 to validate our composite models for the data and derive informative model priors. For a detailed description of general BaNTER validation, we refer the reader to S25. Here, we provide a brief overview of the method in the context of global 21-cm cosmology.

\subsubsection{Global 21-cm data}
\label{Sec:Global21cmData}

Consider a global 21-cm signal data set of the form:
\begin{equation}
    \label{Eq:Data}
    \bm{D} = f(\sTheta_{21}, \sTheta_{\mathrm{Fg}}) + \bm{n} \ ,
\end{equation}
where $f(\sTheta_{21}, \sTheta_{\mathrm{Fg}})$ is the sum of global 21-cm signal and foreground emission in the data, $\bm{n}$ is the noise and $\sTheta_{21}$ and $\sTheta_{\mathrm{Fg}}$ are parameters underlying the physical processes producing the 21-cm signal and foreground emission. The function $f(.)$ captures the generative processes of the signal components, the propagation effects between the source of emission and the instrument (such as ionospheric absorption and refraction), the instrument transfer function, and any corrections applied to the data, such as beam factor chromaticity correction (see Paper I).

In many practical cases, the function $f(\cdot)$ can be approximated as nearly linear over the relevant parameter ranges (see \Cref{Sec:RealisitcBFCCSimulations})\footnote{Ionospheric absorption, as well as instrumental losses, act multiplicatively on both the astrophysical foregrounds and 21-cm signal. However, these effects are typically at the sub-10\% level, and since their fits are dominated by foreground emission 4--6 orders of magnitude brighter than the 21-cm signal, the induced coupling is expected to be negligible in practice.}. Under this approximation, the data model in \Cref{Eq:Data} reduces to:
\begin{equation}
    \label{Eq:Data2}
    \bm{D} \simeq \bm{S}_{21} + \bm{S}_\mathrm{Fg} + \bm{n} \ ,
\end{equation}
where $\bm{S}_{21}$ and $\bm{S}_\mathrm{Fg}$ represent the apparent global 21-cm signal and foreground components in the data, respectively. This linearised form provides intuitive insight into the data structure, although it is not a prerequisite for the validity of the BaNTER framework.

\subsubsection{Foreground, 21-cm signal, and composite models}
\label{Sec:Foreground21cmSignalAndCompositeModels}

Now, assume we have a definitive model for the signal component, denoted $\bm{M}_\mathrm{21}(\sTheta_{21})$, where $\sTheta_{21}$ represents the true values of the model parameters to be determined from the data.\footnote{See S25 for a discussion on how the BaNTER validation framework can be generalised to cases where both $\bm{S}_{21}$ and $\bm{S}_\mathrm{Fg}$ are uncertain.} In this case, the set of models for $\bm{S}_{21}$ is given by:
\begin{equation}
    \label{Eq:CurlyMa}
    \bm{\mathcal{M}}_{21} = \{\bm{M}_\mathrm{21}(\sTheta_{21})\} \ .
\end{equation}

For the foreground component, we define a competing set of models as the union of different foreground model classes, including BFCC, Intrinsic, LinPhys, and MultLin models. Explicitly, the set of all foreground models is:
\begin{equation}
    \label{Eq:CurlyMb}
    \bm{\mathcal{M}}_\mathrm{Fg} = \{\bm{M}_{i} \mid \bm{M}_{i} \in (\bm{\mathcal{M}}_{\mathrm{BFCC}} \cup \bm{\mathcal{M}}_{\mathrm{Intrinsic}} \cup \bm{\mathcal{M}}_{\mathrm{LinPhys}} \cup \bm{\mathcal{M}}_{\mathrm{MultLin}})\} \ ,
\end{equation}
where $\bm{\mathcal{M}}_{\mathrm{BFCC}}$, $\bm{\mathcal{M}}_{\mathrm{Intrinsic}}$, $\bm{\mathcal{M}}_{\mathrm{LinPhys}}$, and $\bm{\mathcal{M}}_{\mathrm{MultLin}}$ are the sets of BFCC, Intrinsic, LinPhys, and MultLin models, respectively. These sets vary in size, ranging from single models in $\bm{\mathcal{M}}_{\mathrm{Intrinsic}}$ and $\bm{\mathcal{M}}_{\mathrm{LinPhys}}$ to multiple models with varying complexity in $\bm{\mathcal{M}}_{\mathrm{BFCC}}$ and $\bm{\mathcal{M}}_{\mathrm{MultLin}}$ (see \Cref{Sec:DataModels}). Each model in $\bm{\mathcal{M}}_\mathrm{Fg}$ is of the form $\bm{M}_{i\mathrm{Fg}}(\sTheta_{i\mathrm{Fg}})$, where $\sTheta_{i\mathrm{Fg}}$ are the parameters of the $i$th foreground model.

Finally, given $\bm{\mathcal{M}}_{21}$ and $\bm{\mathcal{M}}_\mathrm{Fg}$, we define a set of composite models as:
\begin{equation}
    \label{Eq:CurlyMc}
    \bm{\mathcal{M}}_{\mathrm{c}} = \{\bm{M}_{i\mathrm{c}} \mid i = 1, \dots, N_\mathrm{c}\} \ ,
\end{equation}
where each composite model is of the form:
\begin{equation}
    \bm{M}_{i\mathrm{c}} = g(\sTheta_{21}, \sTheta_{i\mathrm{Fg}}) \ .
\end{equation}
Here, $g(\sTheta_{21}, \sTheta_{i\mathrm{Fg}})$ is a model for $f(\sTheta_{21}, \sTheta_{\mathrm{Fg}})$ and $N_\mathrm{c}$ represents the total number of composite models in the set. In the linearised case $\bm{M}_{i\mathrm{c}}$ simplifies to:
\begin{equation}
    \bm{M}_{i\mathrm{c}} = \bm{M}_\mathrm{21} + \bm{M}_{i\mathrm{Fg}} \ ,
\end{equation}
with $\bm{M}_\mathrm{21}(\sTheta_{21})$ being the model for the 21-cm signal component and $\bm{M}_{i\mathrm{Fg}}(\sTheta_{i\mathrm{Fg}})$ being the $i$th model for the foreground component.

\subsubsection{Bayesian null test}
\label{Sec:BayesianNullTest}

The BaNTER validation framework provides a means of separating models in $\bm{\mathcal{M}}_{\mathrm{c}}$ that are able to accurately describe the data with accurate component models from those that provide accurate fits to the data but lead to biased inferences for the parameters of the 21-cm signal model, $\bm{M}_\mathrm{21}$, if present in the data.
In the general (non-linear) case, this is achieved by comparing the Bayesian evidence of a composite model, $\bm{M}_{\mathrm{c}}(\sTheta_{21}, \sTheta_{\mathrm{Fg}})$, against that of a foreground-only model, $\bm{M}_{\mathrm{Fg}}(\sTheta_{\mathrm{Fg}})$, for a foreground-only validation data set of the form:
\begin{equation}
    \label{Eq:ValidationData}
    \bm{D}_\mathrm{v} = \bm{S}_\mathrm{Fg} + \bm{n} \ .
\end{equation}

Since the validation data used in this comparison is constructed to contain only a foreground component, any preference for the composite model over the foreground-only model indicates that the signal model is absorbing residual structure due to an inaccurate foreground model.

In this work, we use high-fidelity simulated foreground-only observations as the $\bm{S}_\mathrm{Fg}$ component of our validation data (see \Cref{Sec:Simulations}).

Given $\bm{D}_\mathrm{v}$ and a composite-foreground pair of models, $\bm{M}_{i\mathrm{c}}$ and $\bm{M}_{i\mathrm{Fg}}$, the BaNTER validation proceeds by fitting each model to the validation data and computing the null-test evidence ratio (Bayes factor):
\begin{equation}
    \label{Eq:lnBcb}
    \ln(\mathcal{B}_\mathrm{cFg}^\mathrm{v}) = \ln \left( \frac{\mathcal{Z}_\mathrm{c}^\mathrm{v}}{\mathcal{Z}_\mathrm{Fg}^\mathrm{v}} \right) \ ,
\end{equation}
where $\mathcal{Z}_\mathrm{c}^\mathrm{v} = \mathcal{P}(\bm{D}_\mathrm{v} \vert \bm{M}_{i\mathrm{c}})$ and $\mathcal{Z}_\mathrm{Fg}^\mathrm{v} = \mathcal{P}(\bm{D}_\mathrm{v} \vert \bm{M}_{i\mathrm{Fg}})$ are the Bayesian evidences for the composite model $\bm{M}_{i\mathrm{c}}$ and its foreground component $\bm{M}_{i\mathrm{Fg}}$, respectively.

When $\ln(\mathcal{B}_\mathrm{cFg}^\mathrm{v}) \geq 0$, the composite model fits the validation data better than the foreground-only model. Since the validation data contains only foregrounds, this preference indicates that the 21-cm component is absorbing systematic residuals from an imperfect foreground model. For example, if a foreground model inadequately describes chromatic instrumental effects, the 21-cm component might fit both the true signal and these residual systematics, yielding a good overall fit to the data in aggregate but biased 21-cm signal estimates.

The validation metric $\ln(\mathcal{B}_\mathrm{cFg}^\mathrm{v})$ becomes large only when the foreground model is insufficiently accurate in describing the foreground component of the data \textit{and} when a spurious fit of the 21-cm signal model absorbs residual structure in the validation data that the foreground model alone cannot fit.

The composite model is deemed to fail the null test if $\ln(\mathcal{B}_\mathrm{cFg}^\mathrm{v}) \ge \ln(\mathcal{B}_\mathrm{threshold}^\mathrm{v})$, where $\ln(\mathcal{B}_\mathrm{threshold}^\mathrm{v})$ is a predefined threshold. We interpret different ranges of $\ln(\mathcal{B}_\mathrm{cFg}^\mathrm{v})$ as follows:
\begin{itemize}
    \item $< 0$: Foreground-only model preferred. This is expected for foreground-only validation data.
    \item $0-3$: Moderate systematic contamination likely to bias signal estimates if a 21-cm signal is present.
    \item $\geq 3$: Severe systematic contamination likely both to bias 21-cm signal recovery, if a 21-cm signal is present, or produce a false detection, if not.
\end{itemize}

Specifically, for $0 \le \ln(\mathcal{B}_\mathrm{cFg}^\mathrm{v}) < 3$, the composite model provides a better fit to the validation data than the foreground-only model. In the context of global 21-cm signal datasets, this suggests that inaccuracies in the foreground model are sufficient to bias estimates of the 21-cm signal, if present. However, under a Bayesian 21-cm detection criterion that requires strong evidence in favour of the composite model over the foreground-only model ($\ln(\mathcal{B}_\mathrm{cFg}) \ge 3$; see \Cref{Sec:21cmSignalDetection}), these inaccuracies are too small to produce a false detection of the 21-cm signal, when it is absent.

For $\ln(\mathcal{B}_\mathrm{cFg}^\mathrm{v}) \ge 3$, the composite model provides a substantially better fit to the validation data than the foreground-only model, indicating that foreground model inaccuracies are severe enough to significantly bias 21-cm signal estimates, if present, or to lead to a false detection, if absent.

In this work, we follow S25 and adopt a conservative approach by setting $\ln(\mathcal{B}_\mathrm{threshold}^\mathrm{v}) = 0$. We treat the prior odds of failed composite models yielding unbiased estimates of the 21-cm signal in a global dataset as negligible when compared to models that successfully pass BaNTER validation.

\subsubsection{Model comparison categorisation}
\label{Sec:ModelComparisonCategorisation}

The possibility exists for composite models that are predictive of the data in aggregate to obtain accurate fits with biased component models. However, in the absence of informative model priors, it is uncertain whether models of this type are included in the set of models under consideration. Following S25, we categorize the Bayesian comparison of composite models in $\bm{\mathcal{M}}_{\mathrm{c}}$ as a \textit{category I} model comparison problem if $\bm{\mathcal{M}}_{\mathrm{c}}$ contains no such models, and a \textit{category II} model comparison problem if such models are present.

For a \textit{category I} model comparison problem, model validation is incidental to the recovery of unbiased 21-cm signal estimates through Bayesian analysis of the data. In contrast, for a \textit{category II} model comparison problem, model validation becomes essential. Thus, understanding which of these categories applies to the Bayesian comparison of a given set of composite models is key to determining whether model validation is necessary for drawing robust inferences with them and, equivalently, for assessing the degree of confidence that can be placed in the conclusions drawn from the comparison of unvalidated models.

We use the BaNTER validation framework to determine whether the Bayesian comparison of the models considered here falls under \textit{category I} or \textit{category II} in \Cref{Sec:Results}.

\subsection{Data likelihood}
\label{Sec:DataLikelihood}

Let the data, vectorised over frequency, be denoted as $\bm{D}$, and define a corresponding vectorised model, parameterised by $\bm{\Theta}$, as $\bm{M}(\sTheta)$.

We assume the noise in the data follows a zero-mean Gaussian distribution, uncorrelated between frequency channels. Consequently, we model the noise covariance matrix, $\mathbfss{N}$, as diagonal, with elements given by:
\begin{equation}
N_{ij} = \left< n_i n_j^* \right> = \delta_{ij} \sigma^{2} \ ,
\end{equation}
where $\left< \cdot \right>$ denotes the expectation value, and $\sigma$ is the root-mean-square (RMS) noise level in the data.

Defining the residuals between the data and model as $\bm{R} = \bm{D} - \bm{M}(\sTheta)$, the Gaussian likelihood function for $\bm{R}$ is given by:
\begin{equation}
\label{Eq:BasicVisLike}
\mathcal{P}(\bm{D} | \bm{\Theta}) = \frac{1}{\sqrt{(2\pi)^{N_{\mathrm{chan}}}\mathrm{det}(\mathbfss{N})}} \exp\left[-\frac{1}{2}\bm{R(\Theta)}^T\mathbfss{N}^{-1}\bm{R(\Theta)}\right] \ .
\end{equation}
When fitting the data in \Cref{Sec:Results}, $\bm{D}$ represents vectorised, simulated beam factor chromaticity-corrected data (see \Cref{Sec:SpectrometerData}). For a spectrum $X(\nu)$, we define the vectorisation operator $\mathrm{vec}(\cdot)$ such that:
\begin{equation}
\mathrm{vec}(X(\nu)) = [X_{0}, X_{1}, \dots, X_{N_{\mathrm{chan}}}]^{T} \ ,
\end{equation}
where $X_{i}$ is the value of $X$ at frequency channel $i$, and $N_{\mathrm{chan}}$ is the total number of channels in the dataset.

\subsection{21-cm signal detection}
\label{Sec:21cmSignalDetection}

We propose that a robust detection of the 21-cm signal should satisfy the following criteria:
\begin{enumerate}
    \item The subcomponents of the model must provide an accurate description of their respective signal components. This requirement prevents errors in one model component from being absorbed by another, which could result in an accurate fit to the data in aggregate but a biased recovery of the 21-cm signal.
    \item The model incorporating the 21-cm signal must provide an accurate description of the data, such that the residuals are consistent with the expected noise level.
    \item There must be strong Bayesian evidence favouring models that include a 21-cm signal component over those that do not.
\end{enumerate}
Criteria (i) and (ii) can be assessed using BaNTER-validated posterior-odds-based model comparison (\Cref{Sec:BayesianInference,Sec:BayesianNullTest}) and an analysis of model residuals (see \Cref{Sec:AccuracyCondition}), respectively. Given observational data $\bm{D}$ and a composite-foreground pair of models, $\bm{M}_{i\mathrm{c}}$ and $\bm{M}_{i\mathrm{Fg}}$, criterion (iii) is satisfied if the following threshold for 21-cm signal detection is met:
\begin{equation}
    \label{Eq:lnBcbData}
    \ln(\mathcal{B}_\mathrm{cFg}) \geq 3 \ .
\end{equation}
Here, $\ln(\mathcal{B}_\mathrm{cFg}) = \ln(\mathcal{Z}_\mathrm{c} / \mathcal{Z}_\mathrm{Fg})$ is the log Bayes factor in favour of the composite model, where $\mathcal{Z}_\mathrm{c} = \mathcal{P}(\bm{D} \vert \bm{M}_{i\mathrm{c}})$ and $\mathcal{Z}_\mathrm{Fg} = \mathcal{P}(\bm{D} \vert \bm{M}_{i\mathrm{Fg}})$ denote the Bayesian evidences for the composite model $\bm{M}_{i\mathrm{c}}$ and its foreground-only counterpart $\bm{M}_{i\mathrm{Fg}}$, respectively. This threshold corresponds to odds of at least 20:1 in favour of the composite model, providing strong evidence for the presence of a 21-cm signal in the data.

To test for the presence of a 21-cm signal in the data, we apply the criteria outlined above in \Cref{Sec:Results} and define the full set of models under consideration as:
\begin{equation}
    \label{Eq:CurlyM}
    \bm{\mathcal{M}} = \{\bm{M}_{i} \mid \bm{M}_{i} \in (\bm{\mathcal{M}}_{\mathrm{c}} \cup \bm{\mathcal{M}}_\mathrm{Fg})\} \ .
\end{equation}

\subsection{Computational techniques}
\label{Sec:ComputationalTechniques}

\subsubsection{Probability densities}

In \Cref{Sec:Results}, when analysing the data, we estimate model evidences and sample from the posteriors on the model parameters, given the data, using nested sampling as implemented by the \textsc{PolyChord} algorithm \citep{2015MNRAS.453.4384H, 2015MNRAS.450L..61H}. Given samples from the posterior distribution of the parameters, $\mathcal{P}(\sTheta\vert\bm{D},\bm{M})$, one can estimate $\mathcal{P}(y\vert\sTheta,\nu,\bm{D},\bm{M})$, the posterior predictive density (posterior PD) of a function $y = f(\sTheta,\nu)$ by calculating the corresponding set of samples from $\mathcal{P}(y\vert \sTheta,\nu,\bm{D},\bm{M})$. We derive contour plots of prior and posterior PDs using the \textsc{fgivenx} software package (\citealt{2018JOSS....3..849H}).

\subsubsection{Summary statistics}
\label{Sec:SummaryStatistics}

Many of the aforementioned parameter posteriors will be characterised by non-Gaussian probability density functions (PDFs). Therefore, following \citet{2025arXiv250409725S}, we use the highest probability density estimates (HPDEs) and highest probability density intervals (HPDIs; e.g. \citealt{b7f71c99-f621-3c7e-a9dd-9d152d4822a4}), $X_\mathrm{HPD}|^{+\sigma_{+}}_{-\sigma_{-}}$ as informative summary statistics of these distributions. Here, $X_\mathrm{HPD}$ is the HPDE value of the PDF of a parameter (or set of parameters), $X$, and $\sigma_{\pm} = |X_\mathrm{HPDI\pm} - X_\mathrm{HPD}|$ characterises its width, with $X_\mathrm{HPDI+}$ and $X_\mathrm{HPDI-}$ the upper and lower bound of the HPDI, respectively.

\section{Simulated data}
\label{Sec:RealisitcBFCCSimulations}

We construct realistic simulations of time-averaged BFCC data following the approach described in Paper I; for a detailed description we refer the reader to that work. In this section, we provide a summary of the approach including the updates to the 21-cm signal component in the simulations used here.

\subsection{Spectrometer data in the snapshot limit}
\label{Sec:SpectrometerData}

Working in the reference frame of the antenna, under the assumption that the integration time, $\Delta t$, is sufficiently short for the measurement to be accurately approximated as an instantaneous snapshot at the central time of the integration ($\Delta t \lesssim 10~\mathrm{mins}$ for EDGES 2), a calibrated autocorrelation spectrum derived from a zenith-pointing antenna, such as EDGES, can be written as (see Paper I):
\begin{equation}
\label{Eq:Tdata}
T_\mathrm{data}(\nu, t) = \int\limits_{\Omega^{+}} B(\nu, \Omega)T_\mathrm{sky}(\nu, \Omega, t)~\mathrm{d}\Omega + n \ .
\end{equation}
Here, $T_\mathrm{sky}(\nu, \Omega, t)$ represents the time-dependent sky brightness temperature distribution above the antenna, $d\Omega$ is a solid angle element, $n$ denotes instrumental noise and $\Omega^{+}$ is the skyward hemisphere centred on zenith. The term $B(\nu, \Omega) = \frac{1}{D_{\Omega^{+}}}D(\nu, \Omega)$ describes the frequency and direction-dependent antenna beam. It is normalised such that the beam pattern integrates to 1 over the skyward hemisphere with $D(\nu, \Omega)$ the antenna directivity pattern and $D_{\Omega^{+}} = \int_{\Omega^{+}} D(\Omega) \mathrm{d}\Omega$.

For an instrument incorporating a large ground plane below the antenna, such as EDGES, the region $\Omega^{+}$ encompasses nearly the full integral antenna directivity. Specifically, for the H2 configuration of the EDGES 2 low-band instrument with a $30~\mathrm{m} \times 30~\mathrm{m}$ sawtooth ground plane, detailed electromagnetic simulations of the antenna directivity (e.g. \citealt{2021AJ....162...38M}) indicate that the fractional directivity towards the nadir-centred hemisphere at a fixed frequency is $1-D_{\Omega^{+}}/D_\mathrm{full} \simeq 10^{-3}$. Here, $D_\mathrm{full} = \int_{0}^{4\pi} D(\Omega) \mathrm{d}\Omega$ is the integral antenna directivity over the full sphere. In this work, we assume that the fractional directivity towards the ground, on the order of $10^{-3}$, has been accurately accounted for through ground-loss correction (e.g. \citealt{2012RaSc...47.0K06R, 2017ApJ...847...64M}). Additionally, the data has been calibrated such that $\int_{0}^{\Omega^{+}} B(\nu, \Omega)\mathrm{d}\Omega = 1$ and $T_\mathrm{data}(\nu, t)$ is an absolute temperature measurement.

\subsection{Beam factor chromaticity correction}
\label{Sec:BFCC}

For the purpose of global 21-cm signal data analysis, we can write the sky brightness temperature in the $10 \lesssim \nu \lesssim 230~\mathrm{MHz}$ frequency range as the sum of two components:
\begin{enumerate}
    \item A bright but spectrally smooth non-21-cm component comprised of synchrotron emission from the Galaxy and extragalactic sources, with a smaller contribution from Galactic free-free emission, and thermal emission from the Earth's ionosphere.
    \item A redshifted 21-cm signal component with less smooth spectral structure determined in detail by the sky-averaged evolution with redshift of the ionization and temperature state of hydrogen and the relative coupling strength of the neutral hydrogen spin temperature to its kinetic temperature and the background radiation temperature.
\end{enumerate}
Additionally, the Earth's ionosphere refracts and absorbs both of these components in a frequency-dependent manner (e.g. \citealt{2014MNRAS.437.1056V, 2021MNRAS.503..344S}).

Despite a dynamic range of several orders of magnitude between these two components, in the absence of instrumental effects and barring a significant level of Faraday rotated polarised foreground emission (e.g. \citealt{2019MNRAS.489.4007S}), the effective foreground after passing through the ionosphere is expected to be spectrally separable from the 21-cm signal. However, instrumental chromaticity, if unaccounted for and in excess of the dynamic range between the foregrounds and 21-cm signal, will eliminate this separation of characteristic spectral scales and will introduce foreground systematics greater than or equal in amplitude to the 21-cm signal of interest, biasing its recovery by spectral means.

The impact of instrumental chromaticity on the separation of the 21-cm signal from the non-21-cm component of the data can be significantly mitigated (although not entirely removed) by dividing the calibrated autocorrelation spectrum by a beam chromaticity correction factor, $B_\mathrm{factor}$, that describes the average spectral structure of the beam weighted by the brightness temperature distribution of the sky at a given reference frequency (see Paper I for details). In the short integration snapshot limit, $B_\mathrm{factor}$ is given by (e.g. \citealt{2017MNRAS.464.4995M, 2019MNRAS.483.4411M}),
\begin{equation}
\label{Eq:CCsnapshot}
B_\mathrm{factor}(\nu, t) = \frac{\int_{\Omega^{+}}  B^\mathrm{m}(\nu, \Omega) T_\mathrm{fg}^\mathrm{m}(\nu_\mathrm{c}, \Omega, t) \mathrm{d}\Omega}{\int_{\Omega^{+}}  B^\mathrm{m}(\nu_\mathrm{c}, \Omega) T_\mathrm{fg}^\mathrm{m}(\nu_\mathrm{c}, \Omega, t) \mathrm{d}\Omega},
\end{equation}
and the BFCC data has the form,
\begin{align}
\label{Eq:Tcorrected}
T_{\rm corrected}(\nu, t) =  T_{\rm data}(\nu, t) / B_\mathrm{factor}(\nu, t).
\end{align}
Time-averaged BFCC data, $T_{\rm corrected}(\nu)$, such as that analysed in B18 and also the subject of the analysis here, is formed by averaging $T_{\rm corrected}(\nu, t)$ over $t$.

Here, as in Paper I, we focus on the effectiveness of BFCC when one has an accurate model for $B^\mathrm{m}$ and $T_\mathrm{fg}^\mathrm{m}(\nu_\mathrm{c}, \Omega, t)$. In upcoming work we will explore how 21-cm signal recovery with the BFCC model derived in Paper I is impacted by realistic deviations from the assumption of an error-free model for $T_\mathrm{fg}^\mathrm{m}(\nu_\mathrm{c}, \Omega, t)$ and $B^\mathrm{m}$.

\subsection{Simulations}
\label{Sec:Simulations}

To construct $T_{\rm corrected}(\nu)$, we first construct simulated time-dependent EDGES-low spectrometer data, $T_{\rm data}(\nu, t)$, following \Cref{Eq:Tdata}, at 120 times, spaced by 6 minute intervals, in the LST range $0 \le LST < 12~\mathrm{h}$, selected to match the LST window of the publicly available EDGES-low data, when the Galactic plane is relatively low in the beam. We simulate data over a $50-100~\mathrm{MHz}$ spectral band, assuming a $1~\mathrm{MHz}$ channel width and observation of a sky model composed of:
\begin{itemize}
    \item foregrounds with realistic spatially dependent spectral structure,
    \item spectrally-dependent absorption by the ionosphere,
    \item ionospheric emission, and
    \item a flattened Gaussian redshifted 21-cm signal profile.
\end{itemize}
We simulate our sky model including the aforementioned components as,
\begin{multline}
\label{Eq:TskySDFgStationaryIonosphere}
T_\mathrm{sky}(\nu, \Omega, t) = \Big[(T_\mathrm{fg}(\nu, \Omega, t) + T_{21} \Big] e^{-\tau_\mathrm{ion}(\nu)} \\
+ T_{\mathrm{e}}(1-e^{-\tau_\mathrm{ion}(\nu)})
\ ,
\end{multline}
where $T_{\mathrm{e}}$ and $\tau_\mathrm{ion}$ are the temperature of electrons and opacity of the ionosphere, respectively. Here, following Paper I, we use, $T_{\mathrm{e}}=450~\mathrm{K}$ and $\tau_\mathrm{ion} = \tau_0(\nu/\nu_\mathrm{c})^{-2}$, with $\tau_0=0.014$ at reference frequency $\nu_\mathrm{c}=75~\mathrm{MHz}$ (e.g. \citealt{2015RaSc...50..130R}). The foreground brightness temperature distribution is given by,
\begin{multline}
\label{Eq:TSDFgStationaryIonosphere}
T_\mathrm{fg}(\nu, \Omega, t) = T_\mathrm{fg}(\nu_\mathrm{c}, \Omega, t)\left(\frac{\nu}{\nu_\mathrm{c}}\right)^{-\beta_{\Omega,t}} +  T_{\gamma} \ .
\end{multline}
Here, $T_\mathrm{fg}(\nu_\mathrm{c}, \Omega, t)$ is the spectral power law component of the foreground brightness temperature at reference frequency $\nu_\mathrm{c}$. $\beta_{\Omega,t}$ is the spatially dependent spectral index distribution characterising the power law structure of that emission. $T_{\gamma}=2.725~\mathrm{K}$ is the CMB temperature and $T_{21}$ is the 21-cm signal in the data.

Following Paper I, we derive $T_\mathrm{fg}(\nu_\mathrm{c}, \Omega, t)$ as a spectral extrapolation from the Haslam 408 MHz all-sky map (\citealt{1981A&A...100..209H, 1982A&AS...47....1H}) reprocessed by \citet{2015MNRAS.451.4311R} and $\beta_{\Omega,t}$ as a spatially dependent spectral index distribution derived from the global sky model (GSM; \citealt{2017MNRAS.464.3486Z}).
Furthermore, we model the global 21-cm signal as a flattened-Gaussian absorption trough, matching the model parametrisation used in B18:
\begin{equation}
\label{Eq:FlattenedGaussian}
T_\mathrm{21}(\nu) = -A\left(\frac{1-e^{-\tau e^{B_{21}}}}{1-e^{-\tau}}\right) \ ,
\end{equation}
where,
\begin{equation}
\label{Eq:FlattenedGaussianB}
B_{21} = \frac{4(\nu-\nu_0)^2}{w^2}\log\left[-\frac{1}{\tau}\log\left(\frac{1+e^{-\tau}}{2}\right)\right] \ ,
\end{equation}
and $A$, $\nu_0$, $w$ and $\tau$ describe the amplitude, central frequency, width and flattening of the absorption trough, respectively.

In the simulated data sets analysed in this work, we incorporate absorption profiles with position and shape parameters: $\nu_{0}=78~\mathrm{MHz}$, $w=19~\mathrm{MHz}$ and $\tau=8$ matching the 21-cm signal shape parameters considered in Paper 1, for ease of comparison\footnote{Given that there are potentially correlated effects that would result from varying both amplitudes and shape simultaneously, by taking this approach we are able to explicitly isolate and test the effect of underlying 21-cm amplitude on signal recovery.}.
For the amplitude of the signal, we consider three cases:
\begin{enumerate}
 \item \textit{A null-amplitude 21-cm signal} ($A=0~\mathrm{mK}$). We use this as a null-test to identify models that lead to spurious signal detection through joint estimation of a 21-cm signal with an insufficiently accurate foreground and ionosphere model.
 \item \textit{A moderate-amplitude 21-cm signal} ($A=150~\mathrm{mK}$), with a signal amplitude consistent with expectations under standard cosmological assumptions regarding cooling of the hydrogen gas during the Dark Ages and a background radiation temperature during CD dominated by the CMB.
 \item \textit{A high-amplitude 21-cm signal} ($A=500~\mathrm{mK}$), consistent with the best-fit recovered in B18 and explainable in a physically motivated manner with additional cooling of the hydrogen gas beyond that due to adiabatic expansion and/or an additional radio background raising the total radio background temperature in excess of the CMB.
\end{enumerate}

We construct our time-dependent beam factor model, $B_\mathrm{factor}(\nu, t)$, and BFCC data, $T_{\rm corrected}(\nu, t)$ , in the manner outlined in \Cref{Sec:BFCC}. For our beam model, $B(\nu, \Omega)$, we use the {\sc{FEKO}} EM simulation of the EDGES-low blade dipole  antenna with a $30~\mathrm{m} \times 30~\mathrm{m}$ sawtooth ground plane from \citet{2021AJ....162...38M}. We calculate our time-averaged BFCC data, $T_{\rm corrected}(\nu)$, by averaging $T_{\rm corrected}(\nu, t) =  T_{\rm data}(\nu, t) / B_\mathrm{factor}(\nu, t)$ over the simulated snapshot spectra. We add noise to the data at a level such that the resultant noise in the BFCC data, after time-averaging, is Gaussian and white, with an RMS amplitude of $20~\mathrm{mK}$ that is comparable to estimates of the noise in the publicly available EDGES-low data (e.g. \citealt{2019ApJ...880...26S}). In all of our simulations, we assume the receiver calibration of the data is unbiased and uncertainty free (see \citealt{2022MNRAS.517.2264M} for a discussion of the impact of uncertainty and bias in the receiver calibration parameter estimation).

\begin{figure*}
	\centerline{
	\begin{subfigure}[t]{0.5\textwidth}
        \caption{}
        \label{Fig:ForegroundBaseMapModel}
        \includegraphics[width=\textwidth]{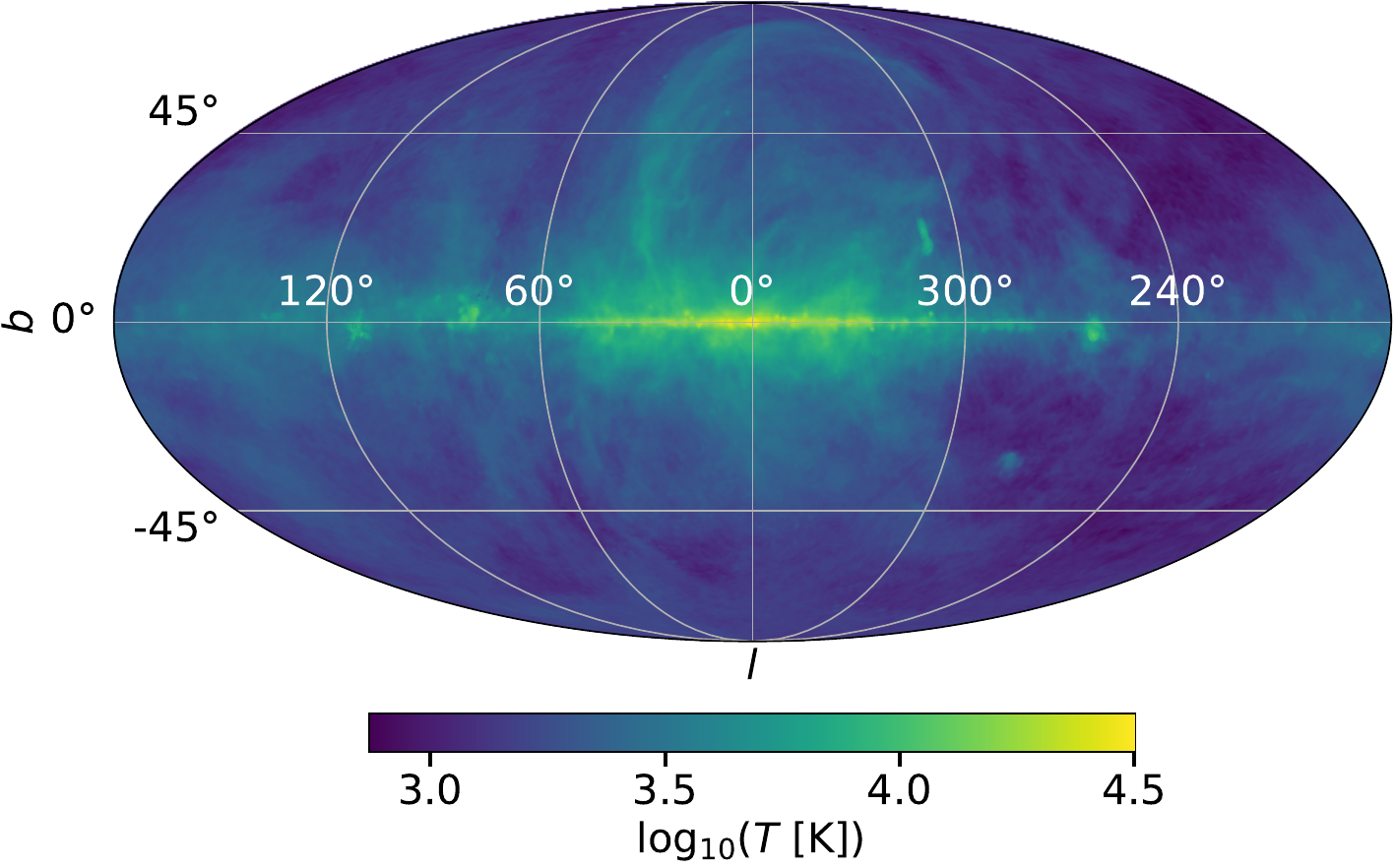}
	\end{subfigure}
	\begin{subfigure}[t]{0.5\textwidth}
        \caption{}
        \label{Fig:SImodel}
        \includegraphics[width=\textwidth]{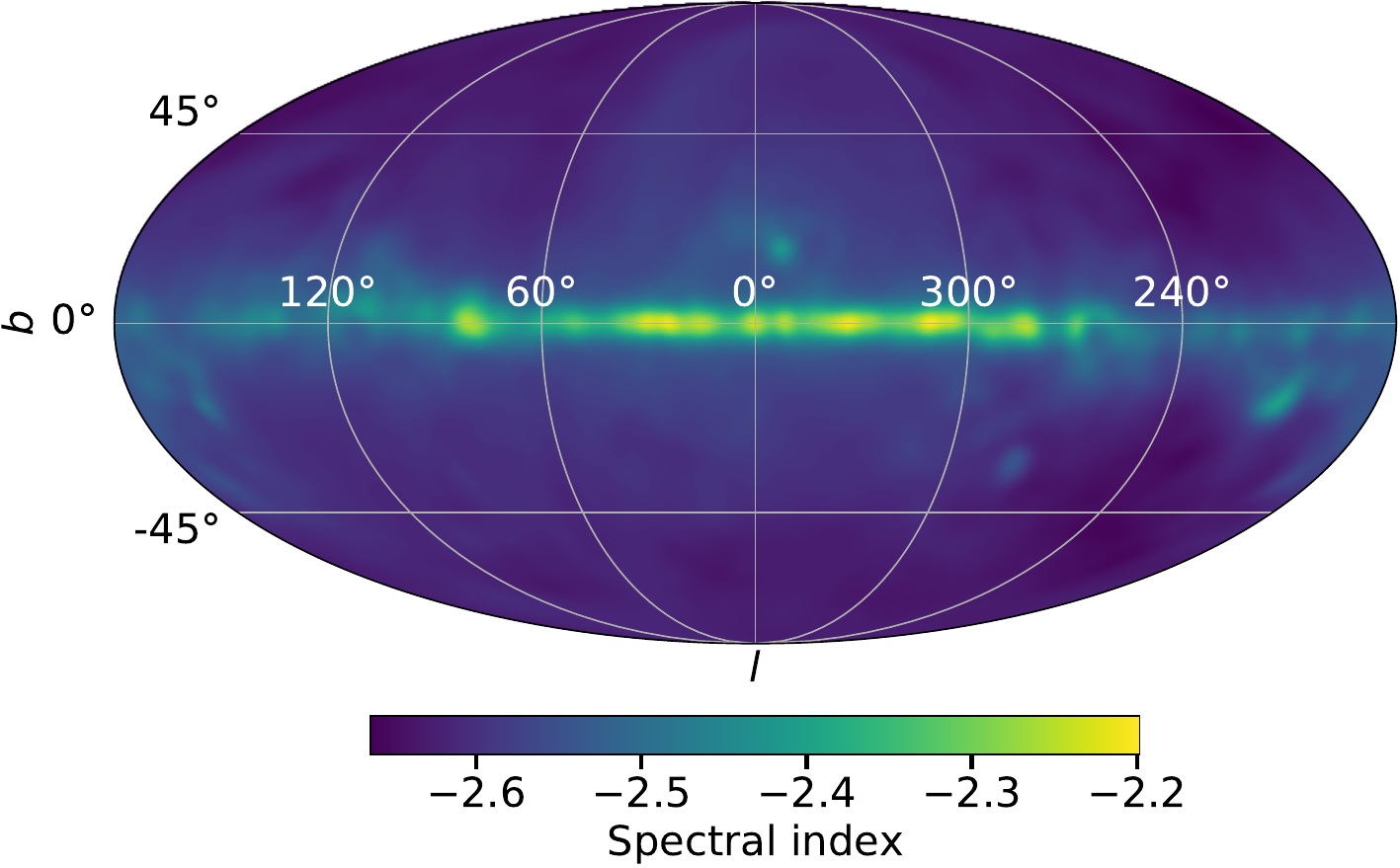}
	\end{subfigure}
	}
	\centerline{
	\begin{subfigure}[t]{0.5\textwidth}
        \caption{}
        \label{Fig:Tcorrected}
        \includegraphics[width=\textwidth]{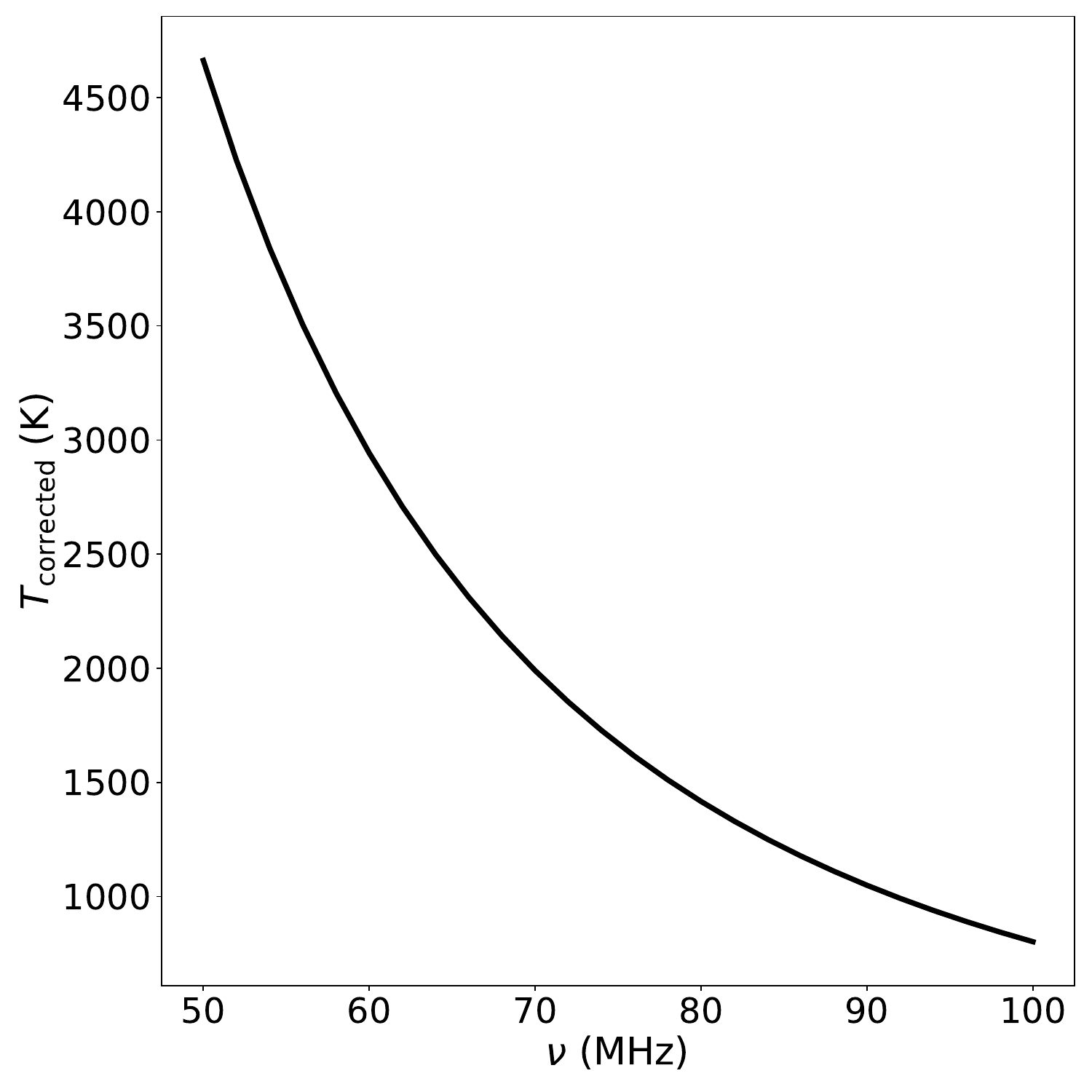}
	\end{subfigure}
	\begin{subfigure}[t]{0.5\textwidth}
        \caption{}
        \label{Fig:T21}
        \includegraphics[width=\textwidth]{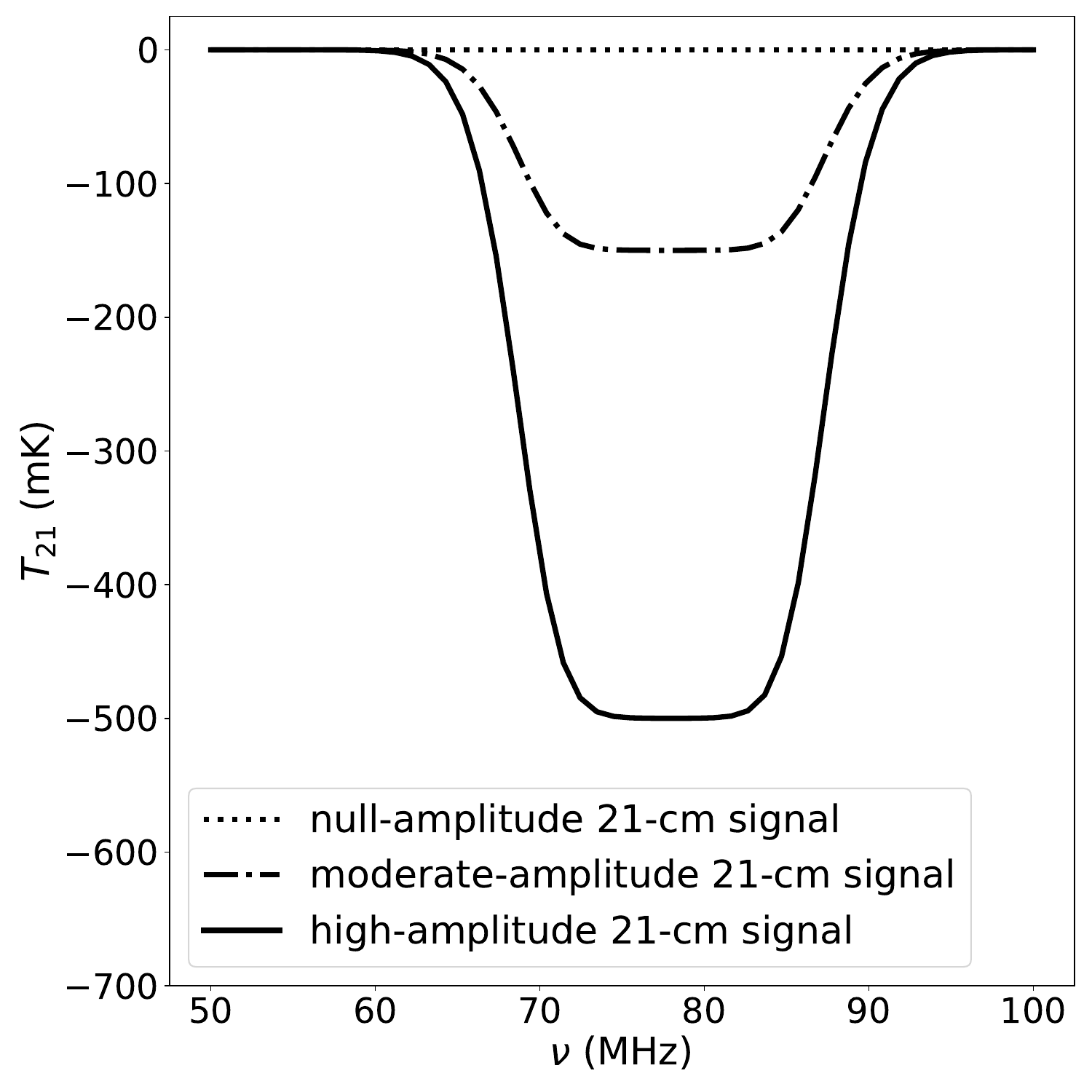}
	\end{subfigure}
	}
\caption{
    Astrophysical components of our simulated data sets.
    \Cref{Fig:ForegroundBaseMapModel}: Our intrinsic foreground brightness temperature distribution model, evaluated at the centre of our simulated spectral band, $T_\mathrm{fg}(75~\mathrm{MHz}, l, b)$.
    \Cref{Fig:SImodel}: The spatially-dependent foreground spectral index distribution $\beta(l, b)$ used when constructing simulated observational data.
    \Cref{Fig:Tcorrected}: Simulated time-averaged, beam factor chromaticity corrected spectrum resulting from time-averaging simulated BFCC EDGES low-band data over 120 simulated snapshot spectra derived at 6 minute intervals in the LST range $0 \le LST < 12~\mathrm{h}$, matching the LST window of the publicly available EDGES low-band data.
    \Cref{Fig:T21}: Input 21-cm signal, in the simulated BFCC data, in the three signal amplitude regimes analysed in \Cref{Sec:Results}.
    }
\label{Fig:SimulatedData}
\end{figure*}

\Cref{Fig:SimulatedData} illustrates the key astrophysical components of our simulated data sets. Our intrinsic foreground brightness temperature distribution model, evaluated at the centre of our simulated spectral band, $T_\mathrm{fg}(75~\mathrm{MHz}, l, b)$, is shown in \Cref{Fig:ForegroundBaseMapModel}. Our model for the foreground spectral index distribution $\beta(l, b)$ is shown in \Cref{Fig:SImodel}. \Cref{Fig:Tcorrected} shows the output simulated time-averaged, beam factor chromaticity corrected spectrum and \Cref{Fig:T21} illustrates the injected 21-cm signals in the data in the three signal amplitude regimes we consider.

\section{Data models \& likelihood}
\label{Sec:DataModels}

We consider four classes of composite models, each assuming the data is composed of 21-cm signal and non-21-cm signal components and instrumental noise (\Cref{Eq:Data}). The non-21-cm signal component models (i) astrophysical foreground emission following propagation through the ionosphere, where it undergoes chromatic absorption, (ii) ionospheric emission, and (iii) any residual instrumental effects in the data not perfectly correct by beam-factor-based chromaticity correction. For brevity, we use `foreground component' as a shorthand for the non-21-cm signal component going forward.

\subsection{Signal model}
\label{Sec:SignalModel}

In each case, following B18, we model the 21-cm signal component of the data as a flattened Gaussian model (\Cref{Eq:FlattenedGaussian}). We assume priors on the amplitude, central frequency, width and flattening of the absorption trough as listed in \Cref{Tab:DataModelSignalPriors}.

When fitting the data, we jointly estimated the 21-cm signal component model with one of four foreground components. The first is the BFCC model, which is the model we developed in Paper I of this series. The second is the Intrinsic model, which is a physically motivated parametrisation of the foreground component of the sky signal after propagation through the ionosphere. The third is the LinPhys model, which is a linear approximation to the Intrinsic model with uninformative priors on the parameters of the model. The fourth is the MultLin model, which is a more general polynomial foreground model also used as the foreground model in some recovered 21-cm signal estimates in B18, again assuming uninformative priors on the parameters of the model. We describe the four models in more detail below.

\begin{table}
\caption{Priors on the parameters of the global 21-cm signal model component of the models fit in \Cref{Sec:Results}.
}.
\centerline{
\begin{tabular}{l l }
\hline
Parameter & Prior     \\
\hline
$A$     & $U(0,1)~\mathrm{K}$ \\
$\nu_0$ & $U(55,95)~\mathrm{MHz}$ \\
$w$     & $U(5,30)~\mathrm{MHz}$ \\
$\tau$  & $U(0,20)$  \\
\hline
\end{tabular}
}
\label{Tab:DataModelSignalPriors}
\end{table}

\subsection{BFCC foreground model}
\label{Sec:BFCCmodel}

In Paper I of this series we derived the BFCC model: a flexible closed-form model for BFCC spectrometer data. The BFCC model explicitly accounts for and models:
\begin{itemize}
    \item the effect of realistic spatially dependent spectral structure of foreground emission,
    \item frequency dependent absorption of the foreground and 21-cm emission while propagating through the ionosphere,
    \item emission by high-temperature electrons in the ionosphere, and
    \item re-weighting of all components of the data, including the 21-cm signal and noise during BFCC.
\end{itemize}
In Paper I, the BFCC model is shown to enable unbiased recovery of a high-amplitude simulated global 21-cm signal. For a detailed derivation of the model, we refer the reader to that work. Here, we quote the final form of the model:
\begin{multline}
\label{Eq:BFCCdataModel}
T_\mathrm{BFCC,fg}^\mathrm{model}(\nu)
= \Bigg[\bar{T}_\mathrm{m_{0}}\left(\frac{\nu}{\nu_\mathrm{c}}\right)^{-\beta_{0}} \left(1 + \sum_{\alpha=1}^{N_\mathrm{pert}} p_{\alpha}\ln\left(\frac{\nu}{\nu_\mathrm{c}}\right)^{\alpha}\right)\\
+ \frac{\left(1-\left(\frac{\nu}{\nu_\mathrm{c}}\right)^{-\beta_{0}}\right) T_{\gamma}}{\bar{B}_\mathrm{factor}(\nu)} + \frac{T_{21}}{\bar{B}_\mathrm{factor}(\nu)}\Bigg] e^{-\tau_\mathrm{ion}(\nu)}\\
+ \frac{T_{\mathrm{e}}}{\bar{B}_\mathrm{factor}(\nu)}(1-e^{-\tau_\mathrm{ion}(\nu)}) \ .
\end{multline}

The first and second terms in \Cref{Eq:BFCCdataModel} describe the spatially-isotropic and -anisotropic subcomponents of the power-law component of the foreground emission, respectively. The third term accounts for the beam-factor weighted (following BFCC) CMB temperature, along with the spatially isotropic subcomponent of the foreground emission where the beam-factor does not cancel during BFCC. The fourth term represents the beam-factor weighted global 21-cm signal temperature. The common product of the terms in square brackets, $e^{-\tau_\mathrm{ion}(\nu)}$, models ionospheric absorption, with the effective ionospheric optical depth modelled as $\tau_\mathrm{ion} = \tau_0(\nu/\nu_\mathrm{c})^{-2}$. Finally, the fifth term models the beam-factor weighted net emission from hot electrons in the ionosphere.

\Cref{Eq:BFCCdataModel} has $N_\mathrm{pert} + 2$ foreground parameters, 2 ionospheric parameters and $N_{21}$ 21-cm model parameters. Of the foreground model parameters, $\bar{T}_\mathrm{m_{0}}$ describes the time- and sky-averaged non-21-cm-signal component of the sky brightness temperature at reference frequency $\nu_\mathrm{c}$. $\beta_{0}$ describes the mean temperature spectral index of the power law component of the foreground emission. $p_{\alpha}$ describes the fractional amplitude of the $\alpha$th log-polynomial model vector for describing spectral fluctuations about the sky-averaged spectrum of the foreground brightness temperature field, normalised to the fractional amplitude of the perturbation relative to the mean brightness temperature at the reference frequency $\nu_\mathrm{c}$. We use Bayesian model comparison to determine the preferred number of log-polynomial model vectors to describe the data, $N_\mathrm{pert}$. The two free parameters of the ionospheric model, $T_{\mathrm{e}}$ and $\tau_0$ describe the temperature of ionospheric electrons and the effective ionospheric optical depth at $\nu_\mathrm{c}$, respectively. For the flattened Gaussian 21-cm absorption trough considered in this work, $N_{21}=4$ and the parameters of the model are the amplitude, $A$, central frequency, $\nu_0$, width, $w$ and flattening, $\tau$, of the absorption trough (see \Cref{Eq:FlattenedGaussian}).

Of the above parameters, one can define physical priors for $\bar{T}_\mathrm{m_{0}}$, $\beta_{0}$, $T_{\mathrm{e}}$ and  $\tau_\mathrm{ion}$ based on existing observations (see Paper I and references therein for details). The $p_{\alpha}$ parameters correspond to the temperatures of individual perturbation spectral model vectors at reference frequency $\nu_\mathrm{c} = 75~\mathrm{MHz}$. The fraction of the antenna temperature described by these terms is expected to be small relative to $\bar{T}_\mathrm{m_{0}}$. In Paper I, it was found that limiting individual perturbation model vectors to $10\%$ absolute fractional perturbations provided sufficient flexibility to accurately model simulated foreground-only BFCC data, without adding a significant degree of superfluous flexibility. We adopt the same range here.

Following Paper I, we incorporate this information when fitting \Cref{Eq:BFCCdataModel} to the simulated data in \Cref{Sec:Results}, in a conservative manner, using broad physical priors on the parameters of the model as listed in \Cref{Tab:DataModelPriors}.

\begin{table*}
\caption{Priors on the parameters of the foreground and ionospheric models defined in \Cref{Sec:DataModels} and fit in \Cref{Sec:Results}.
}.
\centerline{
\begin{tabular}{l l l l }
\hline
Model & Parameter & Model component & Prior     \\
\hline
BFCC & $\bar{T}_\mathrm{m_{0}}$     & foreground & $U(1000,6000)~\mathrm{K}$ \\
& $\beta_{0}$     & foreground & $U(2.0,3.0)$ \\
& $p_{\alpha}$     & foreground & $U(-0.1,0.1)$ \\
& $T_{\mathrm{e}}$     & ionosphere & $U(100,800)~\mathrm{K}$ \\
& $\tau_0$     & ionosphere & $U(0.005,0.025)$ \\
\hline
Intrinsic & $b_{0}$     & foreground & $U(1000,6000)~\mathrm{K}$ \\
& $b_{1}$     & foreground & $U(-0.5,0.5)$ \\
& $b_{2}$     & foreground & $U(0,0.2)$ \\
& $b_{3}$     & ionosphere & $U(0.005,0.025)$  \\
& $b_{4}$     & ionosphere & $U(0.5,20.0)~\mathrm{K}$ \\
\hline
LinPhys & $a_{0}$     & foreground + ionosphere & $U(1000,6000)~\mathrm{K}$ \\
& $a_{1 \cdots 4}$     & foreground + ionosphere & $U(-10^{4},10^{4})$ \\
\hline
MultLin & $c_{0}$     & foreground + ionosphere & $U(1000,6000)~\mathrm{K}$ \\
& $c_{i}$     & foreground + ionosphere & $U(-10^{4},10^{4})$ \\
\hline
\end{tabular}
}
\label{Tab:DataModelPriors}
\end{table*}

\subsection{Intrinsic foreground model}
\label{Sec:IntrinsicForegroundModel}

A detailed description and derivation of the Intrinsic sky model is given in Paper I. Here, we quote the final form of the model:
\begin{multline}
\label{Eq:B18IntrinsicForegroundModel}
T_\mathrm{Intrinsic,fg}^\mathrm{model}(\nu) = b_{0}\left(\frac{\nu}{\nu_\mathrm{c}} \right)^{-2.5 + b_{1} + b_{2}\log\left(\frac{\nu}{\nu_\mathrm{c}} \right)} \mathrm{e}^{-b_{3}\left(\frac{\nu}{\nu_\mathrm{c}} \right)^{-2}} + b_{4}\left(\frac{\nu}{\nu_\mathrm{c}} \right)^{-2}
\ .
\end{multline}

In this construction, $b_{i}$ with $i \in [0,\cdots,4]$ are foreground and ionospheric parameters to be determined in the fit of the model to the data and they acquire direct interpretations in terms of physical properties of the foreground sky and the ionosphere in \Cref{Eq:BFCCdataModel} as follows (see Paper I, Appendix E for details).
\begin{itemize}
  \item $b_0 = \bar{T}_\mathrm{m_{0}} e^{-\tau_{\mathrm{ion}}(\nu)}$ is the (attenuated) mean amplitude of the foreground power-law emission at the reference frequency $\nu_c$.
  \item $b_1 = 2.5 - \beta_0$, with $\beta_0 \approx 2.5$ the power-law spectral index of the radio foreground emission, in the $50 \lesssim \nu \lesssim 190~\mathrm{MHz}$ band, when the Galactic Centre is in the sky (e.g. \citealt{2017MNRAS.464.4995M,2019MNRAS.483.4411M}), and $b_1$ represents a deviation from this value.
  \item $b_2 = \sigma_\beta^2 / 2$, where $\sigma_\beta^2$ is the variance of the spectral index across the sky. This term encodes the amount of foreground spectral curvature generated by averaging over the spatially dependent spectral index distribution visible to the instrument.
  \item $b_3 = \tau_0$ gives the ionospheric opacity at $\nu_c$ (with frequency scaling as $[\nu/\nu_c]^{-2}$).
  \item $b_4 = T_e \tau_0$, with $T_e$ as the electron temperature.
\end{itemize}

The priors we use when fitting \Cref{Eq:B18IntrinsicForegroundModel} are listed in \Cref{Tab:DataModelPriors}. They are set to be equivalent to the priors on $\bar{T}_\mathrm{m_{0}}$, $\beta_{0}$, $T_{\mathrm{e}}$ and $\tau_\mathrm{ion}$ in the BFCC model.

\subsection{LinPhys foreground model}
\label{Sec:LinPhysForegroundModel}

Assuming $b_{i} \ll 1$ with $i \in [1,2,3]$, the linearisation of \Cref{Eq:B18IntrinsicForegroundModel}, over these parameters, will accurately approximate the full non-linear model. Performing this linearisation yields the polynomial foreground model used for recovery of the 21-cm signal in B18 (their Equation 1),
\begin{multline}
\label{Eq:B18LinPhysForegroundModel}
T_\mathrm{LinPhys}^\mathrm{model}(\nu) = a_{0}\left(\frac{\nu}{\nu_\mathrm{c}} \right)^{-2.5} +
a_{1}\left(\frac{\nu}{\nu_\mathrm{c}} \right)^{-2.5}\log\left(\frac{\nu}{\nu_\mathrm{c}} \right) +  \\
a_{2}\left(\frac{\nu}{\nu_\mathrm{c}} \right)^{-2.5}\left[\log\left(\frac{\nu}{\nu_\mathrm{c}} \right)\right]^{2} +
a_{3}\left(\frac{\nu}{\nu_\mathrm{c}} \right)^{-4.5} +
a_{4}\left(\frac{\nu}{\nu_\mathrm{c}} \right)^{-2}
\ .
\end{multline}

A more detailed discussion of the linearisation of \Cref{Eq:B18IntrinsicForegroundModel} can be found in \citet[hereafter H18]{2018Natur.564E..32H}. In brief, this linearisation is performed by Taylor expanding \Cref{Eq:B18IntrinsicForegroundModel} about the point $b_{i} = 0$ with $i \in [1,2,3]$ and retaining terms up to second order in these parameters. The resulting linearised model is a polynomial in $\nu$ with coefficients $a_{i}$, which are related to the coefficients of the non-linear Intrinsic model, $b_{i}$, and thus the physical parameters of the BFCC model, \Cref{Eq:BFCCdataModel}, as follows:
\begin{align}
\label{Eq:LinPhysCoeffDefinitions}
a_{0} &=  b_{0} = \bar{T}e^{-\tau_\mathrm{ion}(\nu)} \ ,\\ \nonumber
a_{1} &= b_{0}b_{1} = (2.5 - \beta_{0})\bar{T}e^{-\tau_\mathrm{ion}(\nu)} \ ,\\ \nonumber
a_{2} &= b_{0}(b_{1}^{2}/2 + b_{2}) = \frac{\bar{T}e^{-\tau_\mathrm{ion}(\nu)} [(2.5 - \beta_{0})^{2} + \sigma_{\beta}^{2}]}{2}\ ,\\ \nonumber
a_{3} &= -b_{0}b_{3} = -\tau_0\bar{T}e^{-\tau_\mathrm{ion}(\nu)} \ ,\\ \nonumber
a_{4} &= b_{4} = T_{\mathrm{e}}\tau_0 \ .
\end{align}
From \Cref{Eq:LinPhysCoeffDefinitions}, it can be seen that for \Cref{Eq:B18LinPhysForegroundModel} to provide a physical model for the emission components it describes, it is necessary that $a_{i}$ with $i \in [0,2,4]$ are strictly positive and $a_{3}$ is strictly negative. These constraints were not imposed when fitting the EDGES-low data in B18. As such, the component of the data fit using \Cref{Eq:B18LinPhysForegroundModel} was not limited to astrophysical and ionospheric effects. The increased flexibility of \Cref{Eq:B18LinPhysForegroundModel} in the absence of physical priors increases the level of correlation between the nominal-foreground and 21-cm components of the model; however, it has the benefit that the LinPhys has some additional flexibility to model systematics such as those expected to arise from imperfect correction of the data for antenna chromaticity (Paper I)\footnote{Receiver calibration error has also been identified as a possible source of systematic structure (e.g. \citealt{2018Natur.564E..35B, 2022MNRAS.517.2264M}); however, we leave more detailed investigation of this possibility to future work.}.

The fact that the maximum likelihood parameters of this nominal-foreground component of the sky model, recovered when jointly fitting it with a flattened Gaussian 21-cm model in B18, do not respect the physicality constraints given above (see e.g. \citealt{2018Natur.564E..32H}) indicates that these parameters are indeed being used, in part, to model systematic structures in the data in the fits presented in B18. However, given that the LinPhys model is not explicitly tailored to fitting the non-intrinsic sky structure remaining after BFCC, it is unclear, a priori, whether this additional flexibility is sufficient to model such systematics. This will be tested in \Cref{Sec:Results}, where we will fit realistic simulated EDGES low-band data with \Cref{Eq:B18LinPhysForegroundModel} using the broad priors listed in \Cref{Tab:DataModelPriors}.

\subsection{MultLin foreground model}
\label{Sec:MultLinForegroundModel}

For a number of 21-cm signal recovery tests, in place of \Cref{Eq:B18LinPhysForegroundModel}, B18 use a more general polynomial model (their Equation 2). This model has the form,
\begin{equation}
\label{Eq:B18MultLinForegroundModel}
T_\mathrm{MultLin}^\mathrm{model}(\nu) = \sum\limits_{n=0}^{N-1}c_{n}\nu^{n-2.5}
\ .
\end{equation}
Here, the exponent $-2.5$ is chosen for the same reason as in the Intrinsic model: to enable more accurate modelling of the dominant synchrotron component of the foreground emission. Additional terms aim to model higher order spectral structure in the foreground emission and can also partially capture some instrumental effects, such as additional spectral structure from chromatic beams or small errors in calibration (B18).

\section{Results}
\label{Sec:Results}

\Cref{Tab:HPDsummaryTable} of \Cref{Sec:HPDsummaryTable} summarises the 21-cm signal detection and parameter inference for the four models -- BFCC, Intrinsic, LinPhys, and MultLin -- across three simulated 21-cm signal amplitude scenarios (null test, moderate amplitude, and high amplitude). We describe the results in detail below.

\subsection{BaNTER validation results}
\label{Sec:BaNTERValidationResults}

\subsubsection{Null test}
\label{Sec:NullTest}

\begin{figure}
	\centerline{
        \includegraphics[width=0.5\textwidth, trim={-2.5cm 0.0 0.5cm 0}, clip]{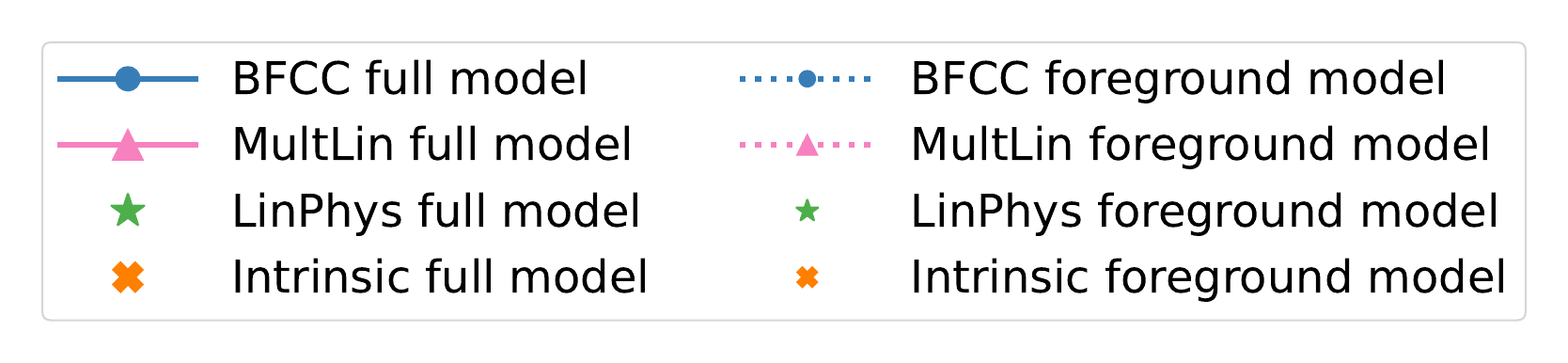}
        }
        \centerline{
            \includegraphics[width=0.5\textwidth]{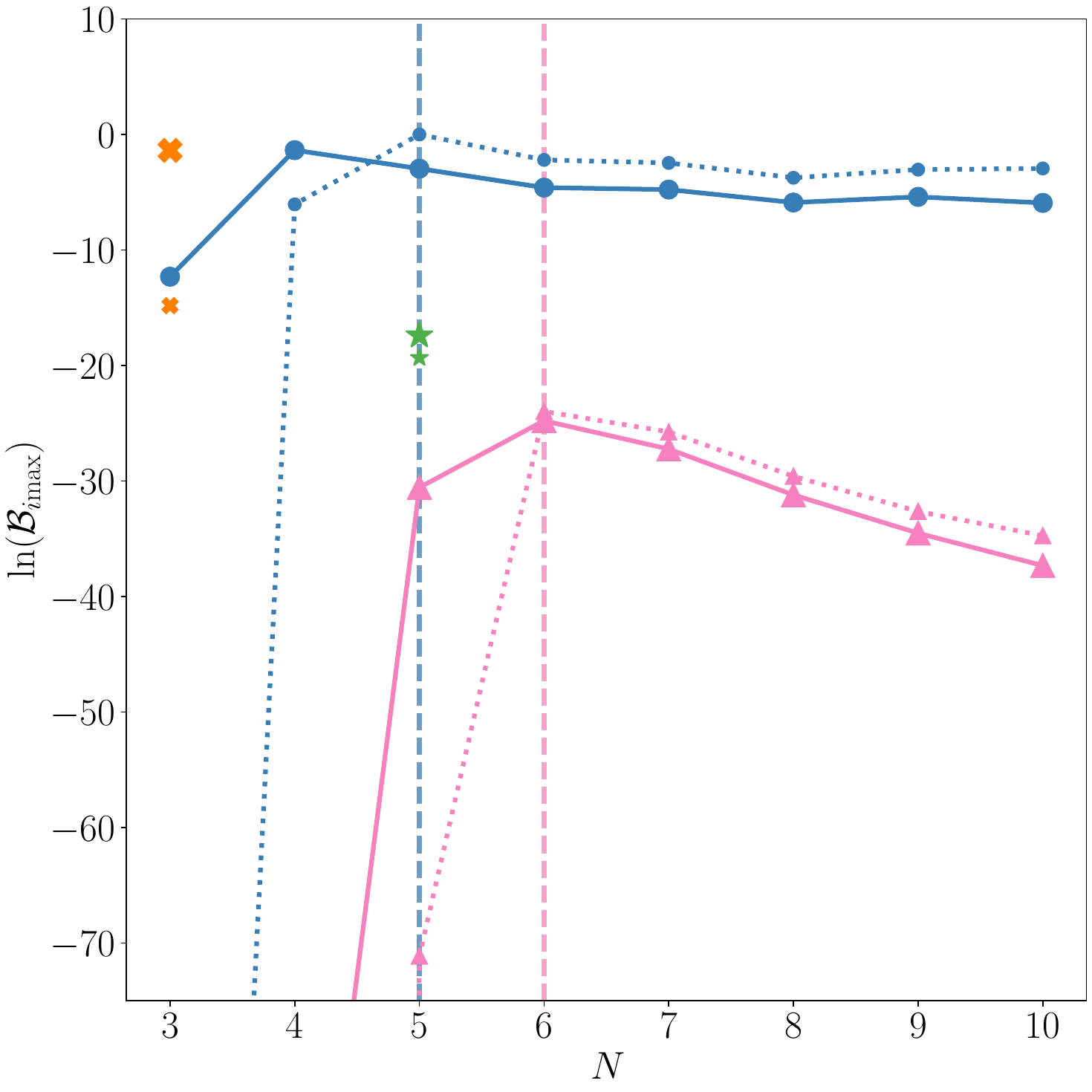}
        }
\caption{
    Bayes factors ($\mathcal{B}_{i\mathrm{max}}$) of model $\bm{M}_{i}$ relative to the highest-evidence model $\bm{M}_{\mathrm{max}}$ for the foreground-only validation data set $\bm{D}_{\mathrm{v}}$. The parameter count $N$ denotes the number of terms in the model component with a priori unknown complexity (see main text). In the flexible-complexity BFCC and MultLin parametrisations, models that include a 21-cm signal are connected by solid lines, and those without a 21-cm signal are connected by dotted lines. In both these and the fixed-complexity Intrinsic and LinPhys models, the presence or absence of a 21-cm signal is also indicated by large and small symbols, respectively (see legend). Models with maximum total evidence in the BFCC and MultLin classes are marked by vertical dashed lines in blue and pink, respectively.
    }
\label{Fig:ForegroundOnlyBayesFactors}
\end{figure}

\begin{table}
    \caption{
        BaNTER model validation results. Bayes factors between $\bm{M}_{i\mathrm{c}}$ and $\bm{M}_{i\mathrm{Fg}}$, as models for the validation data, where $i$ runs over the models defined in \Cref{Sec:DataModels}. A positive $\ln(\mathcal{B}_\mathrm{cFg}^\mathrm{v})$ indicates that the composite model, $\bm{M}_{i\mathrm{c}}$, is preferred over the foreground model, $\bm{M}_{i\mathrm{Fg}}$, for the foreground-only validation data, $\bm{D}_\mathrm{v}$, and the model has failed the BaNTER null test. The reverse is true when $\ln(\mathcal{B}_\mathrm{cFg}^\mathrm{v})$ is negative.
    }
    \centerline{
        \begin{tabular}{l l l l}
            \hline
            Model & $\ln(\mathcal{B}_\mathrm{cFg}^\mathrm{v})$ & Pass/Fail & Comment \\
            \hline
            BFCC $(N=3)$ & 203.3 & Fail & Spurious signal detection \\
            BFCC $(N=4)$ & 4.7 & Fail & Spurious signal detection \\
            BFCC $(N=5)$ & -3.0 & Pass &  \\
            BFCC $(N=6)$ & -2.4 & Pass &  \\
            BFCC $(N=7)$ & -2.3 & Pass &  \\
            BFCC $(N=8)$ & -2.1 & Pass &  \\
            BFCC $(N=9)$ & -2.4 & Pass &  \\
            BFCC $(N=10)$ & -3.0 & Pass &  \\
            MultLin $(N=3)$ & 8396.4 & Fail & Spurious signal detection \\
            MultLin $(N=4)$ & 769.3 & Fail & Spurious signal detection \\
            MultLin $(N=5)$ & 40.6 & Fail & Spurious signal detection \\
            MultLin $(N=6)$ & -0.8 & Pass &  \\
            MultLin $(N=7)$ & -1.5 & Pass &  \\
            MultLin $(N=8)$ & -1.6 & Pass &  \\
            MultLin $(N=9)$ & -1.9 & Pass &  \\
            MultLin $(N=10)$ & -2.6 & Pass &  \\
            LinPhys $(N=5)$ & 1.9 & Fail & Moderate preference for $\bm{M}_{i\mathrm{c}}$ over \\
            & & & $\bm{M}_{i\mathrm{Fg}}$, but below the spurious  \\
            & & & signal detection threshold\\
            Intrinsic $(N=3)$ & 13.4 & Fail & Spurious signal detection \\                        \hline
    \end{tabular}
    }
\label{Tab:lnBcb}
\end{table}

\Cref{Fig:ForegroundOnlyBayesFactors} shows the Bayes factors ($\mathcal{B}_{i\mathrm{max}}$) between model $\bm{M}_{i}$ and $\bm{M}_\mathrm{max}$ for the validation data generated as described in \Cref{Sec:Simulations}. Here, $i$ runs over all models in $\bm{\mathcal{M}}$, we perform BFBMC over models including (solid lines and/or large symbols) and excluding (dotted lines and/or small symbols) a 21-cm signal component, and the validation data contains simulated observations of foreground emission and noise, corresponding to a scenario where no observable 21-cm signal exists within the observation band of interest. Such a situation could arise, for instance, if the first stars had not yet formed during the redshift interval corresponding to the frequency range of the data\footnote{$ z = \frac{\nu_{21}}{\nu_\mathrm{obs}} - 1 $, with $\nu_{21} \simeq 1420.4~\mathrm{MHz}$.}.

Following Paper I, we plot the Bayes factor as a function of $N$, the number of parameters associated with the component of the model whose complexity is a priori unknown. In models that incorporate a foreground component designed to describe a combination of effects that cannot be physically separated (the LinPhys and MultLin models), $N$ corresponds to the complexity of this component. In models where the foreground can be explicitly decomposed into physically motivated ionospheric and astrophysical subcomponents (the Intrinsic and BFCC models), with only the complexity of the astrophysical foreground subcomponent being a priori unknown, $N$ refers to the complexity of the latter subcomponent. $\bm{M}_\mathrm{max}$ is the model with the highest Bayesian evidence for the data, which we find to be the BFCC model with $N=5$ terms and no 21-cm signal component.

\Cref{Tab:lnBcb} lists the corresponding values of $\ln(\mathcal{B}_\mathrm{cFg}^\mathrm{v})$, the Bayes factors between $\bm{M}_{i\mathrm{c}}$ and $\bm{M}_{i\mathrm{Fg}}$ as models for the validation data, $\bm{D}_\mathrm{v}$. For positive $\ln(\mathcal{B}_\mathrm{cFg}^\mathrm{v})$, $\bm{M}_{i\mathrm{c}}$ is preferred over $\bm{M}_{i\mathrm{Fg}}$. Since $\bm{S}_\mathrm{Fg}$ is the only signal component present in $\bm{D}_\mathrm{v}$, a preference for $\bm{M}_{i\mathrm{c}}$ indicates inaccuracy of $\bm{M}_{i\mathrm{Fg}}$ and any detection of $\bm{S}_{21}$ in the validation data is necessarily spurious.

\subsubsection{Spurious signal detection and bias predictions with unvalidated models}
\label{Sec:SpuriousSignalDetection}

Using the composite model validation criteria defined in \Cref{Sec:BayesianNullTest}, we find that the Intrinsic model, LinPhys model, and variants of the BFCC with $N < 5$, as well as variants of the MultLin model with $N < 6$ foreground terms, exhibit evidence in favour of including a spurious 21-cm component when fitting the non-21-cm simulated validation data. These models thus fail the null test.

It follows from this result that, if these models were used to analyse an equivalent data set containing a significant 21-cm signal, they would fit a combination of the true 21-cm signal and the systematics that caused them to fail BaNTER validation. As a result, these models would produce biased estimates of the 21-cm signal. The extent of this bias depends on the degree to which the sum of the true 21-cm signal and the systematics are fittable with the 21-cm model. This, in turn, depends on the level of systematics in the data, the amplitude and shape of the true 21-cm signal, and the flexibility of the 21-cm model.

In \Cref{Tab:lnBcb}, we note that most models failing the BaNTER null test do so with sufficiently large $\ln(\mathcal{B}_\mathrm{cFg}^\mathrm{v})$ values to yield spurious signal detections in a foreground-only data set. The exception is the LinPhys $(N=5)$ model, which fails the BaNTER null test with only a moderate preference for $\bm{M}_{i\mathrm{c}}$ over $\bm{M}_{i\mathrm{Fg}}$, but below the spurious signal detection threshold ($\ln(\mathcal{B}_\mathrm{cFg}^\mathrm{v}) = 3$) defined in \Cref{Sec:BayesianNullTest}. This suggests that we should expect this model to yield biased inferences of the 21-cm signal, but with the level of that bias being lower than for the other models that fail the BaNTER null test.

\subsubsection{BaNTER validation results as binary model priors}
\label{Sec:BaNTERValidationResultsAsModelPriors}

Based on the BaNTER null test results, we judge the Intrinsic and LinPhys models, as well as variants of the BFCC with $N < 5$ and variants of the MultLin model with $N < 6$ foreground terms as inadequate for reliably recovering unbiased estimates of the 21-cm signal. To confirm this conclusion, we perform two model comparison analyses in \Cref{Sec:ModerateAmplitude,Sec:HighAmplitude}:

\begin{enumerate}
    \item \textit{Bayesian model comparison with uninformative model priors:} An unvalidated model comparison analysis is conducted, where the null test is not applied, and all models are treated as equally likely a priori.
    \item \textit{BaNTER-validated posterior-odds-based Bayesian model comparison:} In this case, models that fail the BaNTER null test are assigned negligible prior odds of yielding unbiased 21-cm signal estimates and are excluded from $\bm{\mathcal{M}}$. We denote the resulting validated subset as $\bm{\mathcal{M}}_\mathrm{v}$. Models in $\bm{\mathcal{M}}_\mathrm{v}$ are treated as equally likely a priori and are weighted by their Bayesian evidence as models for the observational data a posteriori.
\end{enumerate}

\subsection{Moderate amplitude 21-cm signal}
\label{Sec:ModerateAmplitude}

\begin{figure*}
	\rightline{
	\includegraphics[width=0.93\textwidth, trim={0cm 0cm 0.7cm 0cm}, clip]{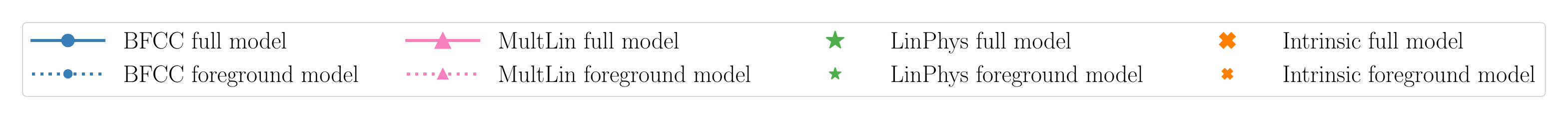}
	}
	\centerline{
        \includegraphics[width=0.5\textwidth]{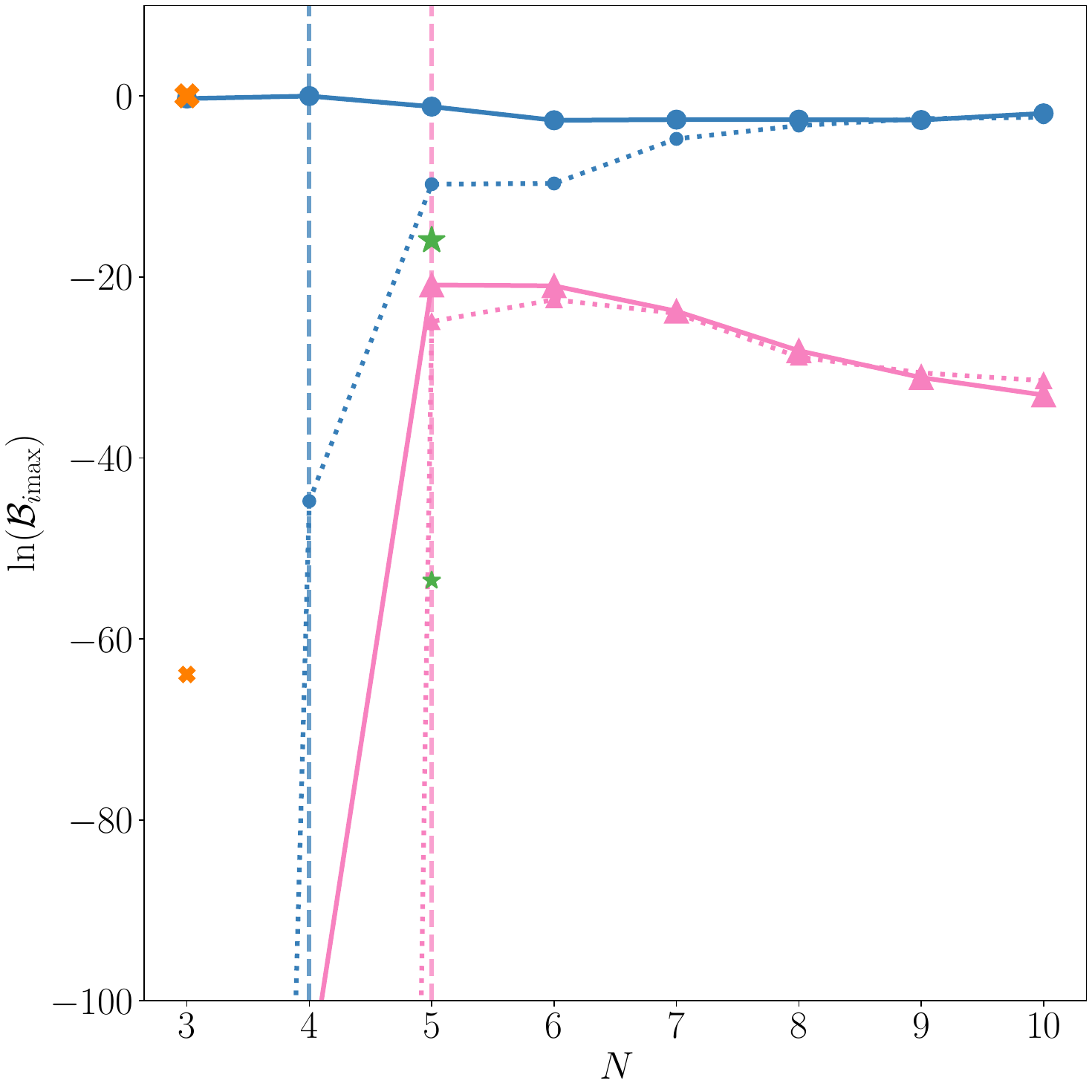}
        \includegraphics[width=0.5\textwidth]{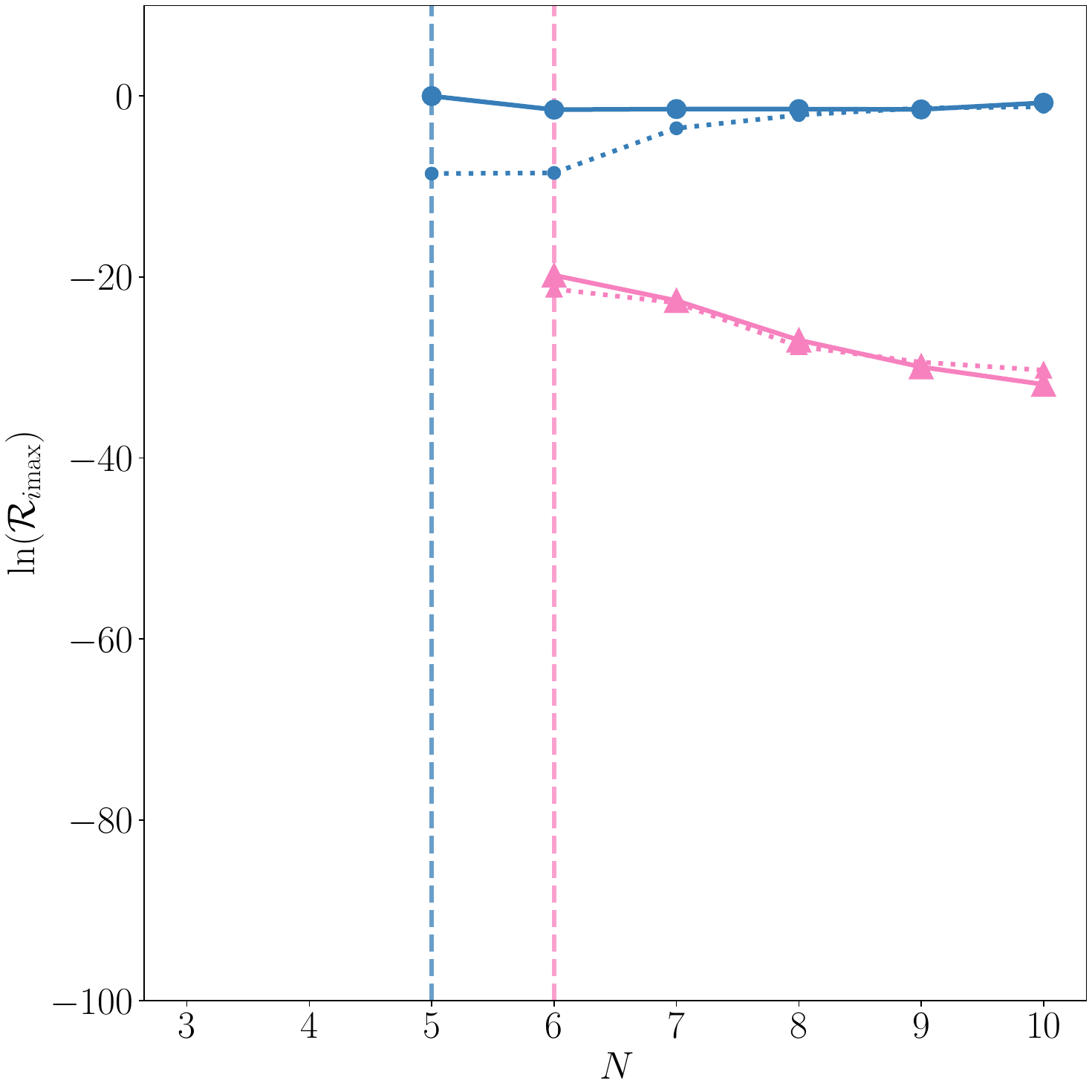}
	}
    \caption{
        Results of the Bayesian comparison of models for simulated data incorporating a moderate amplitude 21-cm signal ($A = 150~\mathrm{mK}$). \textbf{Left}: Bayes factors ($\mathcal{B}_{i\mathrm{max}}$) comparing model $\bm{M}_{i}$ to the maximum evidence model, $\bm{M}_\mathrm{max}$. Here, $i$ runs over all models in the set $\bm{\mathcal{M}}$, which includes both the models that pass and those that fail BaNTER validation). \textbf{Right}: Posterior odds ($\mathcal{R}_{i\mathrm{max}}$) of model $\bm{M}_{i}$ over the validated model $\bm{M}_\mathrm{max, v}$. Here, $\bm{M}_\mathrm{max}$ and $\bm{M}_\mathrm{max, v}$ represent the models with the highest Bayesian evidence and posterior odds, respectively. Symbols and solid and dashed lines have the same meanings as in \Cref{Fig:ForegroundOnlyBayesFactors}. The subset of models that are present in the left panel but are absent in the right represent the set of models that failed the BaNTER null test. The number of foreground terms with the highest evidence (left) and highest posterior odds (right) for these models are indicated by blue and pink vertical, dashed lines, respectively.
    }
\label{Fig:BandRwithAeq150mK}
\end{figure*}

\newtcbox{\redsubfigurebox}[1][]{
    colback=red!5!white,
    colframe=red!50!black,
    rounded corners,
    nobeforeafter,
    box align=top,
    left=0pt,
    right=0pt,
    top=1pt,
    bottom=1pt,
    leftrule=0.01cm,
    rightrule=0.01cm,
    toprule=0.01cm,
    bottomrule=0.01cm,
    #1
    }

\newtcbox{\bluesubfigurebox}[1][]{
    colback=blue!5!white,
    colframe=blue!50!black,
    rounded corners,
    nobeforeafter,
    box align=top,
    left=0pt,
    right=0pt,
    top=1pt,
    bottom=1pt,
    leftrule=0.01cm,
    rightrule=0.01cm,
    toprule=0.01cm,
    bottomrule=0.01cm,
    #1
    }

\begin{figure*}
    \begin{minipage}{\textwidth}
        \redsubfigurebox{
        \begin{subfigure}[t]{0.32\textwidth}
            \caption{\Large{MultLin, $N=3$}}
            \label{Fig:ModerateAmpMultLin3}
            \includegraphics[width=\textwidth]{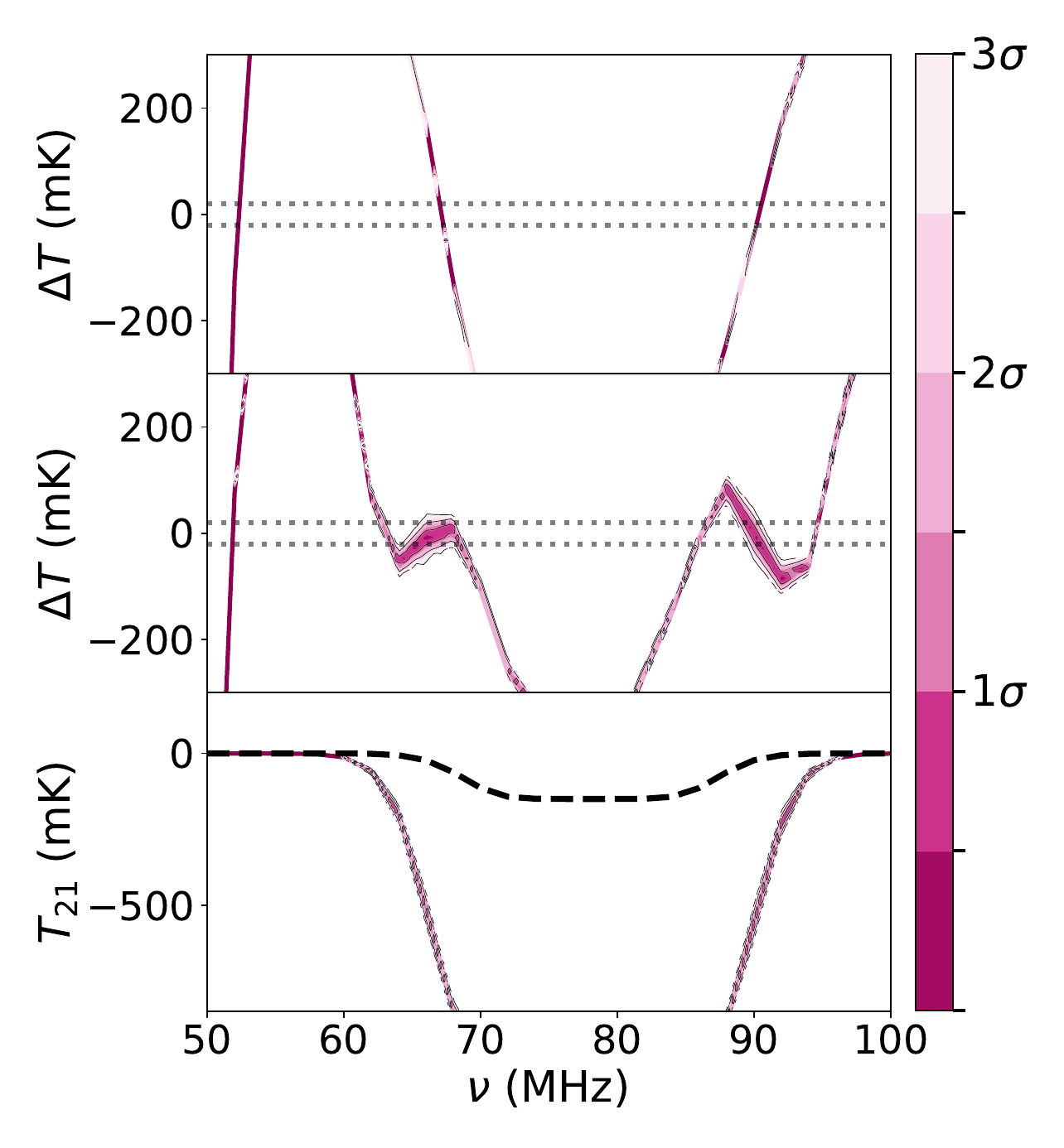}
        \end{subfigure}
        }
        \redsubfigurebox{
        \begin{subfigure}[t]{0.32\textwidth}
            \caption{\Large{MultLin, $N=4$}}
            \label{Fig:ModerateAmpMultLin4}
            \includegraphics[width=\textwidth]{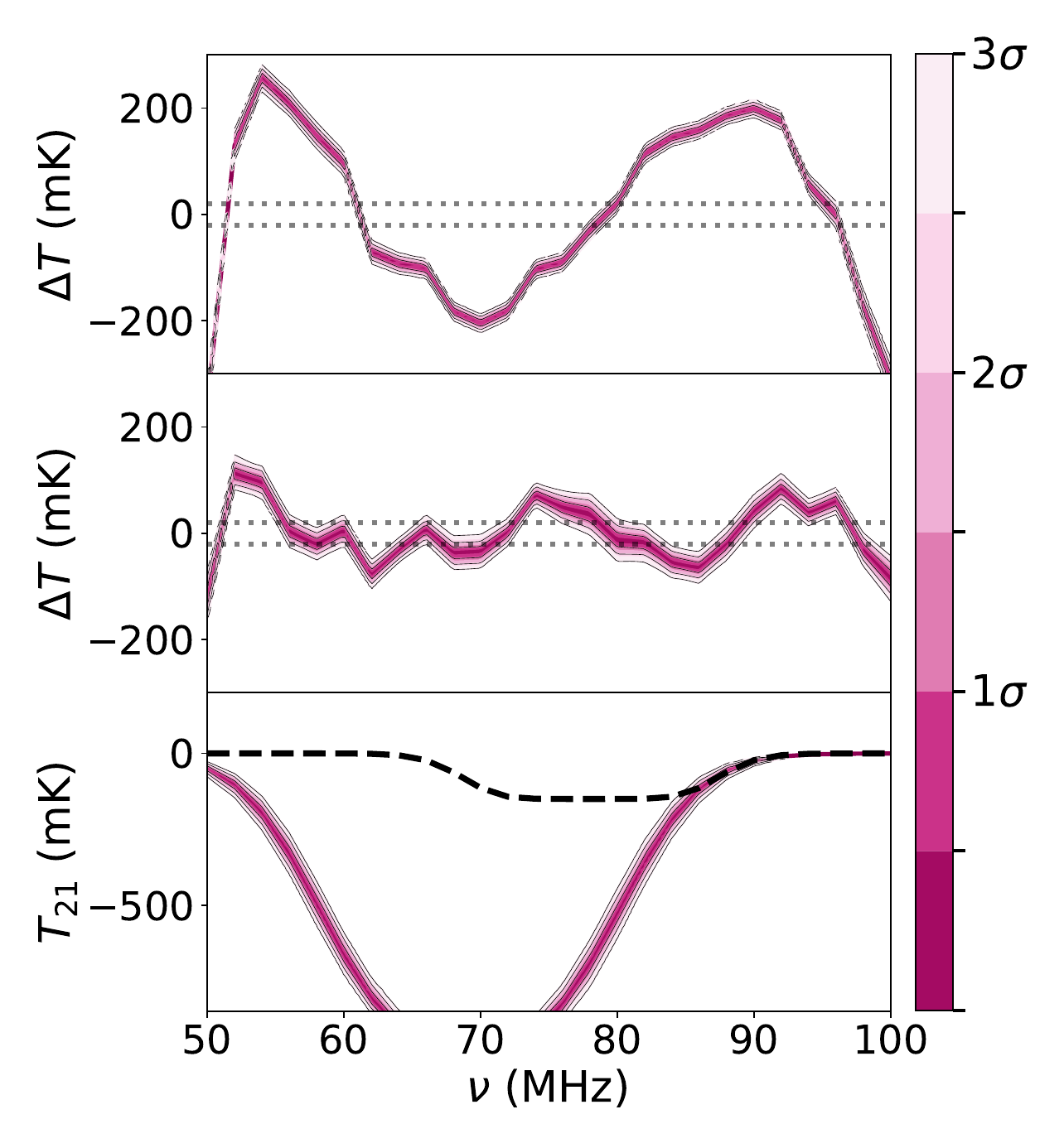}
        \end{subfigure}
        }
        \redsubfigurebox{
        \begin{subfigure}[t]{0.32\textwidth}
            \caption{\Large{MultLin, $N=5$}}
            \label{Fig:ModerateAmpMultLin5}
            \includegraphics[width=\textwidth]{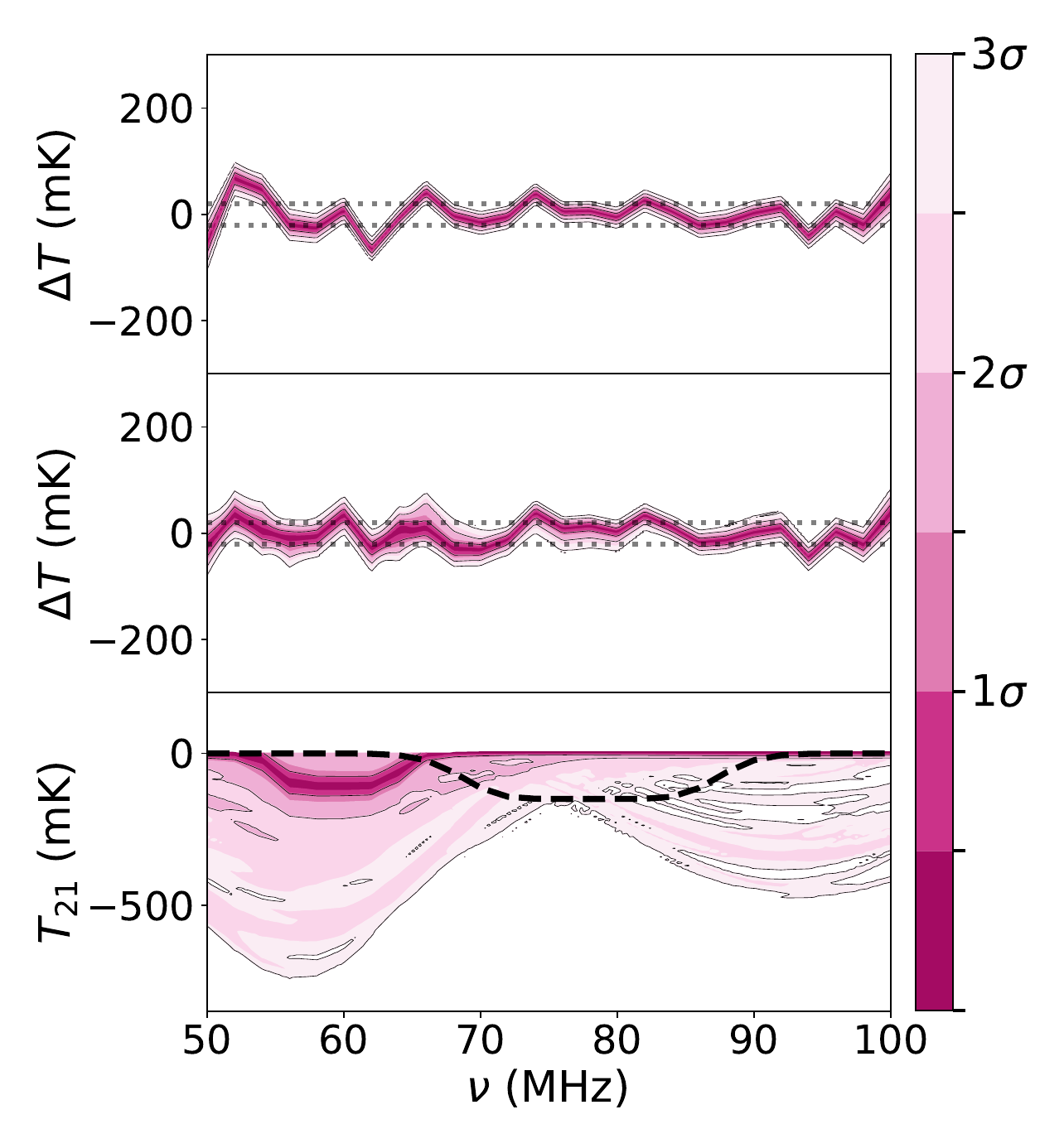}
        \end{subfigure}
        }
        \redsubfigurebox{
            \begin{subfigure}[t]{0.32\textwidth}
                \caption{\Large{LinPhys, $N=5$}}
                \label{Fig:ModerateAmpLinPhys5}
                \includegraphics[width=\textwidth]{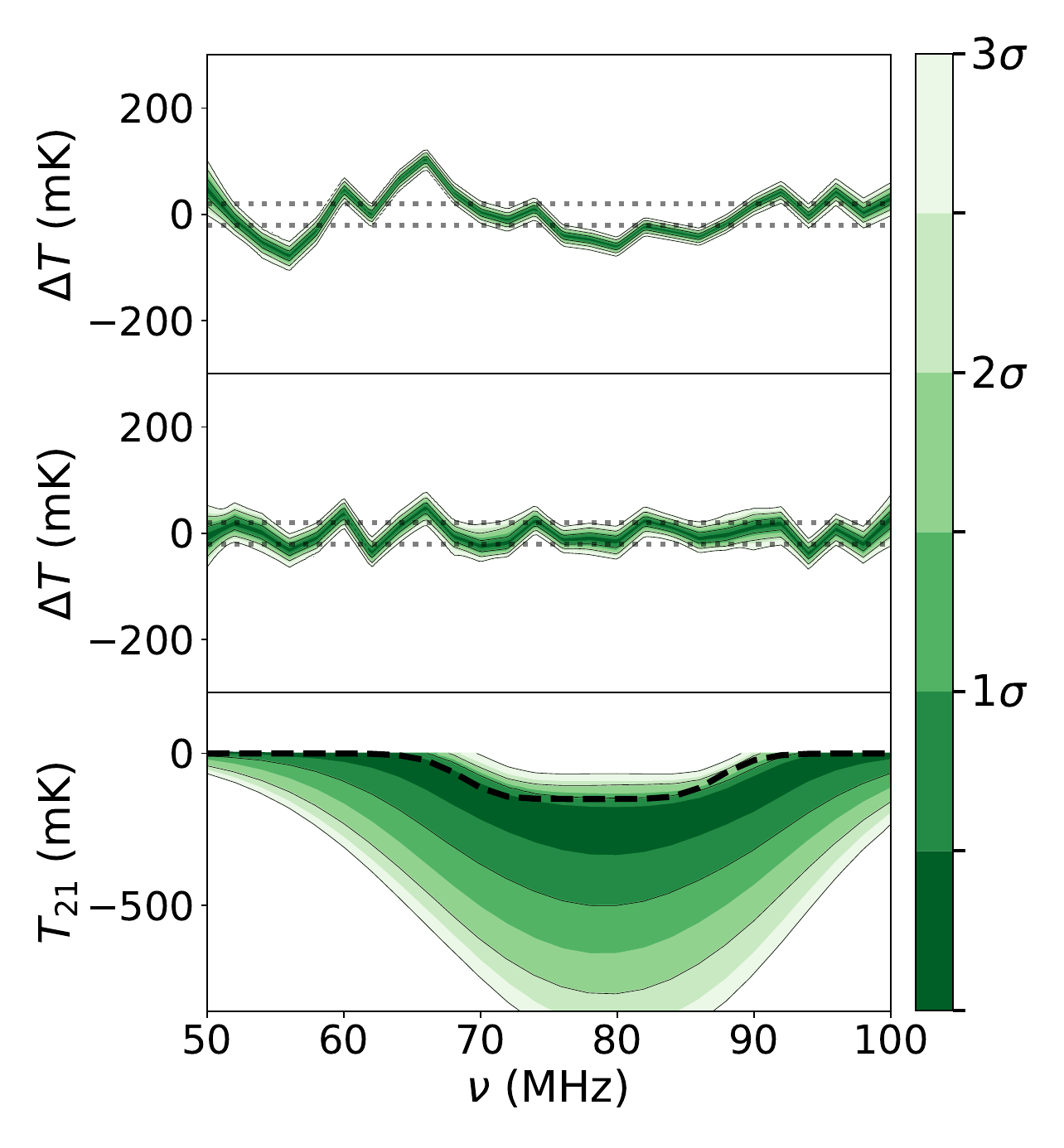}
            \end{subfigure}
        }
        \bluesubfigurebox{
            \begin{subfigure}[t]{0.32\textwidth}
                \caption{\Large{\textbf{\textit{BFCC, $\mathbfit{N=6}$}}}}
                \label{Fig:ModerateAmpBFCC6}
                \includegraphics[width=\textwidth]{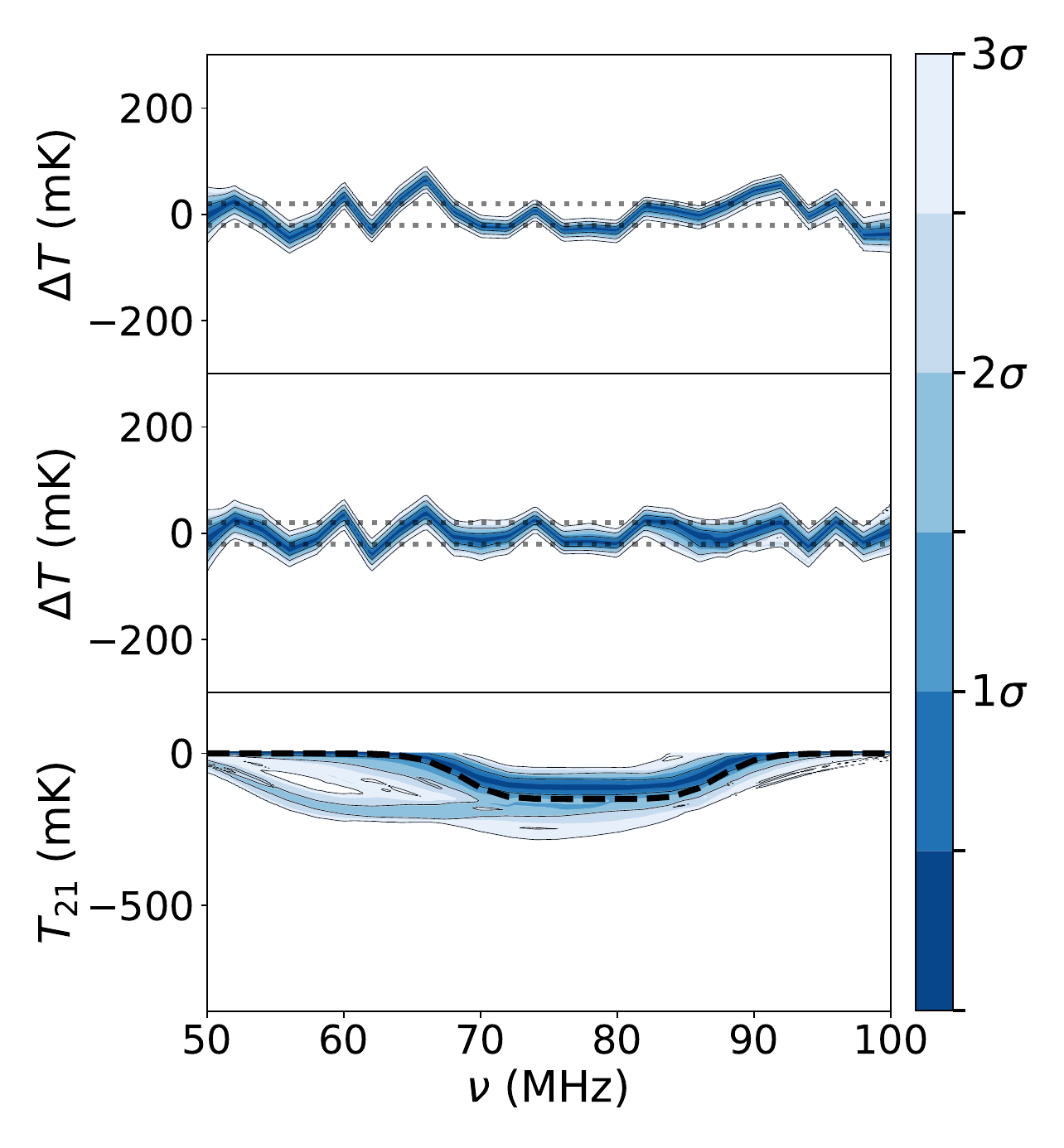}
            \end{subfigure}
        }
        \bluesubfigurebox{
            \begin{subfigure}[t]{0.32\textwidth}
                \caption{\Large{\textbf{\textit{BFCC, $\mathbfit{N=5}$}}}}
                \label{Fig:ModerateAmpBFCC5}
                \includegraphics[width=\textwidth]{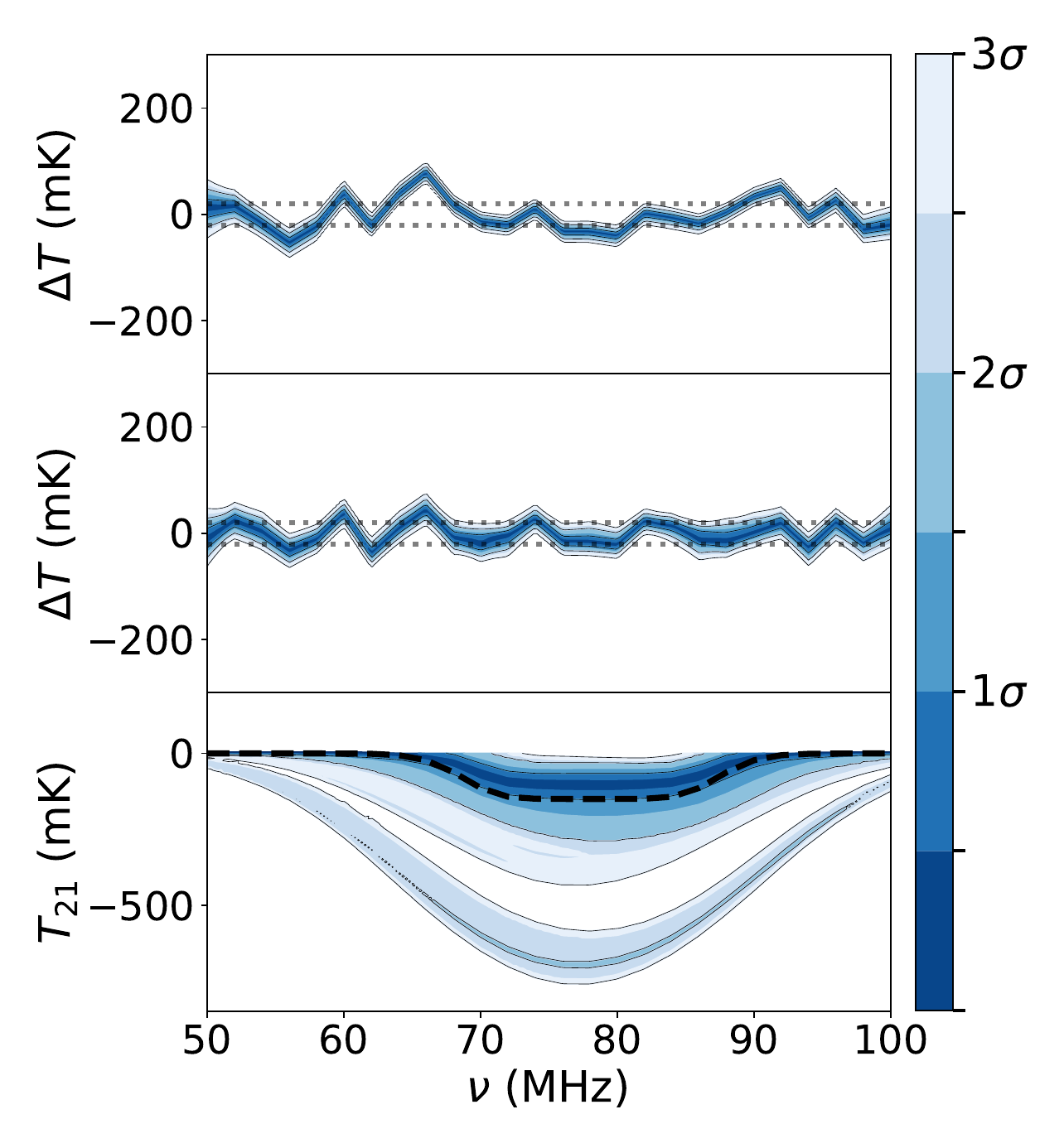}
            \end{subfigure}
        }
        \redsubfigurebox{
            \begin{subfigure}[t]{0.32\textwidth}
                \caption{\Large{\textbf{BFCC, $\mathbf{N=3}$}}}
                \label{Fig:ModerateAmpBFCC3}
                \includegraphics[width=\textwidth]{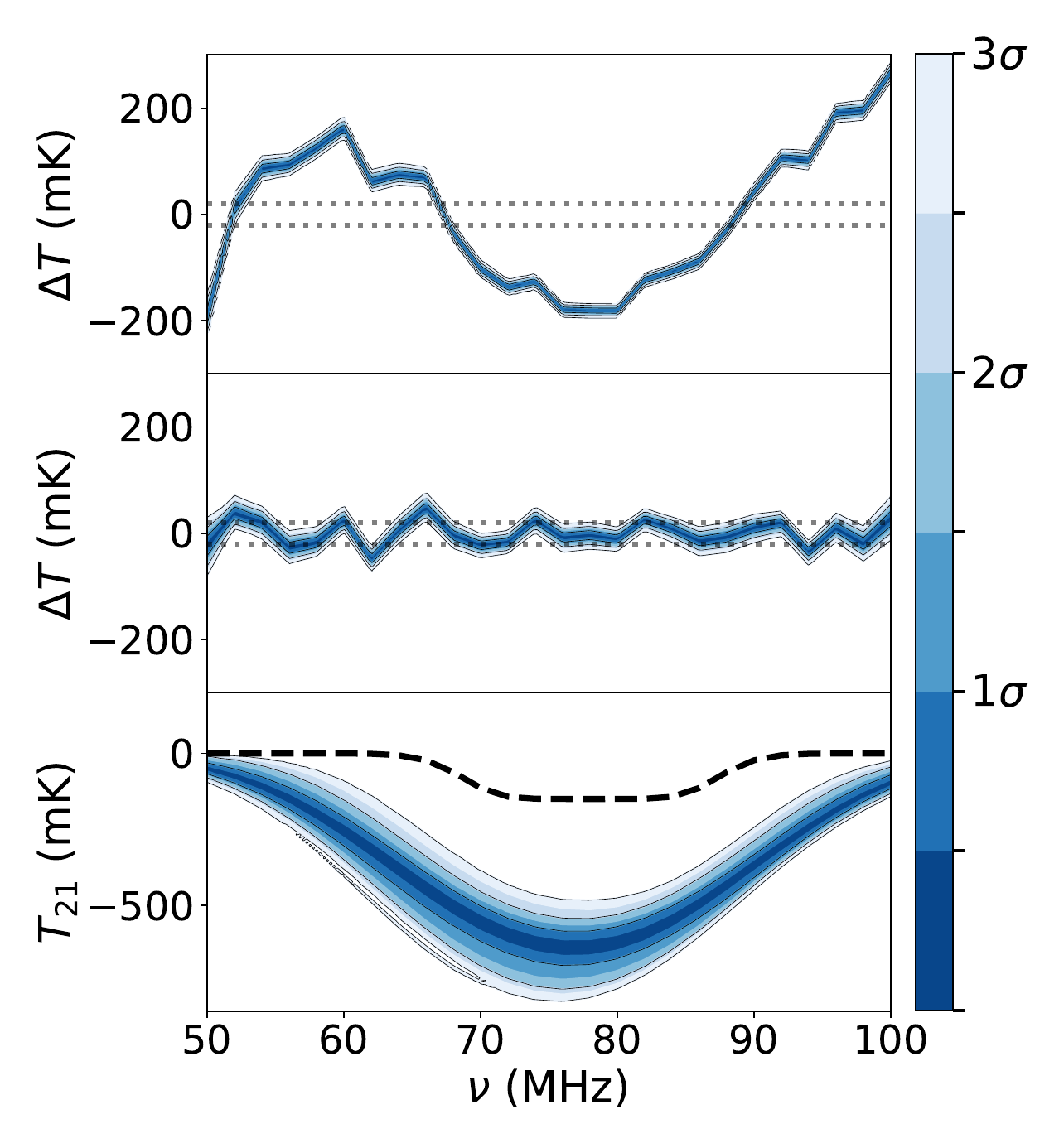}
            \end{subfigure}
        }
        \redsubfigurebox{
            \begin{subfigure}[t]{0.32\textwidth}
                \caption{\Large{\textbf{BFCC, $\mathbf{N=4}$}}}
                \label{Fig:ModerateAmpBFCC4}
                \includegraphics[width=\textwidth]{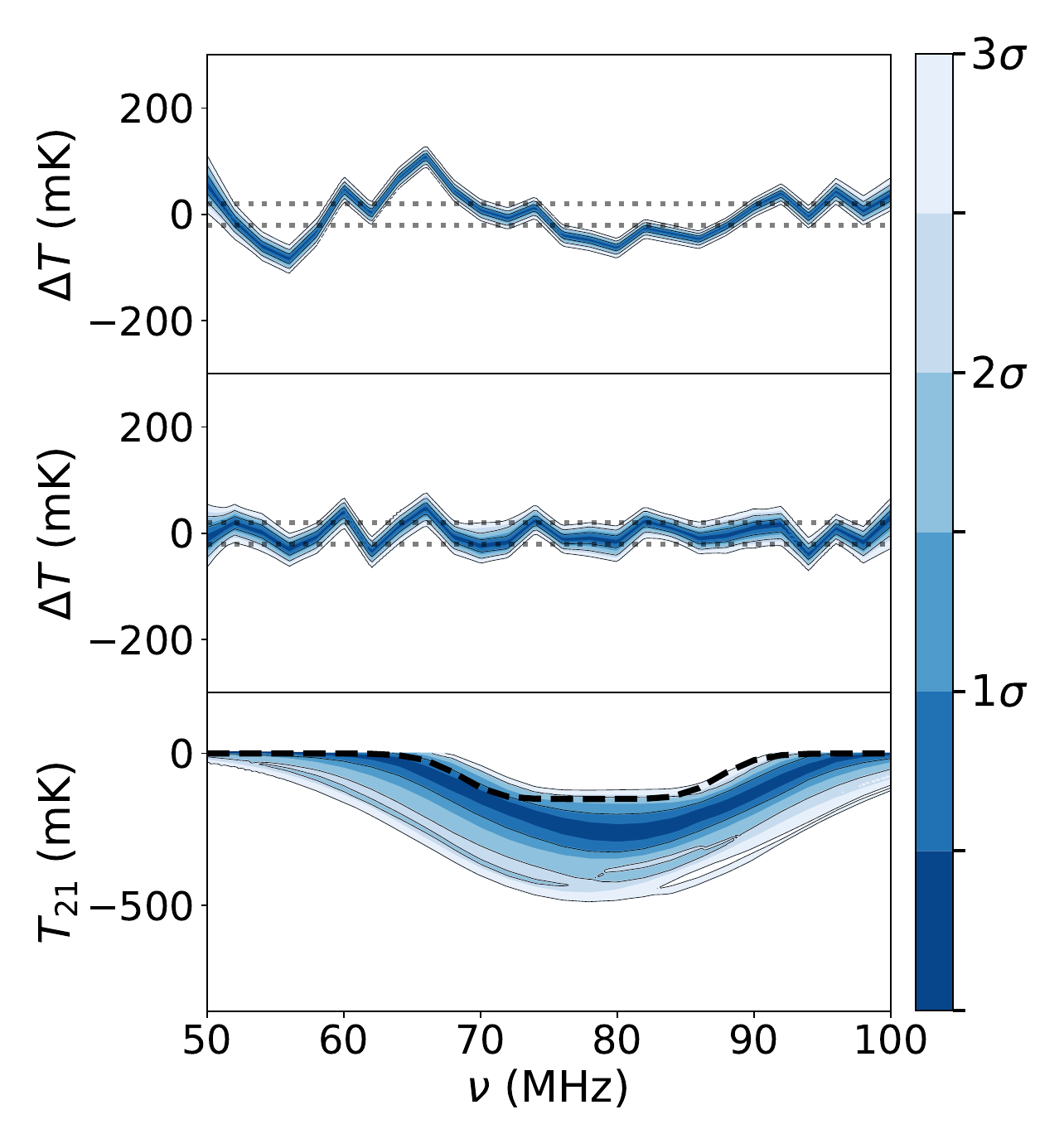}
            \end{subfigure}
        }
        \redsubfigurebox{
            \begin{subfigure}[t]{0.32\textwidth}
                \caption{\Large{\textbf{Intrinsic, $\mathbf{N=3}$}}}
                \label{Fig:ModerateAmpIntrinsic3}
                \includegraphics[width=\textwidth]{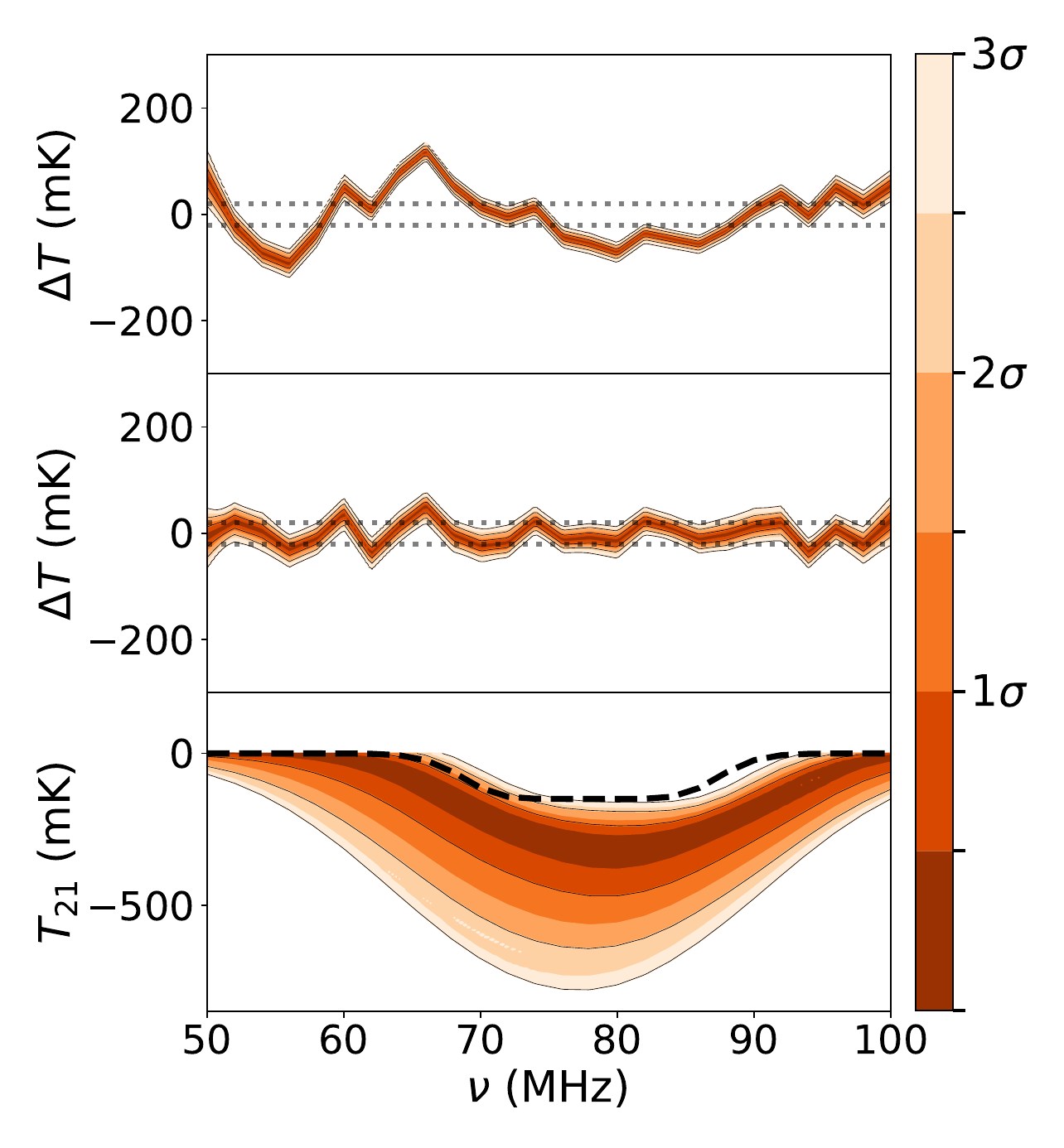}
            \end{subfigure}
        }
    \end{minipage}
    \caption{
        Signal recovery plots for models detecting a 21-cm signal in simulated data containing a moderate amplitude signal ($A=150~\mathrm{mK}$). Each subplot shows posterior probability densities of foreground-only residuals (top), full-model residuals (middle), and recovered 21-cm signal (bottom). Dotted lines in the top and middle panels indicate the noise level; the dashed black line shows the true input signal. Models are arranged by increasing Bayesian evidence (lowest evidence in top-left, highest in bottom-right). Background colours distinguish BaNTER validation results: red backgrounds indicate failed validation and blue backgrounds indicate passed validation. We highlight models using \textbf{bold} for highest evidence models (those with $\ln(\mathcal{B}_{i\mathrm{max}}) \geq -3$) and \textit{italic} for highest BaNTER-validated posterior odds models (those with $\ln(\mathcal{R}_{i\mathrm{max}}) \geq -3$). Models meeting both criteria appear in \textit{\textbf{bold italic}}. Subfigure captions indicate the model type and foreground complexity ($N$).
    }
    \label{Fig:SignalRecoveryAllDetections150mK}
\end{figure*}

\subsubsection{Bayesian model comparison with uninformative model priors}
\label{Sec:ModerateAmplitudeB}

In \Cref{Fig:BandRwithAeq150mK} (left), we present the Bayes factors ($\ln(\mathcal{B}_{i\mathrm{max}})$) comparing models $\bm{M}_{i}$ to the highest-evidence model, $\bm{M}_\mathrm{max}$, for the data set $\bm{T}_{\rm corrected}$. Here, $\bm{T}_{\rm corrected}$ corresponds to the moderate-amplitude 21-cm signal scenario with $A = 150~\mathrm{mK}$, $i$ runs over all models in the set $\bm{\mathcal{M}}$ (i.e. including both the models that pass and those that fail BaNTER validation), and $\bm{M}_\mathrm{max}$ is the BFCC composite model with $N = 4$.

We find that a subset of models -- including the Intrinsic composite model and BFCC composite models with $N = 3$ and $N = 5$--$10$ -- describe the data comparably well to the highest-evidence model ($N = 4$), with no strong Bayesian evidence favouring one over the others. Relative to this subset, the remaining models are decisively disfavoured ($\ln(\mathcal{B}) > 5$ when comparing any model in the high-evidence subset to models outside it; see \Cref{Tab:HPDsummaryTable}).

Among the high-evidence models, the Intrinsic model and BFCC models with $N = 3$ and $4$ are the highest-evidence candidates, with the Intrinsic model and BFCC model with $N = 3$ being only weakly disfavoured relative to $\bm{M}_\mathrm{max}$. All three models strongly support the inclusion of a 21-cm component ($\ln(\mathcal{B}_{\mathrm{cFg}}) > 3.0$).

The remaining models in the preferred subset are moderately disfavoured relative to $\bm{M}_\mathrm{max}$. Two of them provide strong evidence for a 21-cm detection, while the others show no detection.

The LinPhys and MultLin models are decisively disfavoured by the Bayesian evidence. The only model in the LinPhys class and the highest-evidence models in the MultLin class (MultLin $N = 5$ and $6$) yield detections of the 21-cm signal. The remaining MultLin models show a mix of detections and non-detections.

In general, weaker detections or non-detections are more probable in higher-complexity foreground models, as their greater flexibility allows them to fit both the foregrounds and a significant fraction of the 21-cm signal simultaneously. Consequently, the difference in Bayesian evidence between models with and without a 21-cm signal decreases, eventually falling below the detection threshold.

\subsubsection{BaNTER-validated posterior-odds-based model comparison}
\label{Sec:ModerateAmplitudeR}

In \Cref{Fig:BandRwithAeq150mK} (right), we present the posterior odds ($\ln(\mathcal{R}_{i\mathrm{max}})$; c.f. \Cref{Sec:BaNTERValidationResultsAsModelPriors} for our binary prior odds) comparing models $\bm{M}_{i}$ to the highest posterior odds model, $\bm{M}_\mathrm{max,v}$, for the moderate amplitude 21-cm signal data set. The highest posterior odds model for this data set is the BFCC composite model with $N = 5$ foreground terms. We treat the prior odds of models that failed the BaNTER null test as negligible, excluding them from the analysis. Consequently, selecting the highest posterior odds model in $\bm{\mathcal{M}}$ corresponds to selecting the highest evidence model in the validated subset, $\bm{\mathcal{M}}_\mathrm{v}$.

The validated posterior-odds-based model comparison shows that the BFCC composite models with $N = 6$--$10$ describe the data comparably well to $\bm{M}_\mathrm{max,v}$, with no strong preference for one model over the others ($\mathcal{R}_{i\mathrm{max}} < 3$). The remaining models in $\bm{\mathcal{M}}_\mathrm{v}$, outside this subset, are decisively disfavoured.

Among the highest posterior odds models, the BFCC composite models with $N = 5$ and $6$ have the greatest evidence as models for the data. Both yield detections of the 21-cm signal ($\ln(\mathcal{B}_{\mathrm{cFg}}) > 3.0$). The higher complexity BFCC models are weakly to moderately disfavoured relative to $\bm{M}_\mathrm{max,v}$ and do not detect the 21-cm signal.

MultLin is the other model class that passes BaNTER validation; however, these models are decisively disfavoured by BFBMC relative to the BFCC models. None of the validated MultLin models detect the 21-cm signal in the moderate amplitude 21-cm signal data.

When comparing the conclusions drawn from the BaNTER-validated Bayesian model comparison and the Bayesian model comparison with uninformative priors, we observe the following differences in preferred models:
\begin{enumerate}
    \item \textit{Unvalidated workflow:} The Intrinsic model and BFCC models with $N = 3$ and $4$ are the highest-evidence models. All three of these models provide strong support for the inclusion of a 21-cm component ($\ln(\mathcal{B}_{\mathrm{cFg}}) > 3.0$).
    \item \textit{BaNTER-validated workflow:} The three highest-evidence models from the unvalidated workflow fail the BaNTER null test and are thus excluded from the validated model set. The BFCC models with $N = 5$ and $6$ are the highest posterior odds models for the moderate amplitude 21-cm signal data. Both of these models provide strong support for the detection of the 21-cm signal ($\ln(\mathcal{B}_{\mathrm{cFg}}) > 3.0$).
\end{enumerate}

\subsubsection{21-cm signal estimates}
\label{Sec:ModerateAmplitudeParameterEstimation}

Having identified the preferred model using Bayes factors and posterior odds, we now examine whether these preferences align with the ground truth, as determined by the consistency of the recovered parameter posteriors with the true input parameters of the 21-cm signal in the data.
\Cref{Fig:SignalRecoveryAllDetections150mK} illustrates the fit results for all composite models in $\bm{\mathcal{M}}$ that exhibit strong evidence for a 21-cm signal detection. For each model, the figure shows the posterior PDs of:
\begin{enumerate*}
    \item the residuals obtained by fitting the data using only the foreground component of the model,
    \item the residuals obtained by fitting the data using the full model, and
    \item the recovered 21-cm signal derived from fitting the data with the full model.
\end{enumerate*}

The first seven subplots (\Cref{Fig:ModerateAmpMultLin3,Fig:ModerateAmpMultLin4,Fig:ModerateAmpMultLin5,Fig:ModerateAmpLinPhys5,Fig:ModerateAmpLinPhys5,Fig:ModerateAmpIntrinsic3,Fig:ModerateAmpBFCC3,Fig:ModerateAmpBFCC4}) show results for models that failed the BaNTER validation. While the majority of these models achieve reasonable fits to the data (evidenced by the relative consistency between the full model residuals and the expected noise level in the data; see \Cref{Fig:ModerateAmpMultLin5,Fig:ModerateAmpLinPhys5,Fig:ModerateAmpLinPhys5,Fig:ModerateAmpIntrinsic3,Fig:ModerateAmpBFCC3,Fig:ModerateAmpBFCC4}, middle panels), they nevertheless yield substantial (\Cref{Fig:ModerateAmpIntrinsic3,Fig:ModerateAmpLinPhys5,Fig:ModerateAmpBFCC4}) to very substantial (\Cref{Fig:ModerateAmpMultLin3,Fig:ModerateAmpMultLin4,Fig:ModerateAmpMultLin5,Fig:ModerateAmpBFCC3}) biases in their recovered 21-cm signals (amplitudes, location or shape parameters inconsistent with the underlying parameters of the 21-cm signal in the data at 95\% credibility; see \Cref{Tab:HPDsummaryTable}).

Additionally, models that most severely failed the BaNTER null test -- the MultLin model with $N=3$ and $4$, and the BFCC model with $N=3$ (see \Cref{Tab:lnBcb}) -- also exhibit the most biased 21-cm signal recovery. Furthermore, the MultLin model with $N=3$ and $4$ provides poor fits to the data, even with biased 21-cm signal modelling absorbing some structure due to foreground systematics.

In contrast, barring biased recovery of the flatness parameter, $\tau$, the 21-cm signal recovered with the LinPhys model, which failed the BaNTER null test by the smallest margin, shows 21-cm parameter estimates that are consistent with the underlying 21-cm signal in the data at 95\% credibility.

By comparison, models that passed the BaNTER validation and detected the 21-cm signal yield fully unbiased recovery of the 21-cm signal in the data (\Cref{Fig:ModerateAmpBFCC5,Fig:ModerateAmpBFCC6} and \Cref{Tab:HPDsummaryTable}), with all 95\% credibility HPDI parameter estimates consistent with the true signal parameters (\Cref{Tab:HPDsummaryTable}).

Comparing the consistency of the recovered 21-cm signals (represented by the contours in the bottom panels of each subfigure) with the true 21-cm signal data (indicated by the dashed black lines) in \Cref{Fig:SignalRecoveryAllDetections150mK} yields the ground truth efficacy of our models (also see \Cref{Tab:HPDsummaryTable}). By comparing this true efficacy to the expected efficacy based on the BaNTER-validated Bayesian model comparison and the Bayesian model comparison with uninformative priors, we draw the following conclusions regarding these two model comparison methodologies:
\begin{enumerate}
    \item \textit{Unvalidated workflow:} The three highest-evidence models (the Intrinsic model and BFCC models with $N = 3$ and $4$), as determined by the BFBMC analysis, yield biased estimates of the underlying 21-cm signal in the data. This bias demonstrates that comparison of the full set of models considered in the analysis, as judged by BFBMC, is insufficient to identify models that yield unbiased recovery of the 21-cm signal. This conclusion is particularly driven by the inclusion of the Intrinsic model and the BFCC models with $N = 3$ or $4$ in the set of models under consideration (see \Cref{Sec:MCCModerateAmplitude} for details). This result implies that the comparison of models in the unvalidated model set for the moderate amplitude 21-cm signal data constitutes a \textit{category II} model comparison problem for which the conclusions drawn from BFBMC alone are not robust.
    \item \textit{BaNTER-validated workflow:} All models shown to yield biased estimates of the 21-cm signal in \Cref{Fig:SignalRecoveryAllDetections150mK} were correctly identified and excluded from the validated model set by the BaNTER null test. The highest posterior odds models with 21-cm signal detections in the BaNTER-validated posterior-odds-based analysis are correctly identified as the BFCC models with $N = 5$ and $6$. These models are found to be the only ones that yield unbiased estimates of the 21-cm signal in the data in \Cref{Fig:SignalRecoveryAllDetections150mK}.
\end{enumerate}

The excellent agreement between the expected performance of the models, based on the results of BaNTER validation in \Cref{Sec:BaNTERValidationResults}, and the validity of the recovered 21-cm signal estimates demonstrates the necessity and efficacy of the BaNTER-validated posterior-odds-based analysis in the context of the moderate amplitude 21-cm signal scenario.

\subsection{High amplitude 21-cm signal}
\label{Sec:HighAmplitude}

\subsubsection{Bayesian model comparison with uninformative model priors}
\label{Sec:HighAmplitudeB}

\begin{figure*}
	\rightline{
	\includegraphics[width=0.93\textwidth, trim={0cm 0cm 0.7cm 0cm}, clip]{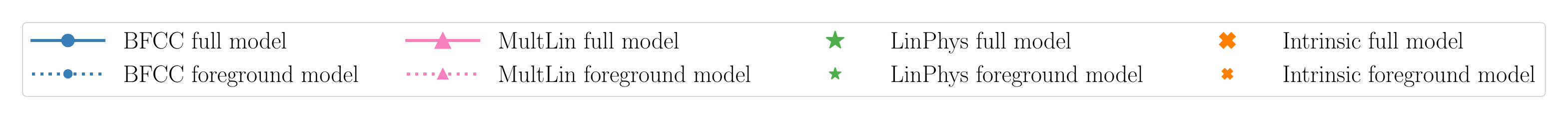}
	}	\centerline{
	\includegraphics[width=0.5\textwidth]{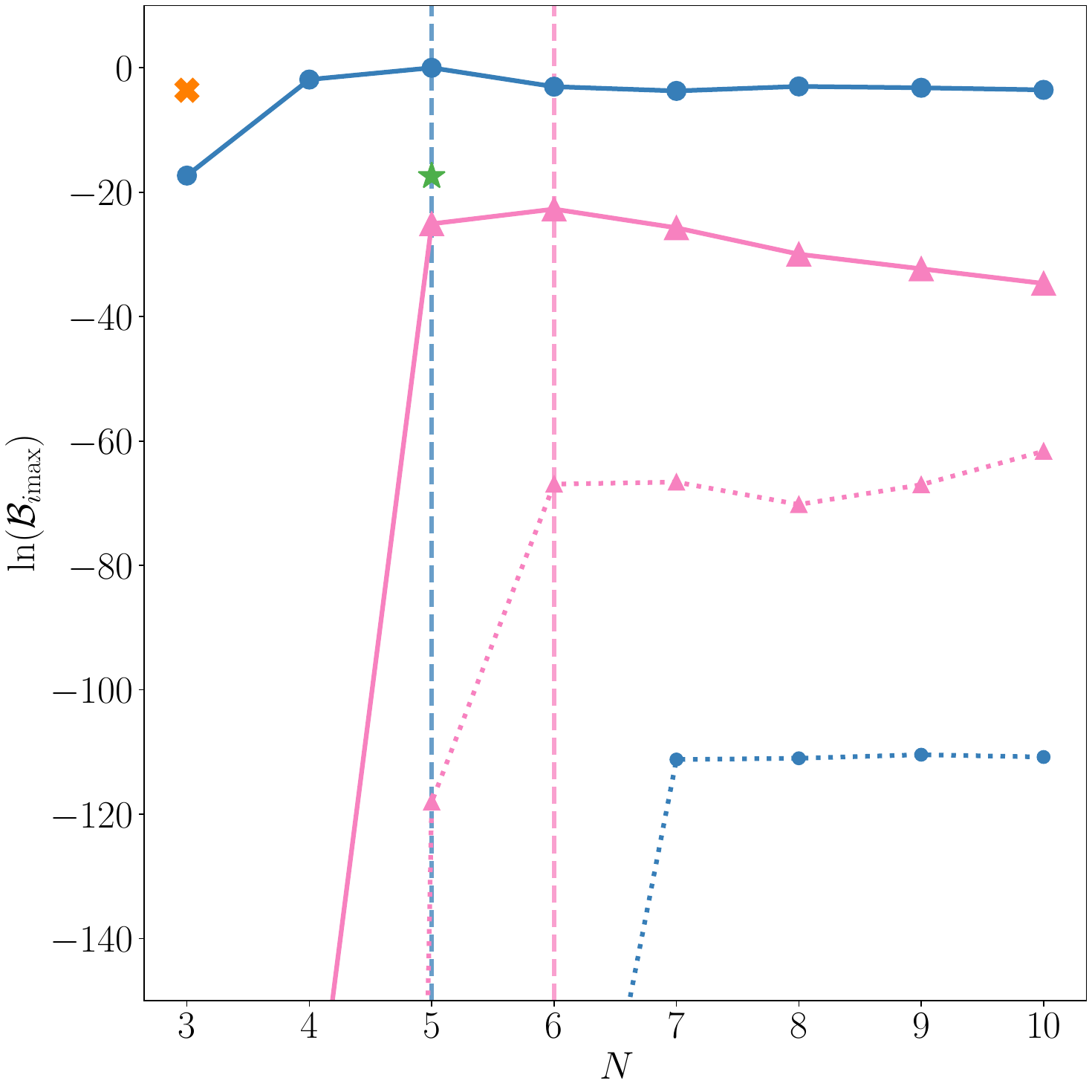}
	\includegraphics[width=0.5\textwidth]{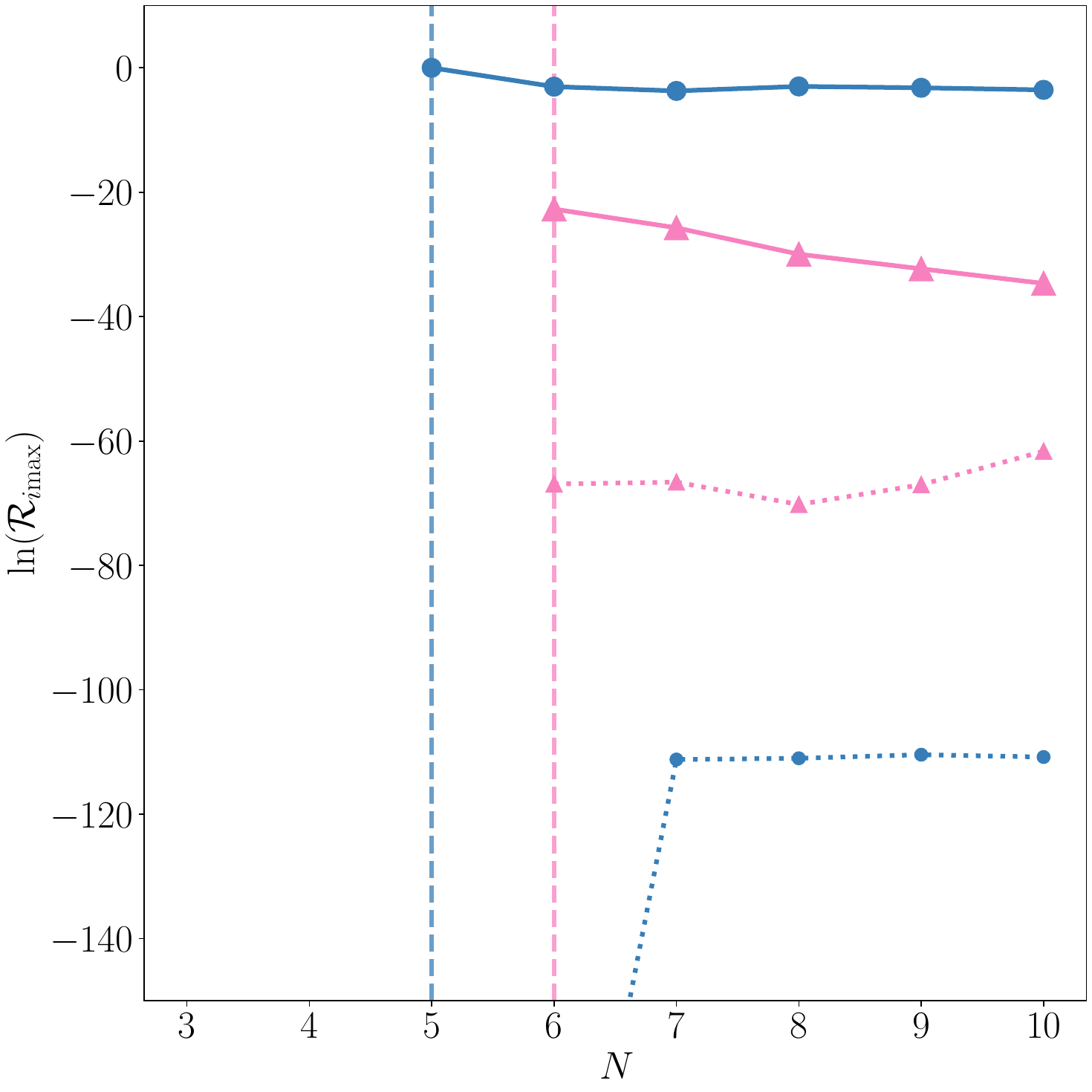}
	}
\caption{
    As in \Cref{Fig:BandRwithAeq150mK} but for the Bayesian comparison of models for simulated data incorporating a high amplitude 21-cm signal model ($A=500~\mathrm{mK}$).
}
\label{Fig:BandRwithAeq500mK}
\end{figure*}

In \Cref{Fig:BandRwithAeq500mK} (left), we show the Bayes factors for the high-amplitude 21-cm signal scenario ($A = 500~\mathrm{mK}$), comparing each model $\bm{M}_{i}$ to the highest-evidence model, $\bm{M}_\mathrm{max}$, for $\bm{T}_{\rm corrected}$. As in the unvalidated moderate-amplitude 21-cm signal analysis, $i$ runs over all models in the set $\bm{\mathcal{M}}$. Here, we find that $\bm{M}_\mathrm{max}$ is the BFCC composite model with $N = 5$.

In contrast to the moderate-amplitude 21-cm signal scenario, all composite models in $\bm{\mathcal{M}}$ show strong support for the inclusion of a 21-cm signal component in the high-amplitude case ($\ln(\mathcal{B}_{\mathrm{cFg}}) \geq 3.0$; see \Cref{Tab:HPDsummaryTable}). Additionally, the subset of highest-evidence models for the high-amplitude data set is smaller. Specifically, similar to the moderate-amplitude regime, the BFCC composite models with $N = 4$ and $5$ remain comparably probable, with no strong Bayesian evidence favouring one over the other ($\ln(\mathcal{B}_{i\mathrm{max}}) < 3$). However, several models that were in the highest-evidence subset for the moderate-amplitude case are now absent. These include the Intrinsic composite model and the BFCC composite models with $N = 3$ and $6$ to $10$.

The decrease in $\ln(\mathcal{B}_{i\mathrm{max}})$ for the BFCC composite models with $N = 6$ to $10$ can be attributed to the Occam penalty associated with their increased complexity (see \Cref{Sec:BayesianModelComparison}), combined with the minimal improvement in their ability to fit the high-amplitude data relative to $\bm{M}_\mathrm{max}$ (see \Cref{Fig:HighAmpBFCC5,Fig:HighAmpBFCC6,Fig:HighAmpBFCC7,Fig:HighAmpBFCC8,Fig:HighAmpBFCC9,Fig:HighAmpBFCC10}, middle panels).

In contrast, the decrease in $\ln(\mathcal{B}_{i\mathrm{max}})$ of the BFCC composite model with $N = 3$ and of the Intrinsic model is driven by their poorer performance in fitting the high-amplitude 21-cm signal data set. The most significant difference in Bayes factor between the two data sets is observed for the BFCC composite model with $N = 3$, which transitions from being in the highest-evidence subset for the moderate-amplitude 21-cm signal data set to being decisively disfavoured for the high-amplitude 21-cm signal data. Specifically, we find that for the high-amplitude signal data, $\ln(\mathcal{B}_{i\mathrm{max}})$  decreases by $\sim 17$ (see \Cref{Tab:HPDsummaryTable}), corresponding to odds in favour this model relative to $\bm{M}_\mathrm{max}$ of worse than $1:10^{7}$.

To understand this difference, recall that the BFCC composite model with $N = 3$ failed the BaNTER null test in \Cref{Sec:BaNTERValidationResults}. This failure indicates that when this model is used to fit a data set containing a non-zero 21-cm signal, the 21-cm component necessarily fits the combined contribution of both the 21-cm signal and the foreground systematics. In the moderate-amplitude regime, while the foreground model imperfectly describes the foregrounds, it also captures a significant fraction of the 21-cm signal. This behaviour allows a biased fit of the 21-cm model component to absorb residual systematics, leading to a relatively better fit to the data in aggregate. In contrast, in the high-amplitude regime, a smaller fraction of the 21-cm signal is described by the foreground model. As a result, the parameters of the 21-cm model are more strongly constrained by the remaining 21-cm component, reducing its flexibility to absorb residual foreground systematics.

It follows from this explanation that one should expect the reduction in Bayes factor to be accompanied by a decrease in the accuracy of the BFCC composite model with $N = 3$ in the high-amplitude 21-cm signal regime relative to the moderate-amplitude case. To quantify this decrease, we use the model accuracy statistic introduced in S25 (see \Cref{Sec:AccuracyCondition}). Indeed, we find that this expectation holds: specifically, in the moderate-amplitude regime, we obtain $Q_{0.999}(\lambda) \simeq 2$, which satisfies the S25 accuracy condition ($Q_{0.999}(\lambda) > 0$). However, in the high-amplitude case, this value drops to $Q_{0.999}(\lambda) \simeq -17$, implying that the fit residuals are inconsistent with the expected noise distribution in the data and are thus contaminated by residual systematics.

The same reasoning applies to the Intrinsic model, which also failed the BaNTER null test in \Cref{Sec:BaNTERValidationResults}. Here, we observe a similar but less pronounced decrease in model accuracy when transitioning from the moderate- to high-amplitude 21-cm signal data sets.

\subsubsection{BaNTER-validated posterior-odds-based model comparison}
\label{Sec:HighAmplitudeR}

In \Cref{Fig:BandRwithAeq500mK} (right), we show the posterior odds ($\ln(\mathcal{R}_{i\mathrm{max}})$), comparing models $\bm{M}_{i}$ to the highest posterior odds model, $\bm{M}_\mathrm{max,v}$, for the high-amplitude 21-cm signal data set. The highest posterior odds model in this case is the BFCC composite model with $N = 5$ foreground terms, which coincides with the highest Bayesian evidence model identified in \Cref{Sec:HighAmplitudeB}. As in the moderate-amplitude regime, models that failed the BaNTER null test are assigned negligible prior odds and thus excluded from the analysis. The validated subset, $\bm{\mathcal{M}}_\mathrm{v}$, contains only the remaining models.

The remaining BFCC composite models in $\bm{\mathcal{M}}_\mathrm{v}$ are strongly disfavoured, while the remaining MultLin models are decisively disfavoured. Within the MultLin class, the $N = 6$ and $7$ foreground complexity composite models have the highest posterior odds, with the remaining MultLin models strongly to decisively disfavoured relative to these two models.

Comparing the conclusions drawn from the BaNTER-validated posterior odds and the Bayesian model comparison with uninformative model priors, in the high amplitude 21-cm signal regime one finds the following similarities and differences in preferred models:
\begin{enumerate}
    \item \textit{Unvalidated workflow:} The BFCC composite models with $N = 4$ and $5$ are the highest-evidence models.
    \item \textit{BaNTER-validated workflow:} BFCC composite models with $N = 4$ fails the BaNTER null test and thus is excluded from the validated model set. The set of the highest posterior odds models for the high amplitude 21-cm signal data contains only the BFCC models with $N = 5$.
\end{enumerate}

\subsubsection{21-cm signal estimates}
\label{Sec:ModerateAmplitudeParameterEstimation}

Having established the preferred models for the data set containing the high-amplitude 21-cm signal using Bayes factors and posterior odds, we now examine whether the recovered parameter posteriors align with these conclusions.

\Cref{Fig:SignalRecoveryAllDetections500mKPt1} illustrates the fit results for all composite models in $\bm{\mathcal{M}}$ that exhibit strong evidence for a 21-cm signal detection. For the high-amplitude 21-cm signal data set, $\bm{\mathcal{M}}$ includes all models considered. For each model, the figure shows the posterior PDs of the foreground (top panels) and of the composite model (middle panels) fit residuals, and the 21-cm signal component of the composite model fit (bottom panels).

\Cref{Fig:HighAmpMultLin3,Fig:HighAmpMultLin4,Fig:HighAmpMultLin5,Fig:HighAmpLinPhys5,Fig:HighAmpIntrinsic3,Fig:HighAmpBFCC3,Fig:HighAmpBFCC4} show results for models that failed BaNTER validation. As in the moderate-amplitude 21-cm signal scenario, we find that while all but two of the failed models achieve good fits to the full data -- evidenced by the consistency between full model residuals and the expected noise level in \Cref{Fig:HighAmpIntrinsic3,Fig:HighAmpBFCC3,Fig:HighAmpBFCC4,Fig:HighAmpMultLin5} -- they all result in biased recovery of the amplitude, location or shape parameters of the underlying 21-cm signal at 95\% credibility (see \Cref{Tab:HPDsummaryTable} for details).

Additionally, mirroring our findings in the moderate-amplitude scenario, the models that most severely failed the BaNTER null test (MultLin models with $N=3$ and $4$ and the BFCC model with $N=3$; c.f. \Cref{Tab:lnBcb}) also yield the most biased 21-cm signal recovery. Conversely, the LinPhys model, which failed the BaNTER null test by the smallest margin, also yields the least biased 21-cm signal parameter estimates of the failed models in the high-amplitude 21-cm signal scenario, with only the central frequency of the recovered signal being inconsistent with the underlying 21-cm signal in the data at 95\% credibility.

By comparison, all models that passed BaNTER validation yield unbiased recovery of the 21-cm signal in the data (see \Cref{Fig:HighAmpMultLin6,Fig:HighAmpMultLin7,Fig:HighAmpMultLin8,Fig:HighAmpMultLin9,Fig:HighAmpMultLin10,Fig:HighAmpBFCC5,Fig:HighAmpBFCC6,Fig:HighAmpBFCC7,Fig:HighAmpBFCC8,Fig:HighAmpBFCC9,Fig:HighAmpBFCC10} and \Cref{Tab:HPDsummaryTable}), with all 95\% credibility HPDI parameter estimates consistent with the true signal parameters (\Cref{Tab:HPDsummaryTable}).

As in \Cref{Sec:ModerateAmplitudeParameterEstimation}, by comparing the consistency of the recovered 21-cm signals in \Cref{Fig:SignalRecoveryAllDetections500mKPt1} with the true 21-cm signal data, one can derive the ground truth efficacy of our models for the high amplitude data set (also see \Cref{Tab:HPDsummaryTable}). By comparing this true efficacy to the expected efficacy based on the BaNTER-validated Bayesian model comparison and the Bayesian model comparison with uninformative priors, we draw the following conclusions regarding these two model comparison methodologies in this regime:
\begin{enumerate}
    \item \textit{Unvalidated workflow:} The subset of highest-evidence models ($\ln(\mathcal{B}_{i\mathrm{max}}) < 3$) for the high-amplitude 21-cm signal data set contains the BFCC composite models with $N = 4$ and $5$. Within this subset, only the BFCC composite model with $N = 5$ yields unbiased estimates of the underlying 21-cm signal, while the BFCC composite model with $N = 4$ results in biased recovery. Thus, Bayesian model comparison with uninformative priors (BFBMC alone) remains insufficient to uniquely identify models that yield unbiased recovery of the 21-cm signal. This result implies that, as in the moderate-amplitude scenario, model comparison in the unvalidated workflow corresponds to a \textit{category II} model comparison problem (see \Cref{Sec:MCCHighAmplitude}). However, the degree of bias in the recovered 21-cm signal estimates is reduced relative to the moderate-amplitude scenario: whereas three-quarters of models yielded biased recovery in the moderate case, this fraction decreases to one-half in the high-amplitude regime.
    \item \textit{BaNTER-validated workflow:} All models that yielded biased estimates of the 21-cm signal in \Cref{Fig:SignalRecoveryAllDetections500mKPt1} were correctly identified a priori and excluded from the validated model set by the BaNTER null test. The highest posterior odds model in the BaNTER-validated posterior-odds-based analysis is the BFCC composite model with $N = 5$, which yields unbiased estimates of the 21-cm signal. Additionally, all remaining models that passed BaNTER validation are found to yield unbiased recovery of the 21-cm signal, albeit generally with lower precision.
\end{enumerate}

The excellent agreement between the expected performance of models, based on the null-test-based Bayesian validation analysis in \Cref{Sec:BaNTERValidationResults}, and the validity of the recovered 21-cm signal estimates further demonstrates the value of BaNTER-validated posterior-odds-based analysis, even in the high-amplitude 21-cm signal scenario. However, the more moderate differences in preferred models between the BaNTER-validated and unvalidated BFBMC workflows indicate that, while the high-amplitude data modelling problem remains a \textit{category II} model comparison problem, BFBMC provides more reliable inferences in this regime than in the moderate-amplitude scenario.

\begin{figure*}
    \begin{minipage}{\textwidth}
        \redsubfigurebox{
        \begin{subfigure}[t]{0.33\textwidth}
                \caption{\Large{MultLin, $N=3$, FV}}
                \label{Fig:HighAmpMultLin3}
                \includegraphics[width=\textwidth]{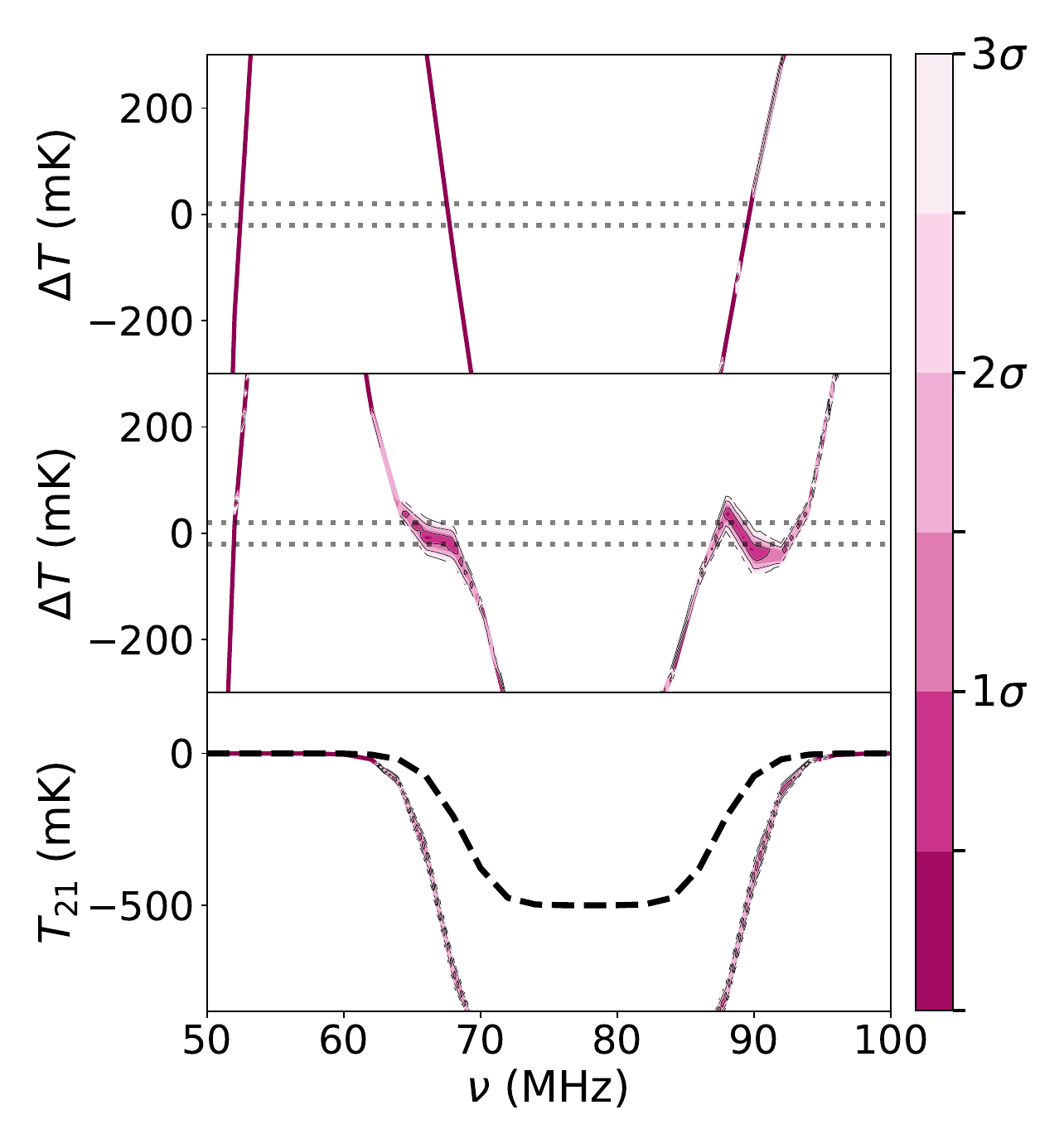}
            \end{subfigure}
        }
        \redsubfigurebox{
        \begin{subfigure}[t]{0.33\textwidth}
            \caption{\Large{MultLin, $N=4$, FV}}
            \label{Fig:HighAmpMultLin4}
            \includegraphics[width=\textwidth]{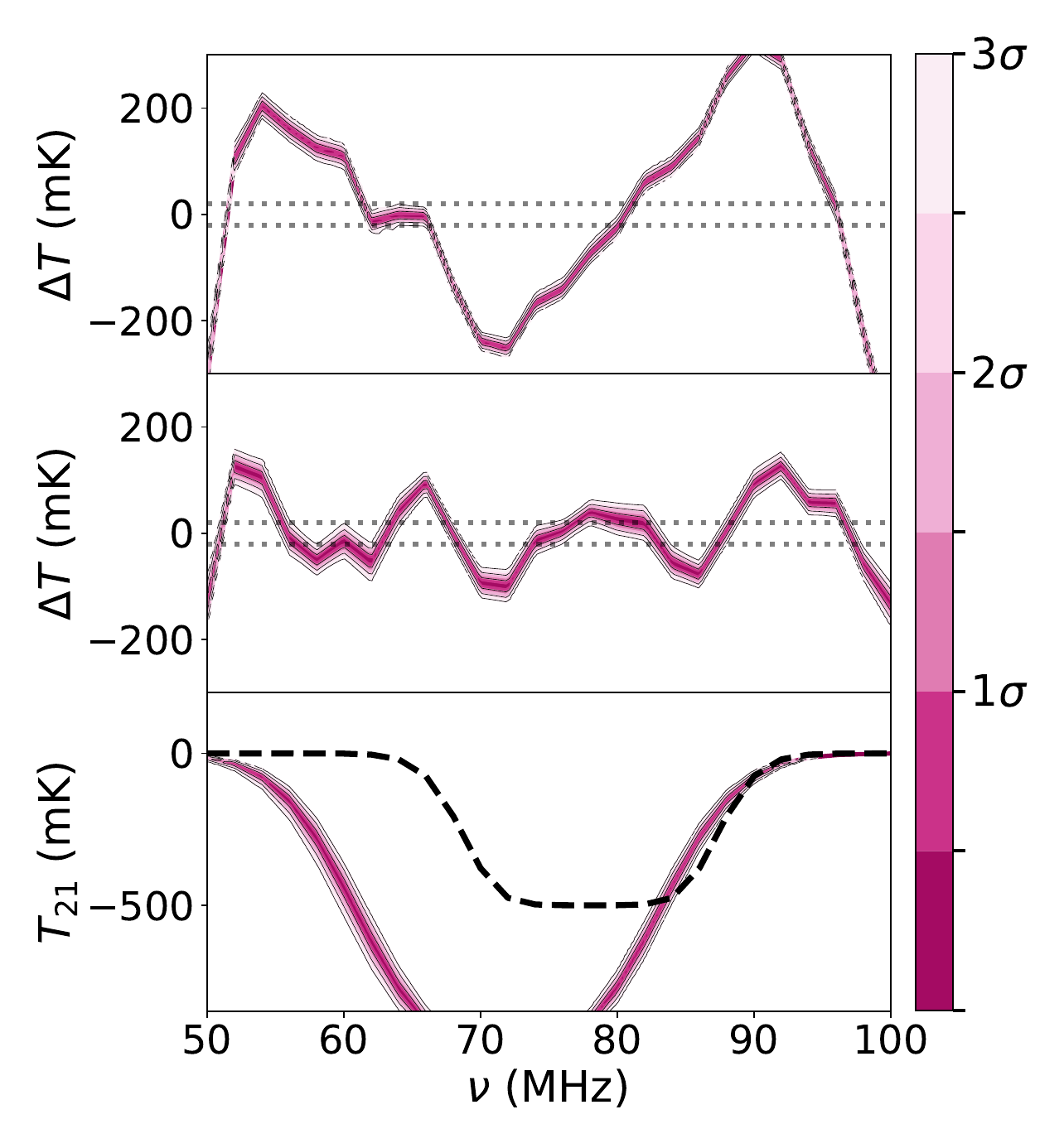}
        \end{subfigure}
        }
        \bluesubfigurebox{
        \begin{subfigure}[t]{0.33\textwidth}
            \caption{\Large{MultLin, $N=10$, PV}}
            \label{Fig:HighAmpMultLin10}
            \includegraphics[width=\textwidth]{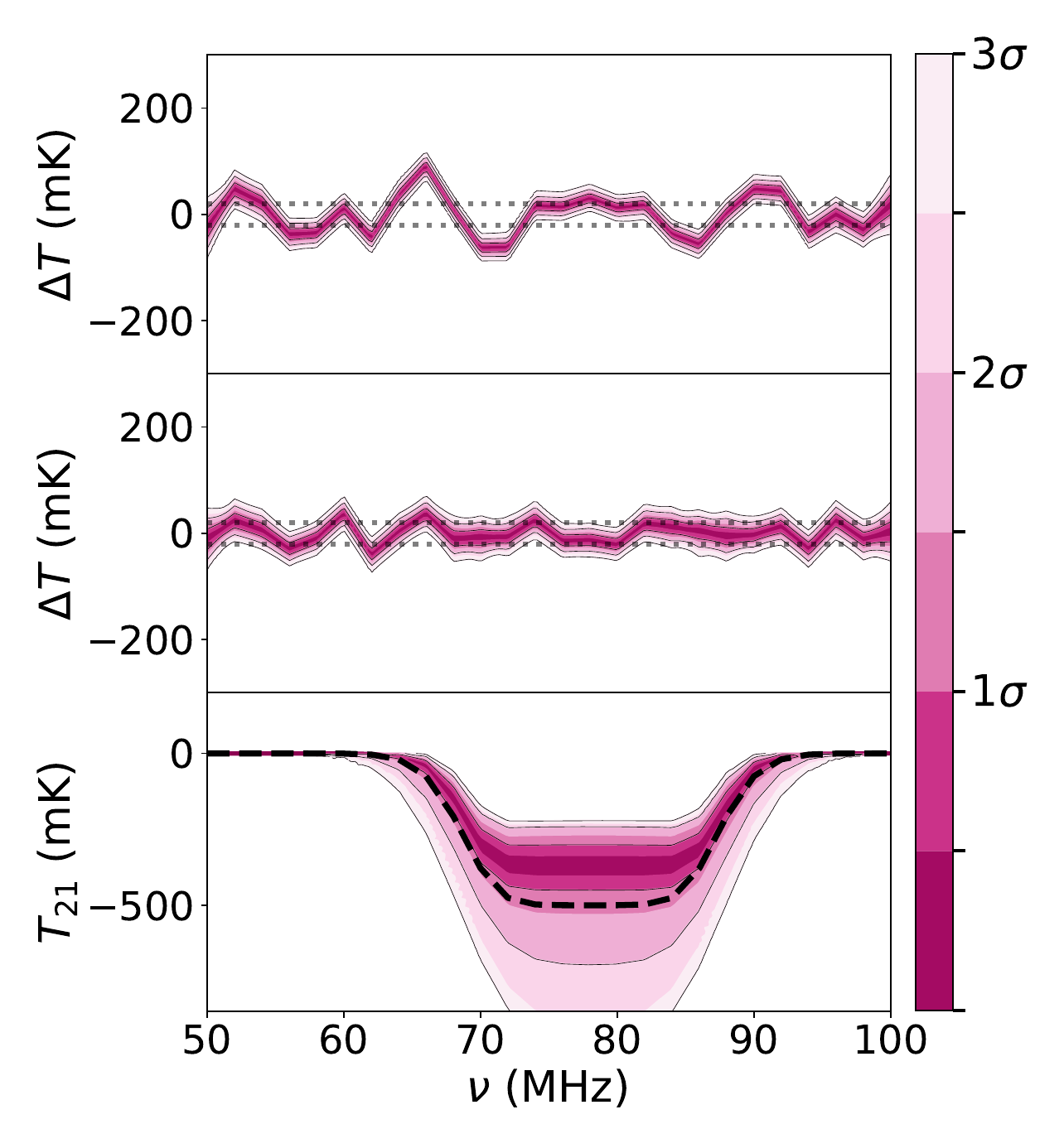}
        \end{subfigure}
        }
        \bluesubfigurebox{
        \begin{subfigure}[t]{0.33\textwidth}
            \caption{\Large{MultLin, $N=9$, PV}}
            \label{Fig:HighAmpMultLin9}
            \includegraphics[width=\textwidth]{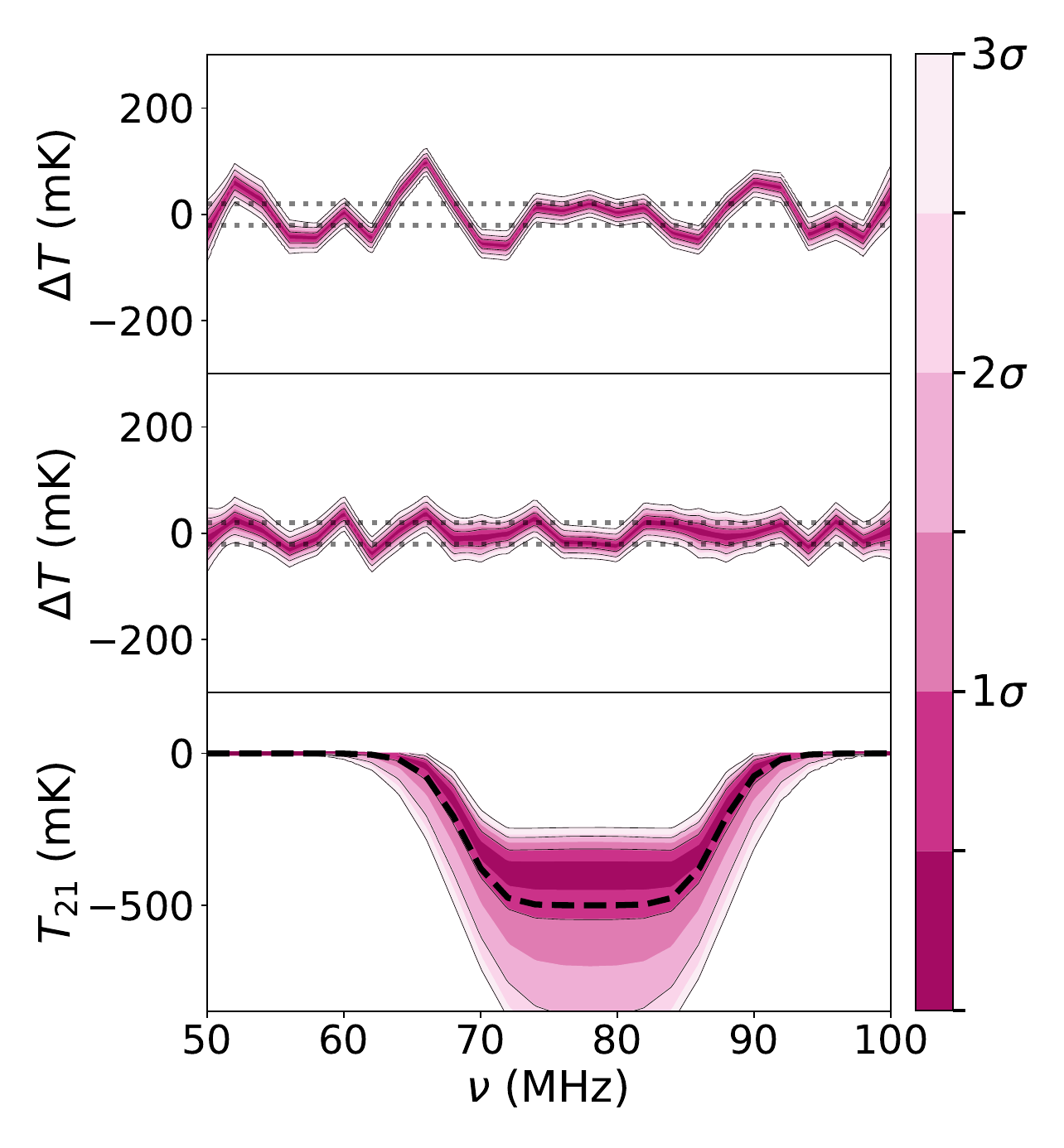}
        \end{subfigure}
        }
        \bluesubfigurebox{
        \begin{subfigure}[t]{0.33\textwidth}
            \caption{\Large{MultLin, $N=8$, PV}}
            \label{Fig:HighAmpMultLin8}
            \includegraphics[width=\textwidth]{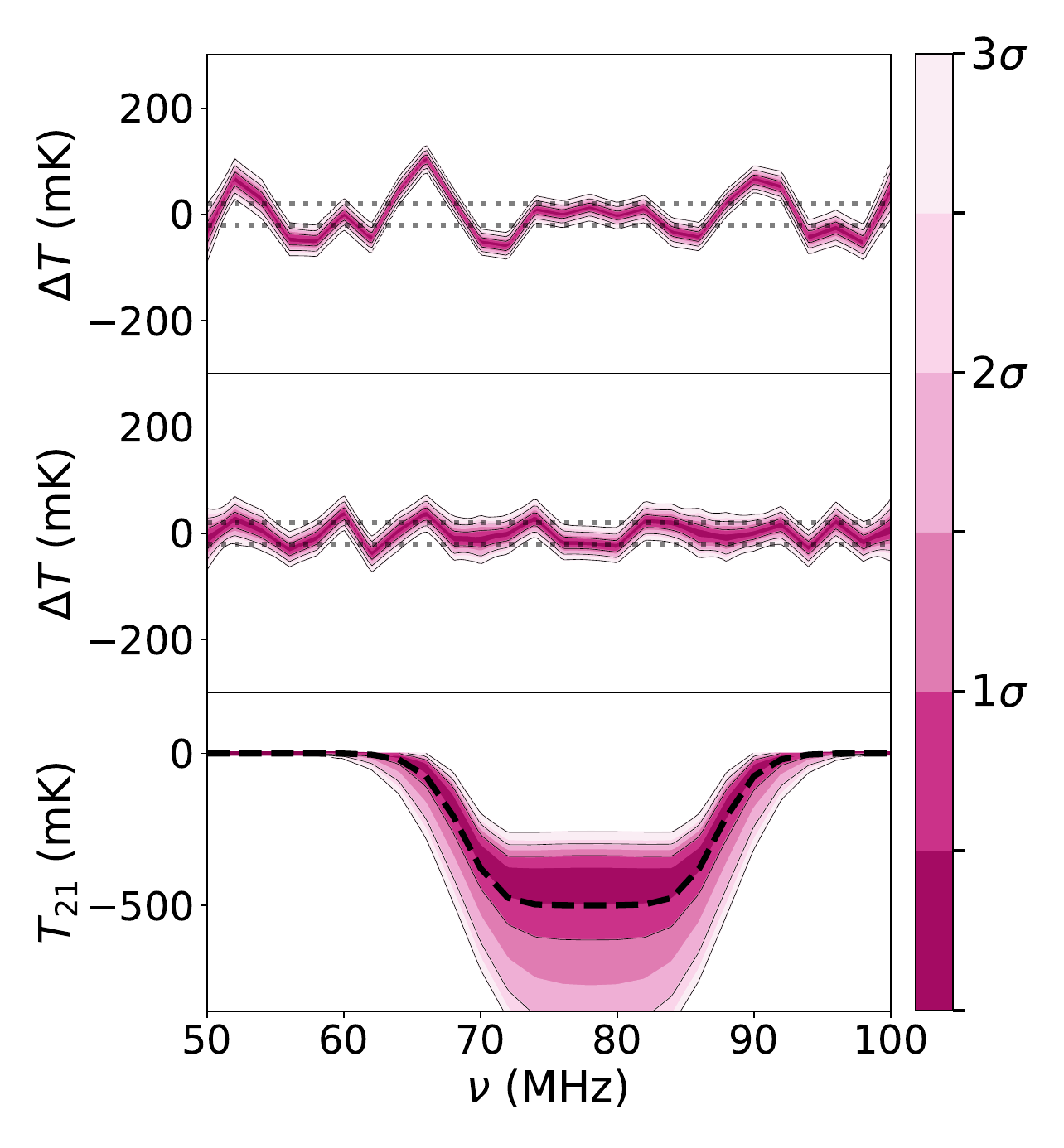}
        \end{subfigure}
        }
        \bluesubfigurebox{
        \begin{subfigure}[t]{0.33\textwidth}
            \caption{\Large{MultLin, $N=7$, PV}}
            \label{Fig:HighAmpMultLin7}
            \includegraphics[width=\textwidth]{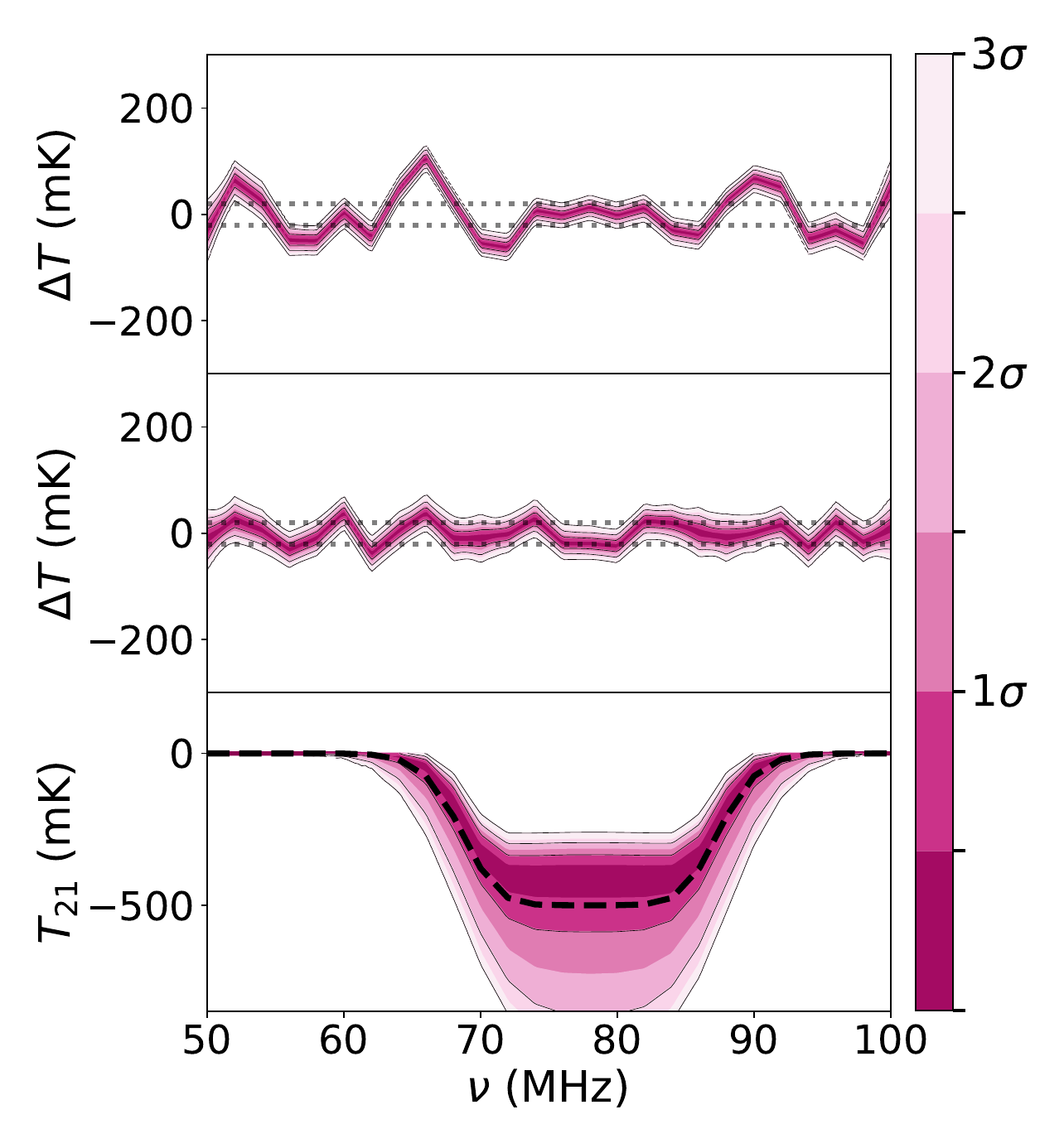}
        \end{subfigure}
        }
        \redsubfigurebox{
        \begin{subfigure}[t]{0.33\textwidth}
            \caption{\Large{MultLin, $N=5$, FV}}
            \label{Fig:HighAmpMultLin5}
            \includegraphics[width=\textwidth]{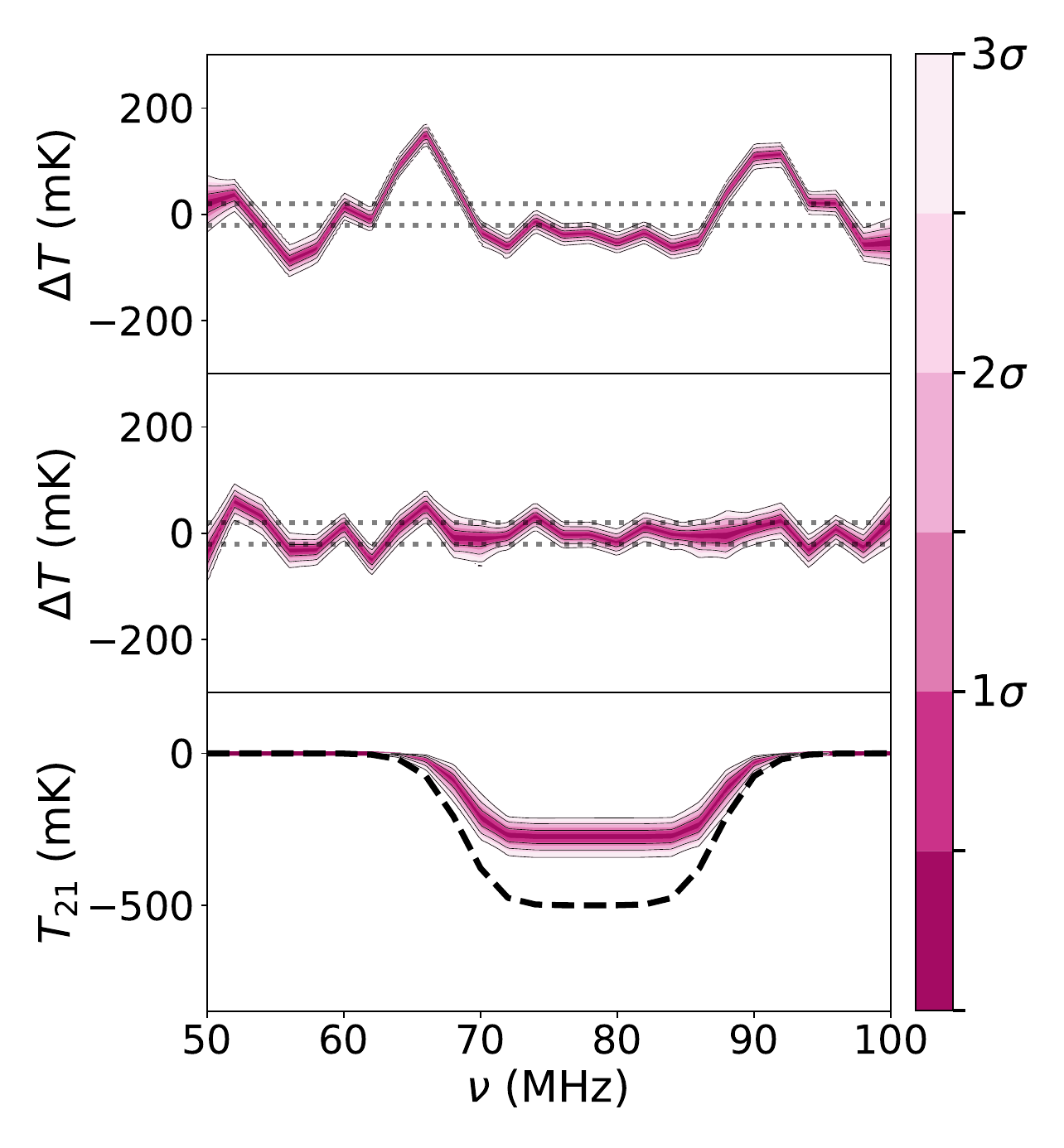}
        \end{subfigure}
        }
        \bluesubfigurebox{
        \begin{subfigure}[t]{0.33\textwidth}
            \caption{\Large{MultLin, $N=6$, PV}}
            \label{Fig:HighAmpMultLin6}
            \includegraphics[width=\textwidth]{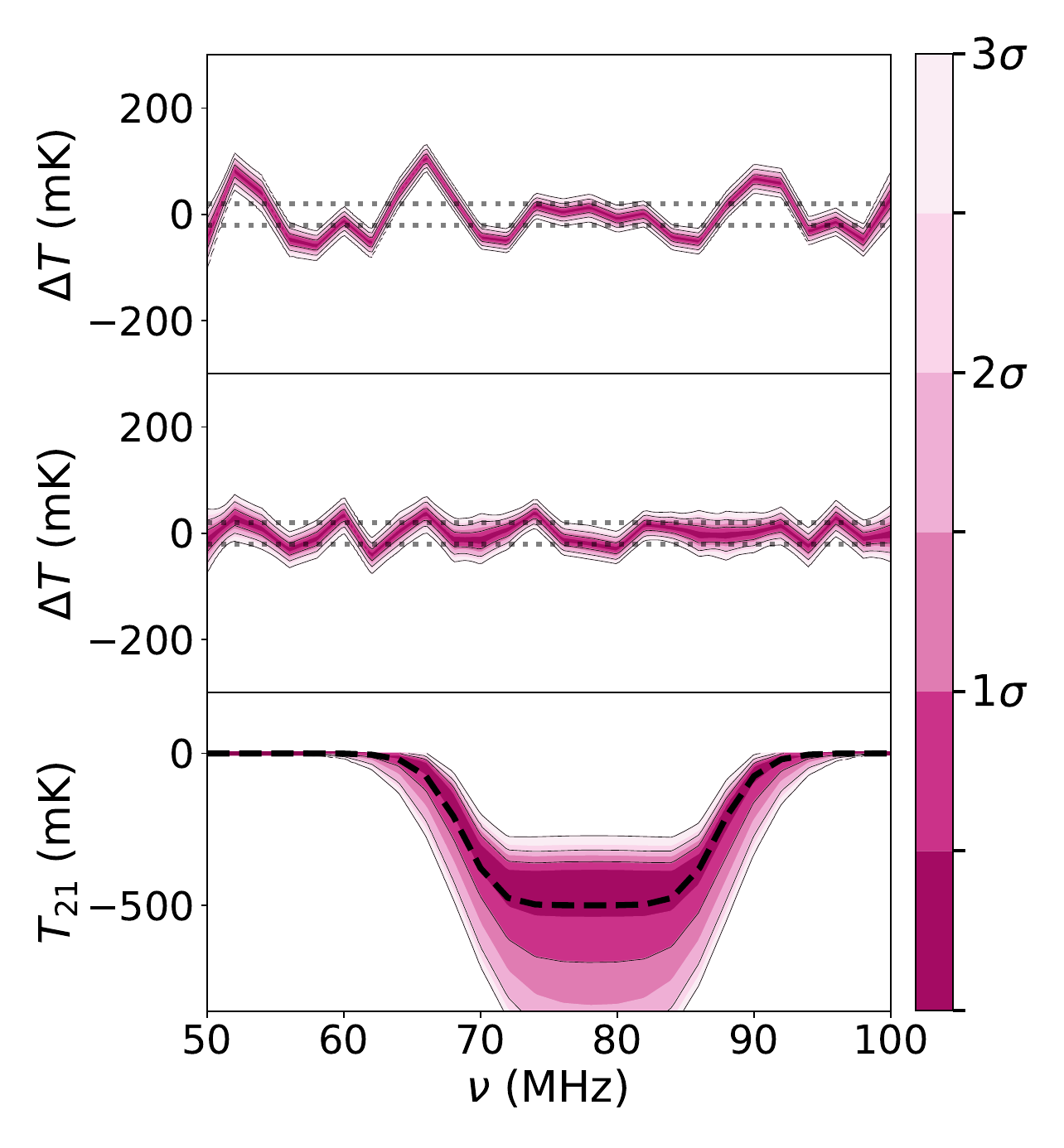}
        \end{subfigure}
        }
        \redsubfigurebox{
        \begin{subfigure}[t]{0.33\textwidth}
            \caption{\Large{LinPhys, $N=5$, FV}}
            \label{Fig:HighAmpLinPhys5}
            \includegraphics[width=\textwidth]{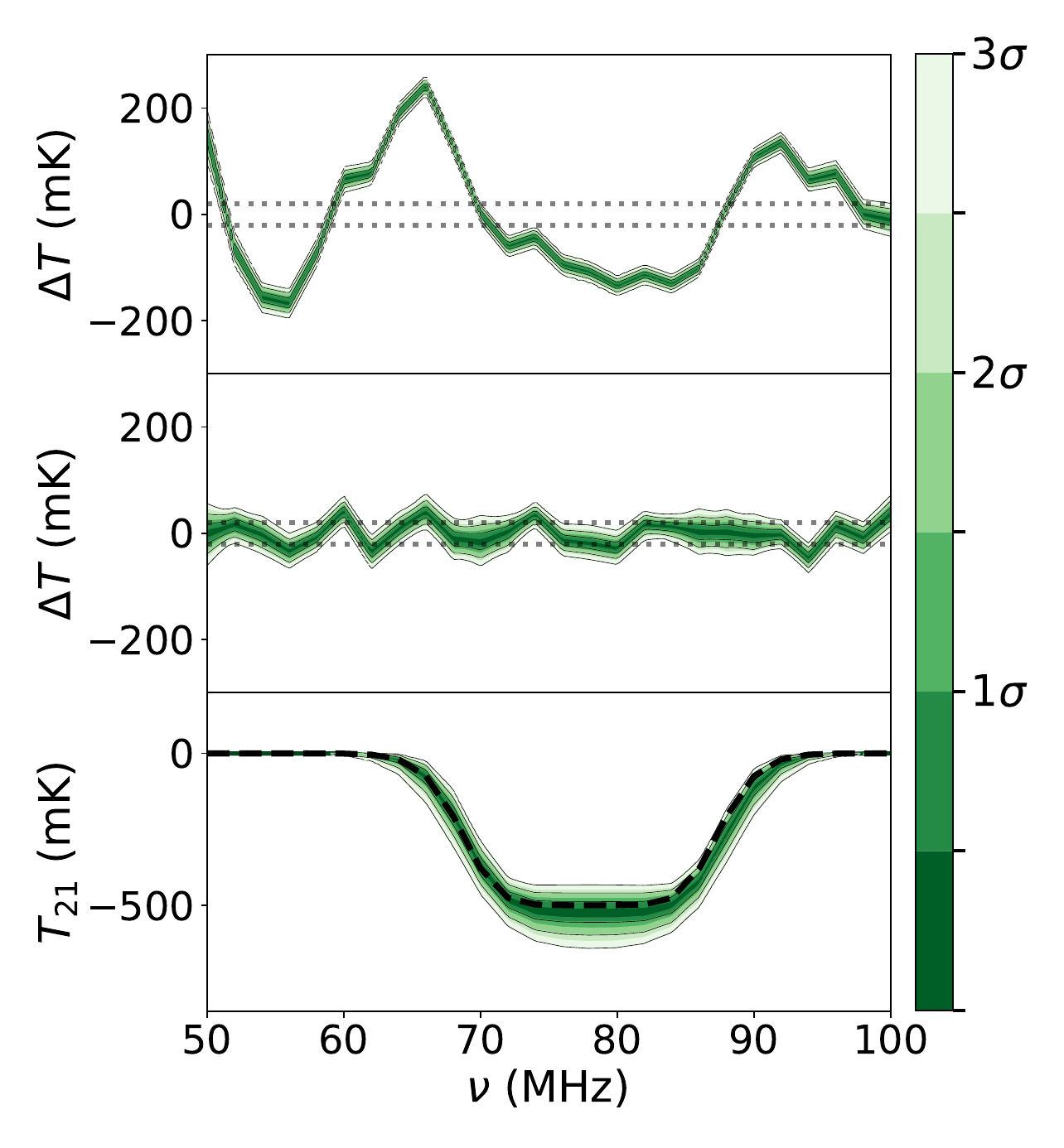}
        \end{subfigure}
        }
    \end{minipage}
    \caption{
        As in \Cref{Fig:SignalRecoveryAllDetections150mK} but for simulated data, $\bm{T}_{\rm corrected}$, containing a high amplitude 21-cm signal ($A=500~\mathrm{mK}$).
    }
\label{Fig:SignalRecoveryAllDetections500mKPt1}
\end{figure*}

\begin{figure*}\ContinuedFloat
    \begin{minipage}{\textwidth}
        \redsubfigurebox{
        \begin{subfigure}[t]{0.33\textwidth}
            \caption{\Large{BFCC, $N=3$, FV}}
            \label{Fig:HighAmpBFCC3}
            \includegraphics[width=\textwidth]{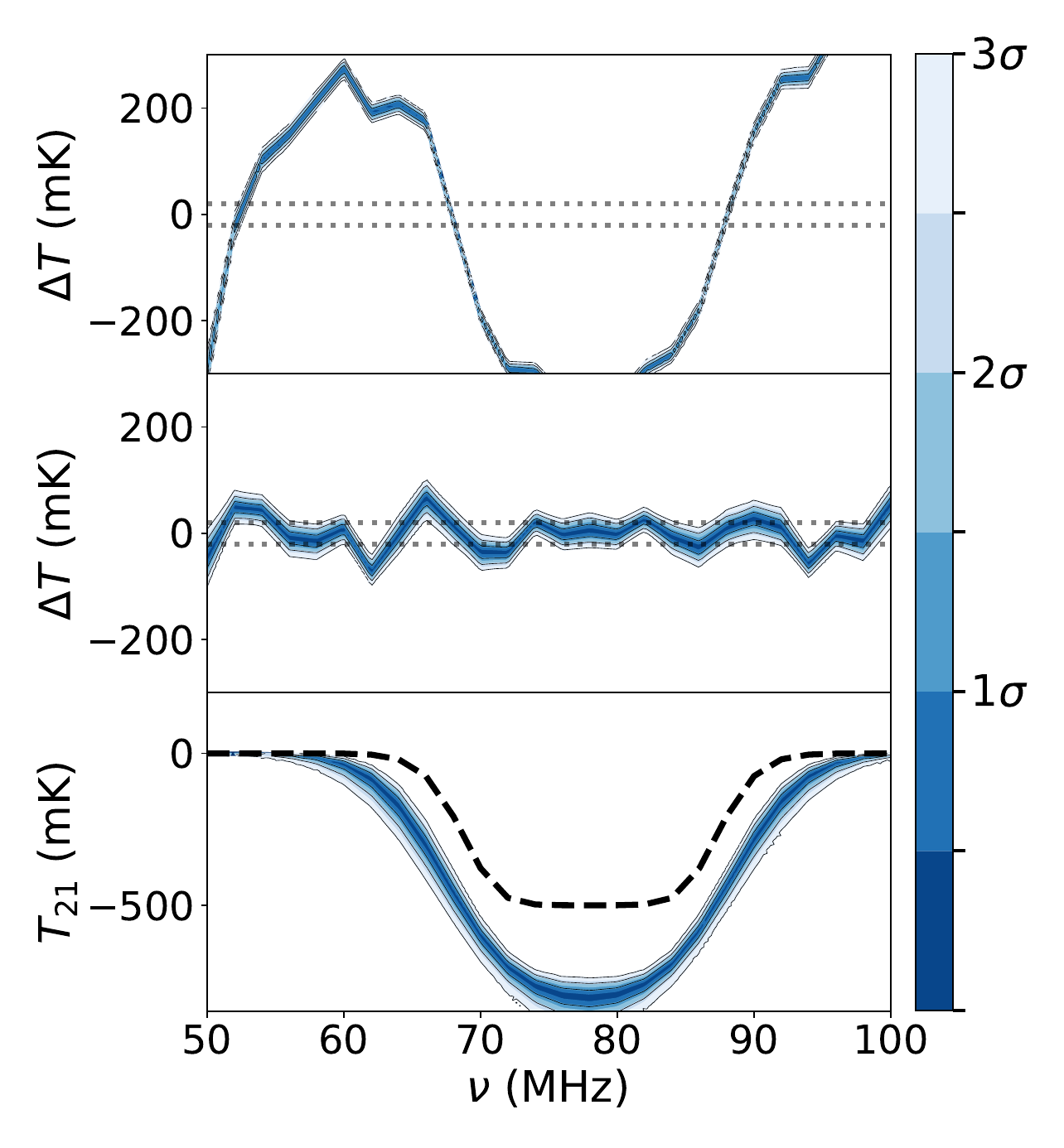}
        \end{subfigure}
        }
        \bluesubfigurebox{
        \begin{subfigure}[t]{0.33\textwidth}
            \caption{\Large{BFCC, $N=7$, PV}}
            \label{Fig:HighAmpBFCC7}
            \includegraphics[width=\textwidth]{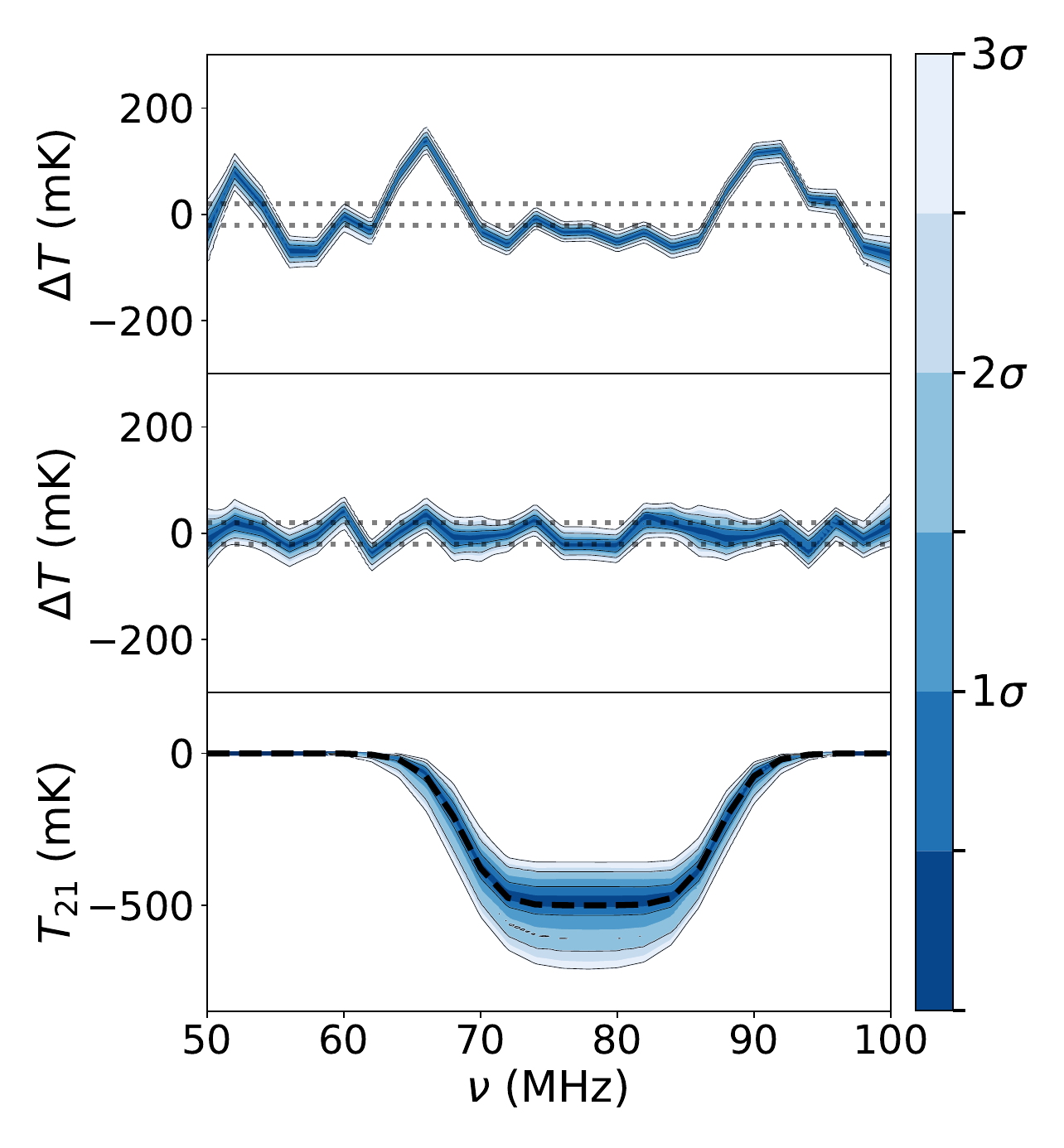}
        \end{subfigure}
        }
        \redsubfigurebox{
        \begin{subfigure}[t]{0.33\textwidth}
            \caption{\Large{Intrinsic, $N=3$, FV}}
            \label{Fig:HighAmpIntrinsic3}
            \includegraphics[width=\textwidth]{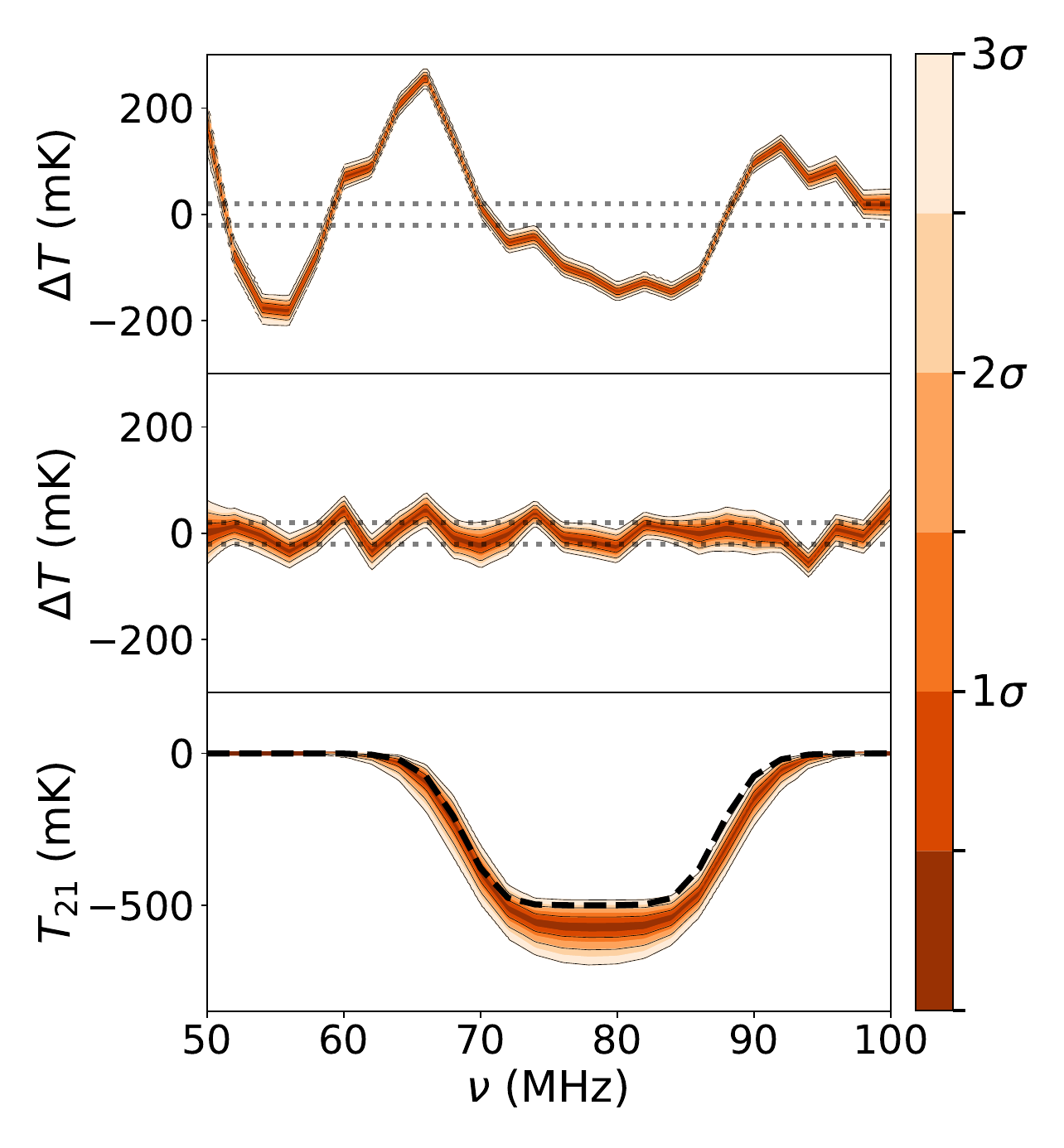}
        \end{subfigure}
        }
        \bluesubfigurebox{
        \begin{subfigure}[t]{0.33\textwidth}
            \caption{\Large{BFCC, $N=10$, PV}}
            \label{Fig:HighAmpBFCC10}
            \includegraphics[width=\textwidth]{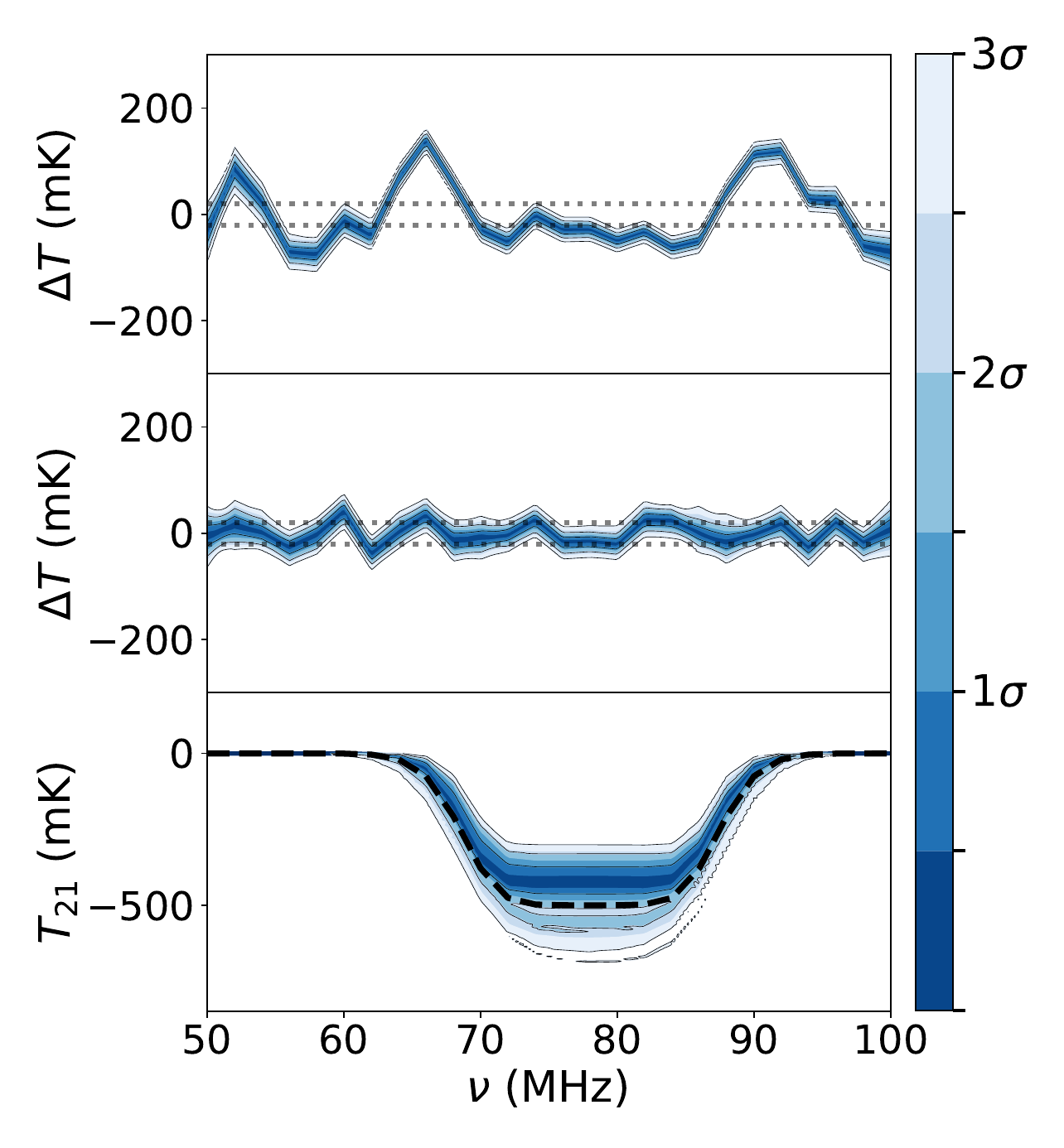}
        \end{subfigure}
        }
        \bluesubfigurebox{
        \begin{subfigure}[t]{0.33\textwidth}
            \caption{\Large{BFCC, $N=9$, PV}}
            \label{Fig:HighAmpBFCC9}
            \includegraphics[width=\textwidth]{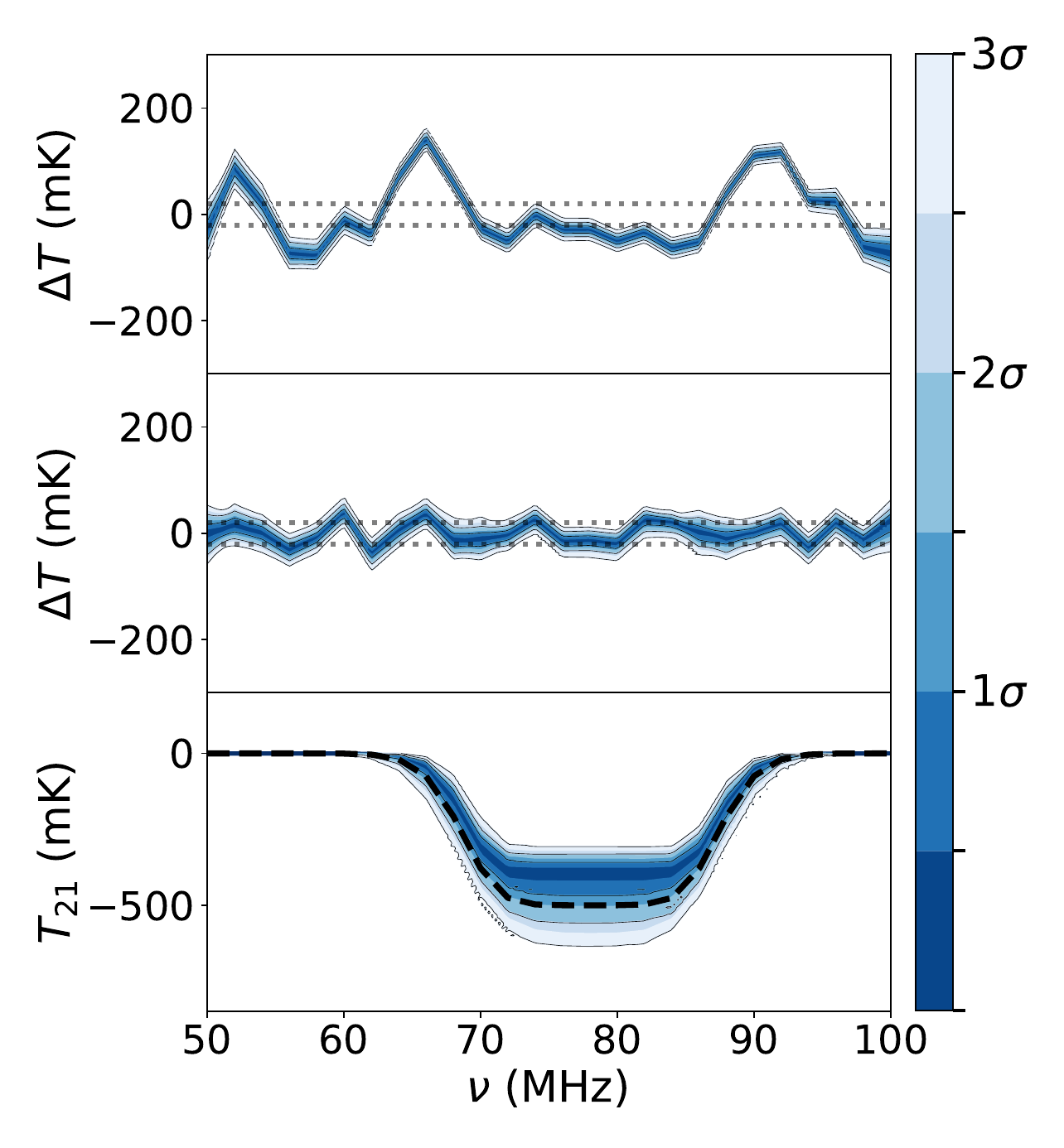}
        \end{subfigure}
        }
        \bluesubfigurebox{
        \begin{subfigure}[t]{0.33\textwidth}
            \caption{\Large{BFCC, $N=6$, PV}}
            \label{Fig:HighAmpBFCC6}
            \includegraphics[width=\textwidth]{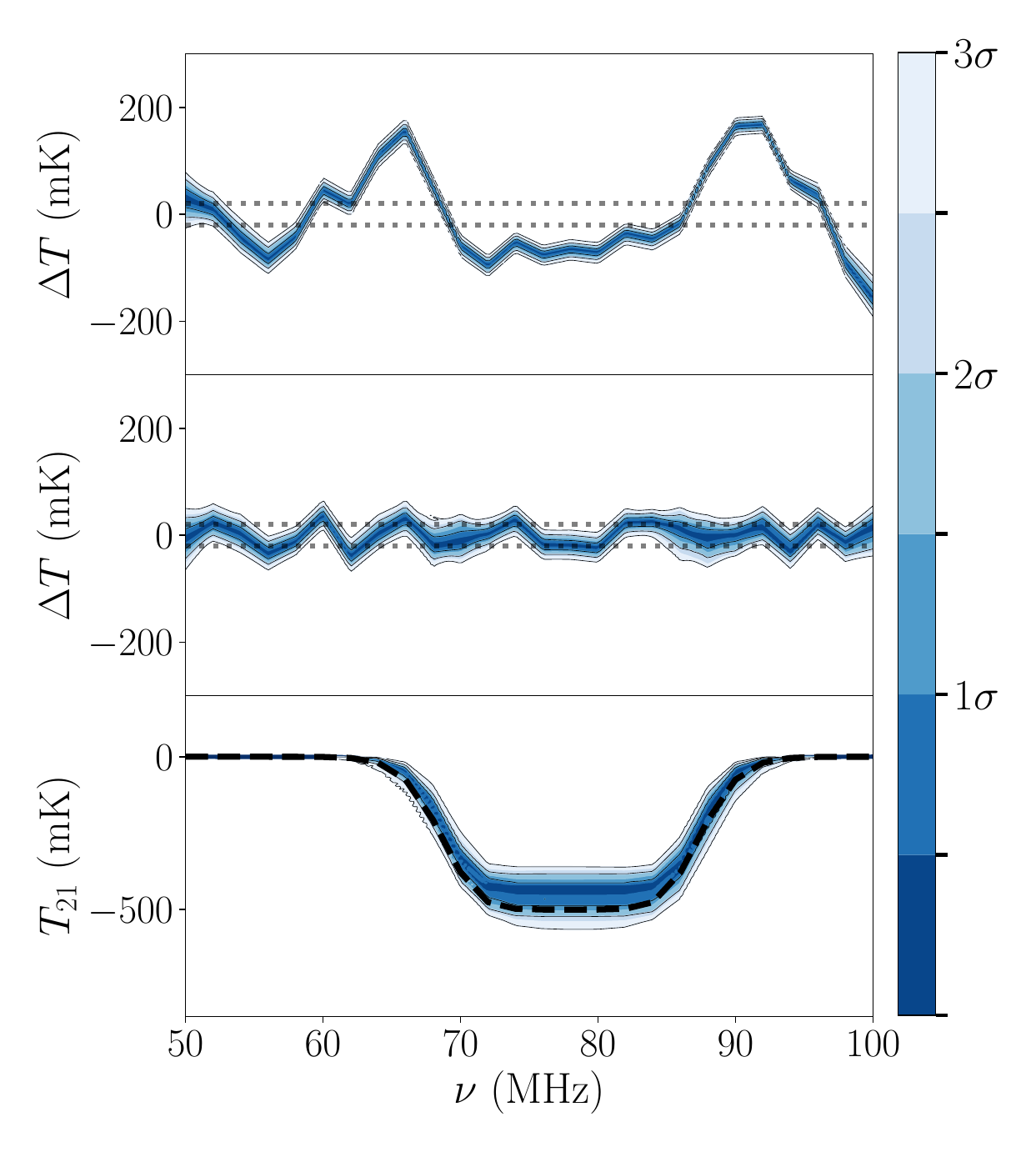}
        \end{subfigure}
        }
        \bluesubfigurebox{
        \begin{subfigure}[t]{0.33\textwidth}
            \caption{\Large{BFCC, $N=8$, PV}}
            \label{Fig:HighAmpBFCC8}
            \includegraphics[width=\textwidth]{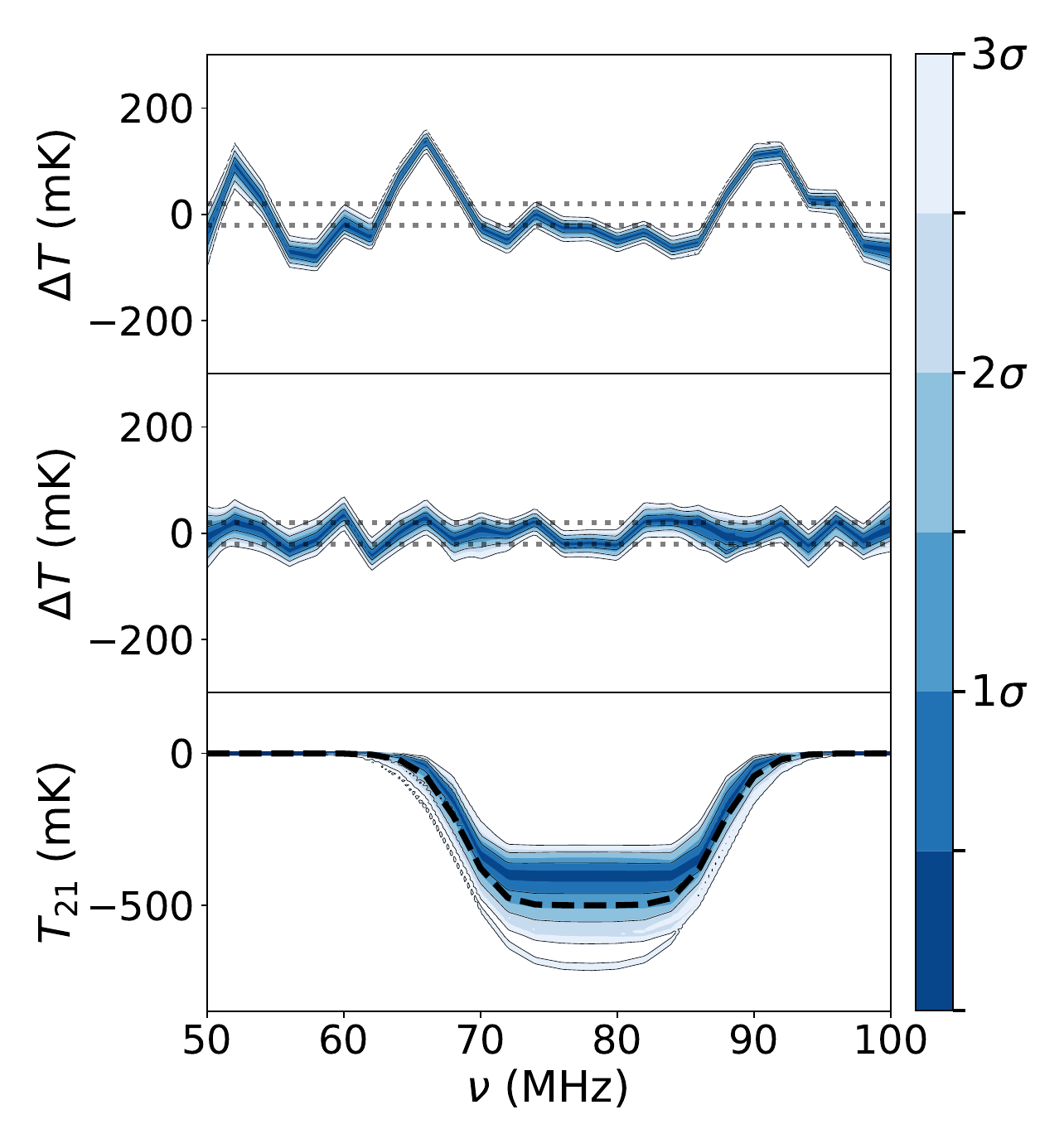}
        \end{subfigure}
        }
        \redsubfigurebox{
        \begin{subfigure}[t]{0.33\textwidth}
            \caption{\Large{\textbf{BFCC, $\mathbf{N=4}$, FV}}}
            \label{Fig:HighAmpBFCC4}
            \includegraphics[width=\textwidth]{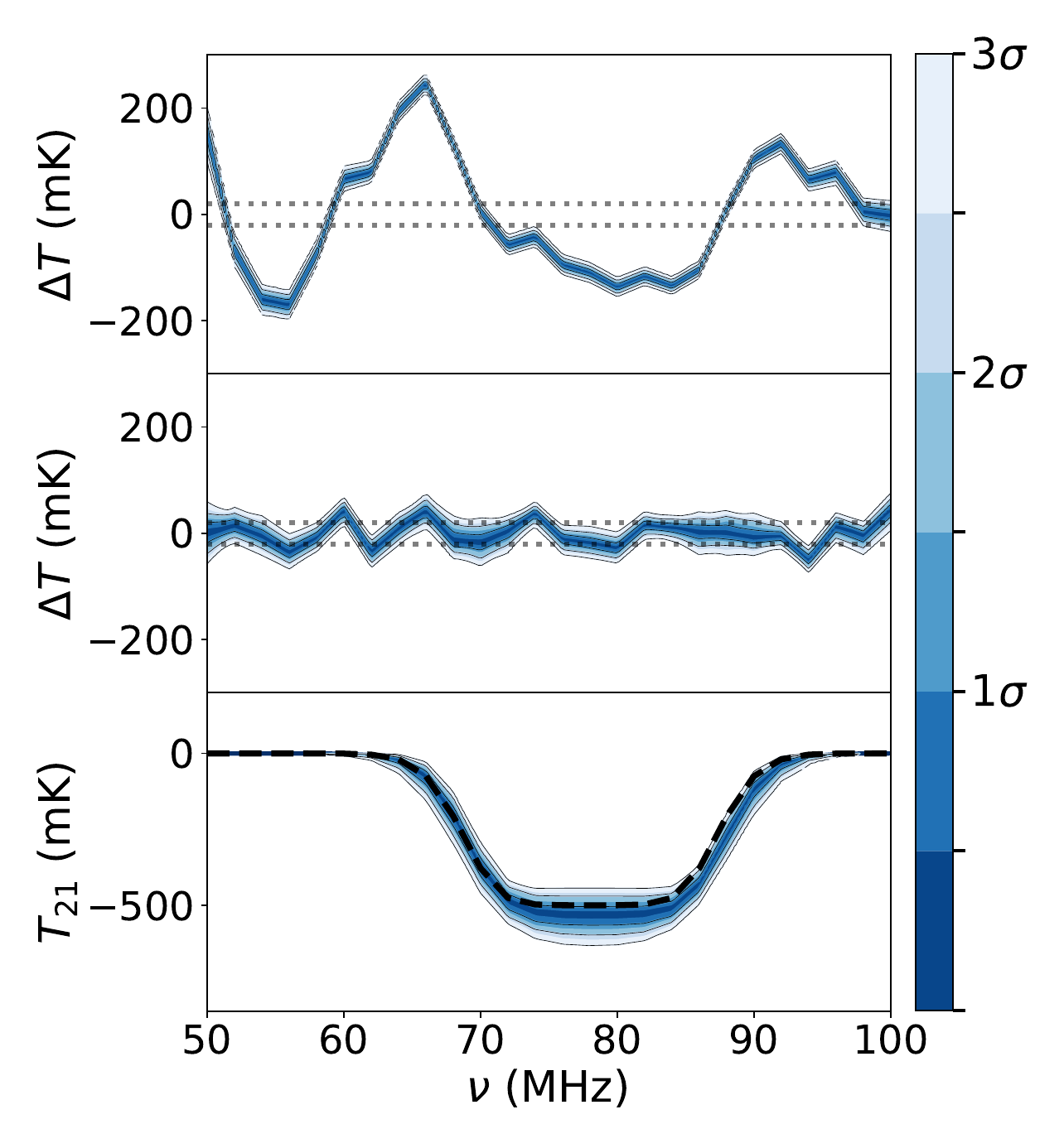}
        \end{subfigure}
        }
        \bluesubfigurebox{
        \begin{subfigure}[t]{0.33\textwidth}
            \caption{\Large{\textbf{\textit{BFCC, $\mathbfit{N=5}$, PV}}}}
            \label{Fig:HighAmpBFCC5}
            \includegraphics[width=\textwidth]{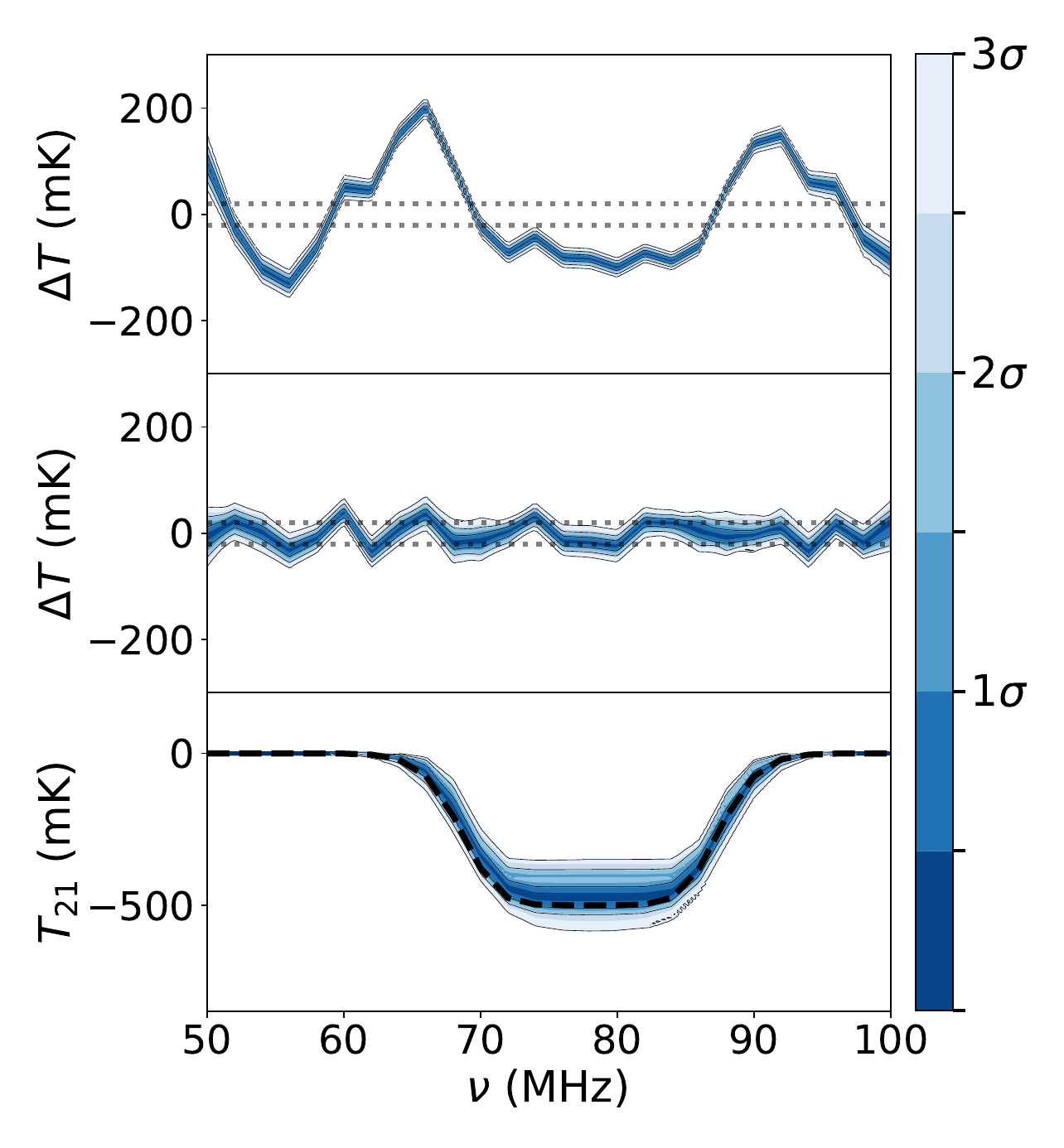}
        \end{subfigure}
        }
    \end{minipage}
    \contcaption{}
\end{figure*}

\section{Discussion}
\label{Sec:Discussion}

\subsection{Model efficacy}
\label{Sec:ModelEfficacy}

The primary goal of global 21-cm signal experiments is to obtain unbiased inferences about the redshifted 21-cm signal in the data. This can be subdivided into two distinct but related sub-goals:
\begin{enumerate*}
    \item the recovery of unbiased estimates of the 21-cm signal when it is present in the data, and
    \item the avoidance of spuriously detecting a 21-cm signal if none is present.
\end{enumerate*}

Our results in \Cref{Sec:Results} demonstrate broad agreement in the models that best achieve sub-goal (i) in the moderate- and high-amplitude 21-cm signal regimes. Specifically, we find that signal detections using the BFCC composite models with $N \ge 5$ and the MultLin composite models with $N \ge 6$ yield unbiased parameter inferences, with 95\% credibility HPDIs consistent with the parameters of the underlying 21-cm signals in the data. In contrast, the Intrinsic and LinPhys composite models, as well as lower-complexity BFCC and MultLin models, produce biased parameter inferences.

A similar distinction holds for models that best achieve sub-goal (ii) in our null-21-cm signal validation dataset. The only difference is that while the LinPhys composite model is preferred over the foreground model for foreground-only validation data, it is not sufficiently preferred for us to consider it a spurious detection of the 21-cm signal in the data ($\ln(\mathcal{B}_{\mathrm{cFg}}) \leq 3$). As such, it satisfies sub-goal (ii) despite failing sub-goal (i).

In a lower-noise observation than considered here, the LinPhys composite model may yield spurious 21-cm signal detection in the null-amplitude regime, failing both sub-goals (i) and (ii). Nevertheless, the fact that LinPhys only fails sub-goal (i) here indicates that the LinPhys foreground model provides only a mildly insufficient description at the $20~\mathrm{mK}$ RMS noise level considered in this work. Therefore, it should be expected to cause only moderate bias in 21-cm signal recovery with that model. This expectation was confirmed by the results in \Cref{Sec:ModerateAmplitude,Sec:HighAmplitude}.

Comparing the BFCC composite models with $N \ge 5$ to the MultLin composite models with $N \ge 6$, we find that while both satisfy sub-goals (i) and (ii), the BFCC models provide more precise parameter inferences. This result suggests the following hierarchy of model efficacy, from most to least effective:
\begin{enumerate}
    \item \textit{BFCC composite models with $N \ge 5$:} These models avoid spurious detection of a 21-cm signal in the null-amplitude regime and yield unbiased and relatively precise estimates of the 21-cm signal in both the moderate- and high-amplitude regimes. These are the only models considered in this work that meet both the model accuracy and constraining power requirements for unbiased recovery of the 21-cm signal in the moderate-amplitude regime.
    \item \textit{MultLin composite models with $N \ge 6$:} These models avoid spurious detection of a 21-cm signal when none is present. In the high-amplitude regime, they have sufficient constraining power to recover unbiased 21-cm signal estimates. However, they yield unbiased yet less precise estimates compared to the BFCC composite models. Nevertheless, they remain viable for global 21-cm cosmology and serve as a consistency check for the more stringent results obtained with BFCC composite models.
    \item \textit{The LinPhys composite model:} This model avoids spurious detection of a 21-cm signal when none is present, but yields biased estimates of the 21-cm signal in the moderate- and high-amplitude 21-cm signal regimes considered here. As such, this model is not recommended.
    \item \textit{The Intrinsic composite model, BFCC models with $N \le 4$, and MultLin models with $N \le 5$:} These models result in the spurious detection of a 21-cm signal when none is present and produce biased estimates of the 21-cm signal when one is present. They are therefore unsuitable for global 21-cm signal analysis.
\end{enumerate}

\subsection{Validated Bayesian model comparison}
\label{Sec:ValidatedBayesianModelComparison}

Our comparison in \Cref{Sec:Results} of the ground truth results to the conclusions drawn from BaNTER-validated Bayesian model comparison and Bayesian model comparison with uninformative model priors demonstrates the necessity of the former and insufficiency of the latter for deriving reliable inferences regarding the 21-cm signal in the null-, moderate-, and high-amplitude regimes.

Specifically, we find that Bayesian model comparison with uninformative model priors fails to uniquely identify the models that yield unbiased estimates of the 21-cm signal in the data across all amplitude regimes. In the null-amplitude regime, two-thirds of the highest-evidence models yield spurious detections of the 21-cm signal. In the moderate-amplitude regime, three-quarters of the highest-evidence models produce biased 21-cm signal recovery. This fraction improves in the high-amplitude regime, dropping to one-half of the highest-evidence models yielding biased recovery.

In contrast, the BaNTER null test effectively identifies and eliminates models that produce poor foreground model fits and/or spurious detections in the null-amplitude regime and biased estimates in both moderate- and high-amplitude regimes. Furthermore, the BaNTER-validated Bayesian model comparison framework assigns the highest posterior odds to the BFCC composite models that yield both accurate and precise estimates of the underlying 21-cm signal in the data. Finally, in both signal amplitude regimes, the remaining validated models used for 21-cm signal detection are also found to yield unbiased recovery of the underlying signal parameters, albeit generally with reduced precision compared to the highest posterior odds models.

\subsubsection{Accounting for imperfectly simulated data}
\label{Sec:ImperfectlySimulatedDataCaveats}

The results of the BaNTER-validated Bayesian model comparison framework applied to the simulated EDGES low-band datasets considered here are highly promising. However, their effectiveness when applied to observational data depends on the accuracy of the simulated data used in model validation. If systematic effects present in real observations are absent from simulations and have amplitudes significant relative to the noise level, certain models that should fail BaNTER validation in an ideal case may instead pass (see \Cref{Sec:PossibleSourcesOfUnmodelledSystematics} for a discussion of possible sources). Such models could then lead to spurious inferences in subsequent data analysis if they also provide an accurate description of the observational data while containing biased component models.

In such a scenario, by eliminating a subset of models that, while accurately describing the observational data, contain biased components, BaNTER validation still enhances the robustness of conclusions compared to those drawn from BFBMC alone. However, ultimately, the goal of validation is to ensure that the simulation is sufficiently accurate to identify \textit{all} models incapable of unbiased 21-cm signal recovery. To strengthen confidence in this criterion, BaNTER \textit{model validation} can be supplemented with additional \textit{simulation validation} tests to evaluate the accuracy of the validation dataset. S25 propose assessing consistency between the modelling complexity required to describe simulated observations and that needed for observational data as a potential approach. We plan to explore this approach further in future work.

\subsubsection{Extending BaNTER validation to other 21-cm cosmology models}
\label{Sec:GeneralApplicability}

In this paper, we have demonstrated the value of BaNTER validation in facilitating robust identification of models that can both accurately describe realistic simulated BFCC EDGES-low spectrometer data and recover unbiased estimates of the 21-cm signal from them. However, we anticipate the framework to be similarly useful for comparing alternative composite models across other datasets.

Analysing time-averaged spectrometer data with alternative models provides an additional potential use case in global 21-cm cosmology. For example, \citet{2021MNRAS.506.2041A} present a forward modelling analysis of spectrometer data, showing that the model enables reliable 21-cm signal detection in data from a relatively smooth conical log spiral antenna but not in data from a more chromatic conical sinuous antenna. By identifying models prone to biased recovery a priori, incorporating BaNTER validation into such analyses could yield benefits similar to those demonstrated for the BFCC and MultLin models in this work.

Fitting time-dependent or multi-instrument data enables one to leverage both angular and spectral information to distinguish between the foregrounds and the 21-cm signal (e.g. \citealt{2013PhRvD..87d3002L, 2020ApJ...897..175T, 2020ApJ...897..132T, 2023ApJ...959..103H, 2023MNRAS.522.1022S, 2023MNRAS.520..850A}), reducing the correlation between anisotropic foregrounds and isotropic 21-cm model components. However, accurately describing the data typically necessitates increased model complexity. Additionally, for a single instrument, the extent of correlation reduction between the foreground and 21-cm signal model components depends on the LST range and time-binning of the data\footnote{Correlation between the foreground and 21-cm signal model components is more likely when jointly modelling data over a shorter LST interval or using coarser time-binning.}.

In cases where time-dependent data modelling weakens but does not entirely eliminate the correlation between the 21-cm signal and other model components, BaNTER validation can be anticipated to similarly improve model selection, thereby facilitating unbiased 21-cm signal recovery.

Finally, BaNTER validation is not limited to composite spectrometer models. For example, the methodology demonstrated here could also be applied to validating Bayesian forward modelling approaches for interferometric 21-cm datasets, which incorporate foregrounds, 21-cm signal fluctuations about the mean, and potential systematics (e.g. \citealt{2006PhR...433..181F}). Potential applications include validating whether forward models possess sufficient accuracy for unbiased signal recovery amid foregrounds and instrumental systematics, particularly in interferometric calibration (e.g. \citealt{2021MNRAS.503.2457B,2022MNRAS.517..910S,2022MNRAS.517..935S}), instrumental modelling (e.g. \citealt{2024RASTI...3..400W}), and 21-cm signal recovery (e.g. \citealt{2016MNRAS.462.3069S,2019MNRAS.484.4152S,2019MNRAS.488.2904S,2023MNRAS.520.4443B,2024MNRAS.535..793B}). Additionally, joint analyses combining spectrometer, interferometric, and other data types could further benefit from BaNTER validation, enhancing model selection and ensuring unbiased 21-cm signal recovery across diverse observational scenarios.

\section{Summary \& Conclusions}
\label{Sec:Conclusions}

In Paper I of this series, we derived the physically motivated flexible-complexity BFCC model for spectrometer data post-processed to suppress instrumentally induced spectral structure using beam-factor-based chromaticity correction (e.g. B18). We demonstrated that the BFCC model, with complexity calibrated using BFBMC, enables unbiased recovery of a flattened Gaussian 21-cm signal consistent with the one reported by B18 from simulated data.

In this work, we applied the BFCC model to the analysis of realistic simulations of chromaticity-corrected EDGES-low spectrometer datasets, considering a broader range of scenarios regarding the 21-cm signal in the data. We analysed data containing 21-cm signals in three amplitude regimes: null ($A=0~\mathrm{mK}$), moderate ($A=150~\mathrm{mK}$), and high ($A=500~\mathrm{mK}$). Additionally, we extended the Bayesian comparison of the BFCC model to three competing model classes previously considered in the literature: the Intrinsic model used in \cite{2018Natur.564E..32H}, as well as the LinPhys model and an extended set of MultLin models applied to 21-cm signal estimation from EDGES data in B18.

By comparing 21-cm parameter posteriors recovered with competing models to the true 21-cm signal parameters in the data, we identify a broad agreement in models that enable unbiased parameter inferences. Our analysis reveals that only BFCC composite models with $N \ge 5$ and MultLin composite models with $N \ge 6$ avoid spurious detections and yield unbiased 21-cm signal estimates, with BFCC models providing superior precision. The complete model efficacy hierarchy is presented in \Cref{Sec:ModelEfficacy}.

Additionally, we investigated the extent to which Bayesian model comparison can identify the models that yield unbiased 21-cm signal estimates in the data. To address challenges arising from systematics that bias the 21-cm signal model fit while still maintaining an accurate fit to the data in aggregate, we employed the BaNTER validation framework introduced in S25. This framework uses a Bayesian null test to identify composite models that are likely to yield biased 21-cm signal estimates. We used BaNTER validation results to derive model priors and conduct a posterior-odds-based Bayesian comparison of the models.

By comparing models that enable unbiased inferences of the underlying 21-cm signal to conclusions drawn from BaNTER-validated posterior-odds-based model comparison and BFBMC alone, we found that the latter fails to reliably identify models yielding unbiased estimates across all amplitude regimes. Using BFBMC alone, we found that 2/3, 3/4, and 1/2 of the highest-evidence models led to spurious 21-cm signal detections or biased estimates in the null, moderate, and high amplitude regimes, respectively.

In contrast, BaNTER validation successfully identified and eliminated models that yield spurious detections in the null-amplitude regime and biased estimates in the moderate- and high-amplitude regimes. Furthermore, the BaNTER-validated posterior-odds-based model comparison framework assigns the highest posterior odds to BFCC composite models that simultaneously provide accurate and precise estimates of the underlying 21-cm signal in the simulated data. Finally, in both signal amplitude regimes, the remaining validated models used to detect the 21-cm signal also yield unbiased recovery of the underlying signal parameters (though generally with reduced precision compared to the highest posterior odds models).

We conclude that the BFCC model holds excellent promise for unbiased inference of the global 21-cm signal from spectrometer data, and we plan to test it on EDGES observations in future work. Moreover, Bayesian validation and model comparison methods, such as those discussed here, provide a powerful framework for identifying optimal models for global 21-cm data sets, ensuring robust signal recovery, and, ultimately, enabling detailed astrophysical insights into the radiative background and structure formation at Cosmic Dawn.

\section*{Acknowledgements}

This work was supported by the NSF through research awards for EDGES (AST-1813850, AST-1908933, and AST-2206766). PHS thanks Irina Stefan for valuable discussions and helpful comments on a draft of this manuscript. This analysis made use of a number of excellent, open-source software packages, including: \textsc{fgivenx} (\citealt{2018JOSS....3..849H}), \textsc{matplotlib} (\citealt{Hunter:2007}), \textsc{numpy} (\citealt{harris2020array}), \textsc{PolyChord} (\citealt{2015MNRAS.453.4384H, 2015MNRAS.450L..61H}) and \textsc{scipy} (\citealt{2020SciPy-NMeth}). EDGES is located at the Inyarrimanha Ilgari Bundara, the CSIRO Murchison Radio-astronomy Observatory. We acknowledge the Wajarri Yamatji people as the traditional owners of the Observatory site. We thank CSIRO for providing site infrastructure and support.

\section*{Data Availability}

The data from this study will be shared on reasonable request to the corresponding author.
Software used in this work to generate simulated data and beam-factors, given an electromagnetic simulation of the beam, is publicly available at \url{https://github.com/edges-collab}.



\bibliographystyle{mnras}
\bibliography{bibliography} 



\appendix


\section{21-cm signal HPD parameter estimates summary}
\label{Sec:HPDsummaryTable}

We summarise in \Cref{Tab:HPDsummaryTable} our 21-cm signal detection and parameter inference results described in \Cref{Sec:Results}.

\begin{table*}
\caption{
    Summary of 21-cm signal detection and parameter inference for the four models (BFCC, Intrinsic, LinPhys and MultLin) and three simulated 21-cm signal amplitude scenarios (21-cm signal null test, moderate amplitude 21-cm signal and high amplitude 21-cm signal). Models in which a 21-cm signal is detected are labelled with a checkmark. Detection of a signal in the 21-cm signal null test scenario corresponds to a failure of the validation null test (see \Cref{Sec:NullTest} for details). In both other scenarios, detection of a 21-cm signal is positive, while its non-detection is associated with significant correlation between the 21-cm signal and the non-21-cm component of the data model.
    For each data set, we list the Bayes factor ($\ln(\mathcal{B}_{i\mathrm{max}})$) between model $\bm{M}_{i}$ and $\bm{M}_\mathrm{max}$. Here, $i$ runs over the models in $\bm{\mathcal{M}}$ and $\bm{M}_\mathrm{max}$ is the highest Bayesian evidence model for the data.
    Additionally, we list the posterior odds ($\mathcal{R}_{i\mathrm{max}}$) between models $\bm{M}_{j}$ and $\bm{M}_\mathrm{max, v}$, where $j$ runs over the models in $\bm{\mathcal{M}}_\mathrm{v}$ and $\bm{M}_\mathrm{max, v}$ is the a posteriori most probable BaNTER validated model.
    We treat models which fail model validation as having negligible probability of facilitating unbiased estimates of the 21-cm signal a priori; these models have log-posterior-odds marked with a '-'. For those models with detected signals ($\ln(\mathcal{B}_{\mathrm{cFg}}) > 3.0$), HPD parameter estimates and uncertainties corresponding to the 95\% HPDI of the posterior distributions are quoted. The input parameters of the flattened Gaussian absorption troughs in the moderate- and high-amplitude signal models are $\nu_{0}=78~\mathrm{MHz}$, $w=19~\mathrm{MHz}$ and $\tau=8$, in both cases, and $A=0.15$ and $0.5~\mathrm{K}$ in the moderate- and high-amplitude signal cases, respectively. The names of models that simultaneously detect the 21-cm signal and recover signal parameters consistent with the underlying signal in the data are highlighted in italic. These are found to exclusively be elements of the BaNTER validated model set $\bm{\mathcal{M}}_\mathrm{v}$ (identifiable by their finite $\mathcal{R}_{i\mathrm{max}}$ values).
}
\begingroup
\renewcommand{\arraystretch}{1.5} 
\centerline{
    \begin{tabular}{l l c c c c c c c c}
    \hline

    Scenario & Model & 21-cm signal & $\ln(\mathcal{B}_{i\mathrm{max}})$ & $\ln(\mathcal{R}_{i\mathrm{max}})$ & $A\ (\mathrm{K})$ & $\nu_0\ (\mathrm{MHz})$ & $w\ (\mathrm{MHz})$ & $\tau$ & Consistent \\
    & & detection & & & & & & &  \\
   \hline
   Foreground-only & & & & & - & - & - & - &  \\
   validation data & & & & & & & & & \\
   \hline
      & BFCC $(N=3)$ & \cmark & -10.9 & -
      & $0.36^{+0.06}_{-0.05}$
      & $76.90^{+1.53}_{-1.70}$
      & $30.00^{+0.00}_{-1.39}$
      & $2.13^{+1.64}_{-1.17}$ & \xmark \\
      & BFCC $(N=4)$ & \cmark & 0.0 & -
      & $0.08^{+0.08}_{-0.05}$
      & $85.25^{+4.44}_{-4.03}$
      & $24.95^{+5.05}_{-6.31}$
      & $3.03^{+14.34}_{-3.03}$ & \xmark \\
      & BFCC $(N=5)$ & &  -1.6 & 0.0
      & & & & & \\
      & BFCC $(N=6)$ & &  -3.2 & -1.7
      & & & & & \\
      & BFCC $(N=7)$ & &  -3.4 & -1.8
      & & & & & \\
      & BFCC $(N=8)$ & &  -4.5 & -2.9
      & & & & & \\
      & BFCC $(N=9)$ & &  -4.0 & -2.4
      & & & & & \\
      & BFCC $(N=10)$ & &  -4.6 & -3.0
      & & & & & \\
      & Intrinsic $(N=3)$ & \cmark & -0.0 & -
      & $0.11^{+0.12}_{-0.04}$
      & $85.19^{+4.32}_{-3.54}$
      & $26.76^{+3.24}_{-6.23}$
      & $2.42^{+14.14}_{-2.42}$ & \xmark \\
      & LinPhys $(N=5)$ & &  -16.0 & -
      & & & & & \\
      & MultLin $(N=3)$ & \cmark & -4342.7 & -
      & $1.00^{+0.00}_{-0.00}$
      & $78.09^{+0.13}_{-0.15}$
      & $25.39^{+0.25}_{-0.28}$
      & $18.44^{+1.56}_{-2.11}$ & \xmark \\
      & MultLin $(N=4)$ & \cmark & -113.7 & -
      & $1.00^{+0.00}_{-0.04}$
      & $67.92^{+0.42}_{-0.40}$
      & $21.10^{+0.55}_{-0.67}$
      & $1.81^{+0.50}_{-0.56}$ & \xmark \\
      & MultLin $(N=5)$ & \cmark & -29.2 & -
      & $1.00^{+0.00}_{-0.43}$
      & $95.00^{+0.00}_{-39.99}$
      & $25.36^{+2.14}_{-5.88}$
      & $0.00^{+0.62}_{-0.00}$ & \xmark \\
      & MultLin $(N=6)$ & &  -23.4 & -21.8
      & & & & & \\
      & MultLin $(N=7)$ & &  -25.9 & -24.3
      & & & & & \\
      & MultLin $(N=8)$ & &  -29.8 & -28.2
      & & & & & \\
      & MultLin $(N=9)$ & &  -33.1 & -31.6
      & & & & & \\
      & MultLin $(N=10)$ & &  -36.0 & -34.4
      & & & & & \\
   \hline

   Moderate amplitude & & & & & 0.15 & 78.0 & 19.0 & 8.0 & \\
   21-cm signal & & & & & & & & & \\
   \hline
      & BFCC $(N=3)$ & \cmark & -0.3 & -
      & $0.65^{+0.12}_{-0.10}$
      & $76.74^{+0.93}_{-1.13}$
      & $30.00^{+0.00}_{-2.95}$
      & $0.72^{+0.59}_{-0.72}$ & \xmark \\
      & BFCC $(N=4)$ & \cmark & -0.0 & -
      & $0.27^{+0.14}_{-0.12}$
      & $79.30^{+1.35}_{-1.01}$
      & $20.12^{+6.35}_{-2.12}$
      & $0.81^{+5.84}_{-0.81}$ & \xmark \\
      & \textit{BFCC} $(N=5)$ & \cmark & -1.2 & 0.0
      & $0.10^{+0.19}_{-0.06}$
      & $78.25^{+1.35}_{-1.01}$
      & $17.85^{+10.67}_{-4.61}$
      & $0.61^{+14.95}_{-0.61}$ & \cmark \\
      & \textit{BFCC} $(N=6)$ & \cmark & -2.7 & -1.5
      & $0.11^{+0.10}_{-0.04}$
      & $77.75^{+1.99}_{-10.37}$
      & $18.65^{+11.34}_{-3.03}$
      & $9.90^{+10.10}_{-9.69}$ & \cmark \\
      & BFCC $(N=7)$ & &  -2.6 & -1.4
      & & & & & \\
      & BFCC $(N=8)$ & &  -2.6 & -1.4
      & & & & & \\
      & BFCC $(N=9)$ & &  -2.7 & -1.5
      & & & & & \\
      & BFCC $(N=10)$ & &  -1.9 & -0.8
      & & & & & \\
   \hline
    \end{tabular}
}
\label{Tab:HPDsummaryTable}
\endgroup
\end{table*}

\begin{table*}
\contcaption{}
\begingroup
\renewcommand{\arraystretch}{1.5} 
\centerline{
    \begin{tabular}{l l c c c c c c c c}
        \hline
        Scenario & Model & 21-cm signal & $\ln(\mathcal{B}_{i\mathrm{max}})$ & $\ln(\mathcal{R}_{i\mathrm{max}})$ & $A\ (\mathrm{K})$ & $\nu_0\ (\mathrm{MHz})$ & $w\ (\mathrm{MHz})$ & $\tau$ & Consistent \\
        & & detection & & & & & & &   \\
       \hline
       Moderate amplitude & & & & & 0.15 & 78.0 & 19.0 & 8.0 & \\
       21-cm signal & & & & & & & & & \\
       \hline
          & Intrinsic $(N=3)$ & \cmark & 0.0 & -
          & $0.31^{+0.32}_{-0.11}$
          & $79.41^{+1.10}_{-2.10}$
          & $22.46^{+7.54}_{-2.06}$
          & $0.00^{+3.01}_{-0.00}$ & \xmark \\
          & LinPhys $(N=5)$ & \cmark & -16.0 & -
          & $0.23^{+0.54}_{-0.11}$
          & $79.25^{+1.32}_{-1.65}$
          & $21.30^{+8.70}_{-1.96}$
          & $0.00^{+7.65}_{-0.00}$ & \xmark \\
          & MultLin $(N=3)$ & \cmark & -5120.2 & -
          & $1.00^{+0.00}_{-0.00}$
          & $78.14^{+0.12}_{-0.14}$
          & $24.37^{+0.25}_{-0.26}$
          & $19.72^{+0.28}_{-2.78}$ & \xmark \\
          & MultLin $(N=4)$ & \cmark & -108.7 & -
          & $1.00^{+0.00}_{-0.03}$
          & $69.21^{+0.45}_{-0.48}$
          & $22.39^{+0.52}_{-0.65}$
          & $2.19^{+0.50}_{-0.47}$ & \xmark \\
          & MultLin $(N=5)$ & \cmark & -20.9 & -
          & $0.11^{+0.23}_{-0.08}$
          & $59.45^{+2.02}_{-2.83}$
          & $9.56^{+12.11}_{-3.03}$
          & $16.77^{+3.23}_{-16.77}$ & \xmark \\
          & MultLin $(N=6)$ & &  -21.0 & -19.8
          & & & & & \\
          & MultLin $(N=7)$ & &  -23.8 & -22.6
          & & & & & \\
          & MultLin $(N=8)$ & &  -28.2 & -27.0
          & & & & & \\
          & MultLin $(N=9)$ & &  -31.1 & -29.9
          & & & & & \\
          & MultLin $(N=10)$ & &  -33.1 & -31.9
          & & & & & \\
       \hline
       High amplitude & & & & & 0.5 & 78.0 & 19.0 & 8.0 & \\
       21-cm signal & & & & & & & & & \\
       \hline
          & BFCC $(N=3)$ & \cmark & -17.3 & -
          & $0.82^{+0.06}_{-0.04}$
          & $77.86^{+0.26}_{-0.27}$
          & $21.11^{+0.72}_{-0.65}$
          & $2.44^{+0.99}_{-0.88}$ & \xmark \\
          & BFCC $(N=4)$ & \cmark & -1.9 & -
          & $0.54^{+0.06}_{-0.06}$
          & $78.40^{+0.26}_{-0.29}$
          & $19.35^{+0.69}_{-0.62}$
          & $6.22^{+3.96}_{-2.34}$ & \xmark \\
          & \textit{BFCC} $(N=5)$ & \cmark & 0.0 & 0.0
          & $0.48^{+0.05}_{-0.09}$
          & $78.23^{+0.26}_{-0.24}$
          & $19.06^{+0.61}_{-0.77}$
          & $8.14^{+9.18}_{-3.71}$ & \cmark \\
          & \textit{BFCC} $(N=6)$ & \cmark & -3.0 & -3.0
          & $0.45^{+0.08}_{-0.05}$
          & $78.21^{+0.25}_{-0.50}$
          & $19.03^{+0.62}_{-0.85}$
          & $9.73^{+8.35}_{-4.18}$ & \cmark \\
          & \textit{BFCC} $(N=7)$ & \cmark & -3.7 & -3.7
          & $0.41^{+0.22}_{-0.04}$
          & $78.15^{+0.34}_{-0.64}$
          & $18.95^{+0.62}_{-0.54}$
          & $13.83^{+1.98}_{-9.88}$ & \cmark \\
          & \textit{BFCC} $(N=8)$ & \cmark & -3.0 & -3.0
          & $0.40^{+0.16}_{-0.05}$
          & $77.97^{+0.42}_{-0.34}$
          & $18.75^{+0.85}_{-0.34}$
          & $13.27^{+5.13}_{-8.31}$ & \cmark \\
          & \textit{BFCC} $(N=9)$ & \cmark & -3.2 & -3.2
          & $0.40^{+0.16}_{-0.05}$
          & $78.03^{+0.39}_{-0.32}$
          & $19.03^{+0.47}_{-0.77}$
          & $7.59^{+9.16}_{-2.52}$ & \cmark \\
          & \textit{BFCC} $(N=10)$ & \cmark & -3.6 & -3.6
          & $0.43^{+0.14}_{-0.09}$
          & $77.94^{+0.43}_{-0.30}$
          & $18.90^{+0.68}_{-0.72}$
          & $11.84^{+3.44}_{-6.71}$ & \cmark \\
          & Intrinsic $(N=3)$ & \cmark & -3.6 & -
          & $0.58^{+0.06}_{-0.07}$
          & $78.50^{+0.26}_{-0.27}$
          & $19.49^{+0.70}_{-0.63}$
          & $5.14^{+3.11}_{-1.83}$ & \xmark \\
          & LinPhys $(N=5)$ & \cmark & -17.4 & -
          & $0.52^{+0.07}_{-0.06}$
          & $78.34^{+0.27}_{-0.27}$
          & $19.32^{+0.62}_{-0.66}$
          & $5.96^{+5.43}_{-2.17}$ & \xmark \\
          & MultLin $(N=3)$ & \cmark & -7418.5 & -
          & $1.00^{+0.00}_{-0.00}$
          & $78.20^{+0.12}_{-0.10}$
          & $22.57^{+0.22}_{-0.23}$
          & $20.00^{+0.00}_{-0.88}$ & \xmark \\
          & MultLin $(N=4)$ & \cmark & -179.0 & -
          & $1.00^{+0.00}_{-0.01}$
          & $71.98^{+0.42}_{-0.47}$
          & $22.70^{+0.58}_{-0.65}$
          & $2.43^{+0.47}_{-0.41}$ & \xmark \\
          & MultLin $(N=5)$ & \cmark & -25.1 & -
          & $0.27^{+0.04}_{-0.04}$
          & $78.25^{+0.44}_{-0.47}$
          & $18.85^{+0.89}_{-0.80}$
          & $20.00^{+0.00}_{-8.64}$ & \xmark \\
          & \textit{MultLin} $(N=6)$ & \cmark & -22.7 & -22.7
          & $0.43^{+0.50}_{-0.11}$
          & $78.25^{+0.37}_{-0.35}$
          & $19.10^{+0.74}_{-0.74}$
          & $4.33^{+13.05}_{-2.05}$ & \cmark \\
          & \textit{MultLin} $(N=7)$ & \cmark & -25.7 & -25.7
          & $0.41^{+0.44}_{-0.11}$
          & $78.10^{+0.42}_{-0.44}$
          & $19.02^{+0.72}_{-0.72}$
          & $4.83^{+15.17}_{-2.01}$ & \cmark \\
          & \textit{MultLin} $(N=8)$ & \cmark & -30.0 & -30.0
          & $0.42^{+0.46}_{-0.12}$
          & $78.08^{+0.46}_{-0.46}$
          & $18.92^{+0.82}_{-0.66}$
          & $4.59^{+13.60}_{-1.99}$ & \cmark \\
          & \textit{MultLin} $(N=9)$ & \cmark & -32.3 & -32.3
          & $0.40^{+0.46}_{-0.12}$
          & $78.09^{+0.47}_{-0.44}$
          & $19.11^{+0.67}_{-0.78}$
          & $4.74^{+15.26}_{-1.65}$ & \cmark \\
          & \textit{MultLin} $(N=10)$ & \cmark & -34.7 & -34.7
          & $0.37^{+0.30}_{-0.12}$
          & $78.12^{+0.47}_{-0.45}$
          & $19.37^{+0.77}_{-0.86}$
          & $18.38^{+1.62}_{-13.34}$ & \cmark \\
       \hline
    \end{tabular}
}
\endgroup
\end{table*}

\section{Model comparison categorisation}
\label{Sec:ModelComparisonCategorisationResults}

In \Cref{Sec:BaNTERValidationResults}, we found that several models failed the BaNTER null test. These include the Intrinsic model, the LinPhys model, BFCC variants with $N = 3$ or $4$, and MultLin variants with $N = 3$, $4$, or $5$ foreground terms.

Composite models that fail BaNTER validation cannot be considered credible for 21-cm cosmology, as they are prone to yielding spurious detections or biased estimates of the 21-cm signal (if present). However, the extent to which their inclusion in the set of considered models biases conclusions from BFBMC-alone depends on their Bayesian evidence relative to accurate and predictive composite models with accurate and predictive components (see S25 for details).

When the Bayesian evidence of failed models is substantially lower than that of the highest-evidence models, BFBMC naturally downweights these erroneous models. Consequently, their inclusion does not meaningfully influence Bayesian model-averaged conclusions or the selection of the most probable model. This corresponds to the \textit{category I} model comparison scenario defined in S25.

In contrast, the inclusion of models that fail the BaNTER null test yet have Bayesian evidence comparable to that of the highest-evidence models is more problematic. Because these models pass Bayesian selection criteria despite failing BaNTER validation, they introduce systematic biases that cannot be corrected by BFBMC alone. In the absence of model validation, their inclusion in the model set risks significantly biasing conclusions. This corresponds to the \textit{category II} model comparison scenario defined in S25.

\subsection{Moderate amplitude 21-cm signal}
\label{Sec:MCCModerateAmplitude}

In the moderate amplitude analysis in \Cref{Sec:ModerateAmplitude}, we determined that, among the models that detect the 21-cm signal, only the BFCC composite models with $N = 5$ and $6$ yield unbiased recovery of the 21-cm signal parameters (see \Cref{Fig:SignalRecoveryAllDetections150mK}). From \Cref{Fig:BandRwithAeq150mK} (or \Cref{Tab:lnBcb}), it can be seen that, among the models that failed BaNTER validation, only the LinPhys model and BFCC models with $N = 3$ and $4$ have Bayesian evidence comparable to that of the highest posterior odds model, the BFCC model with $N = 5$.

Thus, including the LinPhys model and MultLin models with $N = 3$, $4$, or $5$ -- all of which are decisively disfavoured relative to the BFCC model with $N = 5$ -- does not significantly bias BFBMC conclusions. If only this subset of models is included alongside accurate and predictive composite models with accurate and predictive components in $\bm{\mathcal{M}}$, Bayesian comparison of these models would constitute a \textit{category I} model comparison, for which BFBMC is sufficient.

In contrast, the inclusion of the Intrinsic model or the BFCC model with $N = 4$ or $5$ in the set of models under consideration means that, in the absence of model validation, BFBMC applied to the moderate amplitude data will yield biased 21-cm inferences. This represents a \textit{category II} model comparison problem, for which BaNTER validation is essential for unbiased recovery of the 21-cm signal.

\subsection{High amplitude 21-cm signal}
\label{Sec:MCCHighAmplitude}

In the high amplitude analysis in \Cref{Sec:HighAmplitude}, we determined that the BFCC composite models with $N \geq 5$ and MultLin models with $N \geq 6$ yield unbiased recovery of the 21-cm signal parameters (see \Cref{Fig:SignalRecoveryAllDetections500mKPt1}). From \Cref{Fig:BandRwithAeq500mK}, it can be seen that, among the models that failed BaNTER validation (see \Cref{Tab:lnBcb}), only the BFCC model with $N = 4$ has Bayesian evidence comparable to that of the highest posterior odds model, the BFCC model with $N = 5$.

Thus, including the BFCC model with $N = 3$, the Intrinsic model, the LinPhys model, and MultLin models with $N = 3$, $4$, or $5$ -- all of which are strongly or decisively disfavoured relative to the BFCC model with $N = 5$ -- does not significantly bias BFBMC conclusions. If only this subset of models were included alongside accurately predictive composite models in $\bm{\mathcal{M}}$, Bayesian comparison of these models would constitute a \textit{category I} model comparison, for which BFBMC is sufficient.

In contrast, the inclusion of the BFCC model with $N = 4$ in the set of models under consideration means that, in the absence of model validation, BFBMC applied to the high amplitude data will yield biased 21-cm inferences. This represents a \textit{category II} model comparison problem, for which BaNTER validation is essential for unbiased recovery of the 21-cm signal.

\section{S25 accuracy condition}
\label{Sec:AccuracyCondition}

Following the model-validated Bayesian inference workflow introduced in S25 we apply the Bayesian null-test described in \Cref{Sec:BayesianNullTest} a priori and use the results in combination with the relative evidences of the models to derive the posterior odds we ascribe to the models using \Cref{Eq:PrMiGivenD}. This approach yields a validated set of models that, of the set of models under consideration, are a posteriori most probable for recovering unbiased estimates of the global 21-cm signal in the data.

However, edge-case possibility remains that despite being the most probable models of those under consideration these models nevertheless provide insufficiently accurate descriptions of the data to be credible models for unbiased recovery of the component signals. To ensure that this is not the case, we test each of these models using the absolute accuracy condition introduced in S25.

Given the dataset $\bm{D}$, the noise covariance matrix $\mathbfss{N}$, and composite model $\bm{M}_{i\mathrm{c}}$, we define the median a posteriori likelihood of $\bm{M}_{i\mathrm{c}}$ as $\ln(\overline{\mathcal{L}}_{i})$. Writing the data likelihood $\mathcal{L}(\bm{r}_{i\mathrm{c}}(\sTheta_{i\mathrm{c}}))$ as a function of the residual vector, $\bm{r}_{i\mathrm{c}}(\sTheta_{i\mathrm{c}}) = [\bm{D} - \bm{M}_{i\mathrm{c}}(\sTheta_{i\mathrm{c}})]$ (see \Cref{Sec:DataLikelihood}), the likelihood distribution for an ideal model (one that describes the data perfectly, excluding noise) can be sampled by substituting $\bm{r}_{i\mathrm{c}}(\sTheta_{i\mathrm{c}})$ in the likelihood expression with noise realizations drawn from the covariance matrix $\mathbfss{N}$. We denote this ideal model likelihood distribution as $\mathcal{L}_\mathrm{noise}$.
We define the model's accuracy parameter as the logarithm of the ratio of the median a posteriori likelihood of $\bm{M}_{i\mathrm{c}}$ to the ideal model likelihood distribution:
\begin{equation}
    \label{Eq:lambdaLR}
    \lambda_{i} = \ln\left( \frac{\overline{\mathcal{L}}_{i}}{\mathcal{L}_\mathrm{noise}} \right) \ .
\end{equation}
When the distribution of $\lambda_{i}$ is consistent with zero, this implies $\overline{\mathcal{L}}_{i}$ is comparable to typical values of $\mathcal{L}_\mathrm{noise}$ and $\bm{M}_{i\mathrm{c}}$ is accurate. In contrast, when most of the probability mass of $\lambda_{i}$ is negative, $\bm{M}_{i\mathrm{c}}$ is comparably inaccurate.

Qualitatively, we define an accurate composite model as one with a fit likelihood that is credibly drawn from the ideal model likelihood distribution. Quantitatively, we classify $\bm{M}_{i\mathrm{c}}$ as accurate if it satisfies the following generalised accuracy condition:
\begin{equation}
    \label{Eq:AccuracyCondition}
    Q_{q_\mathrm{threshold}}(\lambda_{i}) \ge 0 \ .
\end{equation}
Here, $Q(.)$ is the quantile (or inverse cumulative distribution) function, defined such that for a random variable $X$, $Q_q(X)$ is the value of $x$ such that $P(X \leq x) = q$. The closer $q_\mathrm{threshold}$ is to unity, the further $\overline{\mathcal{L}}_i$ can fall towards the lower end of the $\mathcal{L}_\mathrm{noise}$ distribution while still being classified as an element of $C$. In this work, we use $q_\mathrm{threshold} = 0.999$ meaning that $\bm{M}_{i\mathrm{c}}$ will fail the accuracy condition if its median posterior likelihood is exceeded by 99.9\% of the probability mass of the $\mathcal{L}_\mathrm{noise}$ distribution.

\section{Sources of possible unmodelled systematics}
\label{Sec:PossibleSourcesOfUnmodelledSystematics}

Examples of potential sources of systematic effects that in this work are approximated as being sufficiently small to be neglected without impacting 21-cm signal inference include:
\begin{itemize}
    \item uncertainties in the antenna beam and foreground models (e.g. \citealt{2013PhRvD..87d3002L,2018ApJ...853..187T,2020ApJ...897..175T,2020ApJ...897..174R,2020ApJ...905..113H,2021MNRAS.506.2041A,2021ApJ...923...33B,2021MNRAS.503..344S,2021AJ....162...38M,2022RaSc...5707558R,2022MNRAS.515.1580S,2023ApJ...959..103H,2024MNRAS.527.5649P,2024MNRAS.527.2413P,2024ApJ...961...56M,2024ApJ...974..137A}), which may impact the effectiveness of instrumental chromaticity correction;
    \item receiver calibration uncertainties (e.g. \citealt{2017ApJ...835...49M,2021MNRAS.505.2638R,2021ApJ...915...66T,2022MNRAS.517.2264M,2024arXiv241214023K,2025ExA....59....7R});
    \item antenna\footnote{We include here resistive losses in antenna panels, the balun, and connectors.} and/or ground\footnote{Resulting from partial absorption by the ground of radio emission visible to the antenna due to its non-zero beam directivity below the horizon.} loss correction uncertainties (e.g. \citealt{2017ApJ...847...64M, 2024ApJ...961...56M});
    \item spectral chromaticity induced by ionospheric effects beyond those captured by a static, isotropic, time-averaged ionospheric model (e.g. \citealt{2014MNRAS.437.1056V,2016ApJ...831....6D,2021MNRAS.503..344S});
    \item polarised foreground emission (e.g. \citealt{2019MNRAS.489.4007S}).
\end{itemize}


\bsp	
\label{lastpage}
\end{document}